\documentstyle[12pt,epsf]{article}

\newcommand{\reseteqnum}{\setcounter{equation}{0}}

\newcommand{\ds}{\displaystyle}
\newcommand{\ba}{\begin{array}}
\newcommand{\ea}{\end{array}}
\newcommand{\nn}{\nonumber \\}
\newcommand{\mm}{\medskip\\}
\newcommand{\be}{\begin{equation}}
\newcommand{\ee}{\end{equation}}
\newcommand{\bea}{\begin{eqnarray}}
\newcommand{\eea}{\end{eqnarray}}
\newcommand{\beann}{\begin{eqnarray*}}
\newcommand{\eeann}{\end{eqnarray*}}
\newcommand{\bd}{\begin{description}}
\newcommand{\ed}{\end{description}}
\newcommand{\eq}[1]{eq.~(\ref{eq: #1})}
\newcommand{\eqs}[1]{eqs.~(\ref{eq: #1})}
\newcommand{\num}[1]{(\ref{eq: #1})}

\newcommand{\vev}[1]{\langle0\vert #1 \vert0\rangle}

\newcommand{\vevqq}{\vev{\overline qq}}
\newcommand{\bra}[1]{\langle #1 \vert}
\newcommand{\ket}[1]{\vert #1 \rangle}
\newcommand{\Bra}[1]{\Big\langle #1 \Big\vert}
\newcommand{\Ket}[1]{\Big\vert #1 \Big\rangle}
\newcommand{\bbra}[1]{\langle\kern-3.5pt\langle #1 \vert}
\newcommand{\kket}[1]{\vert #1 \rangle\kern-3.5pt\rangle}
\newcommand{\iprod}[2]{\langle #1 \vert #2 \rangle}
\newcommand{\iiprod}[2]{\langle\kern-3.5pt\langle #1 \vert #2
 \rangle\kern-3.5pt\rangle}

\renewcommand{\slash}[2]{#1\kern-#2pt\mbox{\it/}}
\newcommand{\sle}{\slash{\epsilon}{5.2}}
\newcommand{\slp}{\slash{p}{6.0}}
\newcommand{\slbp}{\slash{\mbox{\boldmath $p$}}{6.0}}
\newcommand{\slq}{\slash{q}{5.8}}
\newcommand{\slv}{\slash{v}{5.8}}
\newcommand{\slvp}{\slv\,{}'}
\newcommand{\slG}{(\epsilon \cdot G)}
\newcommand{\slqh}{\slash{\widehat{q}}{5.8}}
\newcommand{\Sp}[1]{\frac{1}{4}{\rm tr}\left[\: #1 \:\right]}
\newcommand{\gf}{\gamma_5}
\newcommand{\gmu}{\gamma^\mu}
\newcommand{\gmd}{\gamma_\mu}
\newcommand{\gnu}{\gamma^\nu}

\newcommand{\tr}{{\rm tr}}
\newcommand{\trace}[1]{{\rm tr}\left[\, #1 \,\right]}
\newcommand{\Lpm}{\Lambda_\pm}
\newcommand{\Lmp}{\Lambda_\mp}
\newcommand{\Lp}{\Lambda_+}
\newcommand{\Lm}{\Lambda_-}

\newcommand{\ta}{\frac{\tau^a}{2}}
\newcommand{\tb}{\frac{\tau^b}{2}}
\newcommand{\GVp}{\Gamma_\mu^a(p;q)}

\newcommand{\GAp}{\Gamma_{5\mu}^a(p;q)}
\newcommand{\GAk}{\Gamma_{5\mu}^a(k;q)}

\newcommand{\bp}{{\mbox{\boldmath $p$}}}
\newcommand{\bkk}{{\mbox{\boldmath $k$}}}
\newcommand{\bq}{{\mbox{\boldmath $q$}}}
\newcommand{\whq}{\widehat q}
\newcommand{\hq}{v}
\newcommand{\lqcd}{\Lambda_{\rm QCD}}
\newcommand{\oh}{{\scriptstyle\frac{1}{2}}}
\newcommand{\SI}{iS_F^{-1}}
\newcommand{\N}{{\cal N}}
\newcommand{\SS}{{\cal S}}
\newcommand{\PP}{{\cal P}}
\newcommand{\QQ}{{\cal Q}}
\newcommand{\RR}{{\cal R}}
\newcommand{\pdk}{(p\cdot k)_{_E}}
\newcommand{\pck}{(p\times k)_{_E}}
\newcommand{\wh}{\widehat}
\newcommand{\bB}{\vert B(q)\rangle}
\newcommand{\kB}{\langle B(q)\vert}
\newcommand{\T}{{\rm T}}
\newcommand{\llra}{\longleftrightarrow}
\newcommand{\mom}{\bigg(\frac{m}{M}\bigg)}
\newcommand{\mi}{{(-1)}}
\newcommand{\ze}{{(0)}}
\newcommand{\on}{{(1)}}

\newcommand{\p}{{(+)}}

\newcommand{\pv}{p\!\cdot\!\hq}
\newcommand{\vpv}{v'\!\cdot\! v}
\newcommand{\Vcb}{|V_{\rm cb}|}
\newcommand{\KM}[1]{V_{\rm #1}}
\newcommand{\mmcg}{p'\!-\!\eta' q'\!=\!p\!-\!\eta q}
\newcommand{\mmc}{p'\!-\!\eta q'\!=\!p\!-\!\eta q}
\newcommand{\mmcl}{p'\!-\!mv'\!=\!p\!-\!mv}
\newcommand{\ct}{\cos\theta}
\newcommand{\spc}{\rule[-4pt]{0mm}{16pt}}
\renewcommand{\c}{\!\cdot\!}
\newcommand{\iru}{\lambda_U}
\newcommand{\irx}{\lambda_X}
\newcommand{\uvu}{\Lambda_U}
\newcommand{\uvx}{\Lambda_X}
\newcommand{\vma}{\Pi_{VV}(q^2)-\Pi_{AA}(q^2)}

\newcommand{\intdk}{\int\!\!\frac{d^4k}{(2\pi)^4i}}
\newcommand{\intdp}{\int\!\!\frac{d^4p}{(2\pi)^4i}}
\newcommand{\intdpp}{\int\!\!\frac{d^4p'}{(2\pi)^4i}}
\newcommand{\intdt}{\int_0^\pi\!\!d\theta\sin^2\theta}

\begin{document}

\begin{titlepage}

% Preprint Numbers
\begin{flushright}
KUNS-1350 \\
HE(TH)~95/10 \\
hep-ph/9506294 \\
%\today
\end{flushright}

% Title
\vspace*{1cm}
\begin{center} \LARGE
Chiral Symmetry, Heavy Quark Symmetry
\end{center}
\begin{center} \LARGE
 and
\end{center}
\begin{center} \LARGE
Bound States
\footnote{ Doctoral thesis }
 \\
\end{center}
\bigskip
\bigskip

% Author(s)
\begin{center} \Large
Yuhsuke Yoshida%
%\footnote{e-mail address :
%{\tt yoshida@gauge.scphys.kyoto-u.ac.jp}} \\
\end{center}
\bigskip

% Address
\begin{center} \large \it
Department of Physics, Kyoto University \\
Kyoto 606-01, Japan \\
\end{center}

% Date
\begin{center}
January 5, 1995
\end{center}
\bigskip

%\smallskip
\begin{center} \Large \bf
Abstract
\end{center}

% Abstract
I investigate the bound state problems of lowest-lying mesons and
heavy mesons.
Chiral symmetry is essential when one consider lowest-lying mesons.
Heavy quark symmetry plays an central role in considering the
semi-leptonic form factors of heavy mesons.
Various properties based on the symmetries are revealed using
Bethe-Salpeter equations.
\end{titlepage}

\tableofcontents

\newpage
\section{Introductions}
\reseteqnum

\begin{flushleft}
\underline{ Chiral Symmetry }
\end{flushleft}

The masses of $u$ and $d$ quarks are sufficiently smaller than the
typical interaction scale $\lqcd$ of the strong interaction.
It is a good picture that the system possesses the $SU(2)_L\times
SU(2)_R$ chiral symmetry.\cite{Nambu,NJL}

There are two important low-energy phenomena in the system.
First one is that quarks form bound states.
In the first part of this review, we consider lowest-lying mesons
using Bethe-Salpeter equations.
Second one is the dynamical breaking of the chiral symmetry.
The physical manifestation of the chiral symmetry is the presence of
the Nambu-Goldstone bosons such as pions.\cite{Nambu,NJL}
The actual pion, even though it has non-zero mass, is recognized to be
the Nambu-Goldstone boson of the chiral symmetry.
The pion is a pseudoscalar bound state of quarks, and is described by
a homogeneous Bethe-Salpeter equation.
The Bethe-Salpeter equation of the pion is solved in the chiral limit
with the help of the axial Ward-Takahashi identity.\cite{ABKMN,a:JM}

The simplest order parameter of the chiral symmetry breaking is the
vacuum expectation value of the quark bilinear $\vev{\overline q q}$
which is calculated using the Schwinger-Dyson equation in the improved
ladder approximation.
The vacuum expectation value depends on a renormalization point.
The low-energy parameter $L_{10}$ which is introduced by
Gasser-Leutwyler\cite{b:GL}, or so-called $S$ parameter\cite{PT} in
QCD, is best and well-defined order parameter of the chiral
symmetry breaking, because it is a finite quantity which is free from
ultraviolet divergences.
We have the value $S=0.43\sim 0.48$ in the improved ladder exact
approximation.\cite{HaYo}
This means the dynamical symmetry breaking of the symmetry in that
approximation.

The Nambu-Goldstone bosons, as well as lowest-lying mesons, control
many physical processes under low energy theorems.
The difference of the vector and axial-vector
two-point functions, which we call `V-A' two-point function, is
exactly saturated by the contribution of the massless pion and its
derivative, the QCD $S$ parameter, is dominated by the contributions
of lowest-lying mesons such as $\rho$ and $a_1$ mesons in the
low-energy limit.

\begin{flushleft}
\underline{ Heavy Quark Symmetry }
\end{flushleft}

Recently, there has been many investigations on the $B$ meson in
order to define or verify the detailed structure of the Standard Model
and to revile new physics beyond it.
When we extract the absolute value of one component of the
Cabbibo-Kobayashi-Maskawa (CKM) flavor mixing matrix
$\Vcb$,\cite{Ca,KoMas}
we use the exclusive semi-leptonic decay process of the $B$
meson $B\!\rightarrow\! D^{(*)} l \overline \nu$.
This decay process is intensively studied by using the heavy quark
effective theory.
When we consider mesons containing a heavy quark (such as the $b$ or
$c$ quark), a new approximate symmetry called heavy quark spin-flavor
symmetry appears.\cite{IWa,IWb}

According to the heavy quark effective theory\cite{Georgi,EH},
when the heavy quark mass goes to infinity,
every form factor of the semi-leptonic $B$ decay is expressed in terms
of a single universal function, the Isgur-Wise function $\xi(t)$.
It is known that the Isgur-Wise function is normalized to unity at the
kinematical end point $\xi(1)=1$.
The notion of universal form factor was discussed in Ref.\cite{DCM}
first.
The normalization of the semi-leptonic decay form factors at the
kinematical end point was first suggested in Ref.\cite{NW}.
The inclusion of $1/M_Q$ corrections is studied in Refs.\cite{EH,FGL}.

The measurement of the semi-leptonic decay $B\!\rightarrow\! D^{(*)} l
\overline\nu$ enables us to know its differential decay rate
which is proportional to $\Vcb^2$ and the relevant form factors
squared.
{}From the condition $\xi(1)=1$ the differential decay rate divided by
some phase volume is nothing but $\Vcb^2$ at the kinematical end
point.
Then, we can extract $\Vcb$ without knowing the dynamics of the strong
interaction in the $B$ meson.

However, the kinematical end point is actually the end point in phase
space, and the volume of the phase space is almost zero.
The event is very rare, and we have relatively large statistical
error.
Moreover, there is a serious difficulty arising from the successive
decay $D^*\!\rightarrow\! D$ with soft pion emitted.
This increases the systematic error.
In order to extract $\Vcb$ we have to know the Isgur-Wise function
away from the kinematical end point so that we extrapolate the
experimental data to the kinematical end point.
For this purpose we have to solve the QCD dynamics in the $B$ mesons.

Under these circumstances we deal with this problem by using the
Bethe-Salpeter equation, and we calculate the Isgur-Wise function
$\xi(t)$ and extract $\Vcb$.
To say more concretely we solve the pseudoscalar BS equation for the
heavy-light quark system in the three approximations; i) constant
quark mass approximation, ii) improved ladder approximation, iii) the
heavy quark limit.
The Isgur-Wise function is calculated from the resultant BS
amplitudes.

\newpage
\part{Chiral Symmetry and Lowest-lying Mesons}

\section{Dynamical Breaking of the Chiral Symmetry}
\reseteqnum

\subsection{Schwinger-Dyson Equation}

The exact Schwinger-Dyson (SD) equation is derived from the identity
that the functional integration over a total derivative
vanishes\cite{I-Z}
\be
\int\!\! DA^a_\mu D\psi D\overline\psi ~
\frac{\delta}{\delta\overline\psi(x)}
\exp i\Big\{S[A^a_\mu,\psi,\overline\psi] + \overline\zeta\psi +
\overline\psi\zeta + J^{a\mu}A_\mu^a\Big\} \equiv 0 ~,
\label{eq: functional SD}
\ee
where we suppress the $B$-field and ghosts.
\begin{figure}[hbtp]
\begin{center}
\ \epsfbox{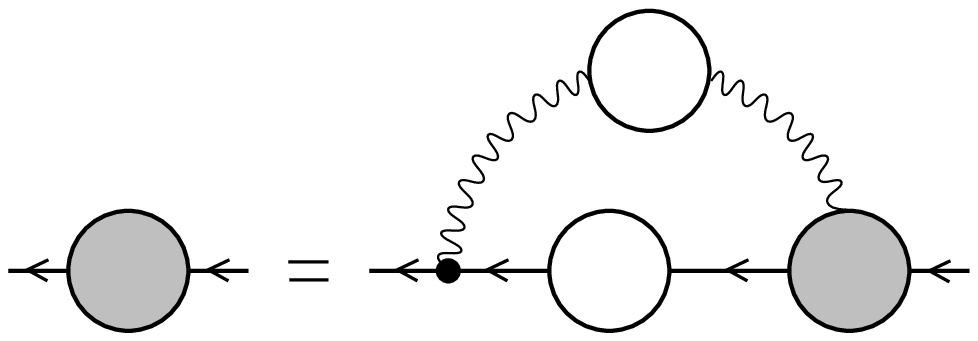}
\vspace{-15pt}
\caption[]{
The exact Schwinger-Dyson equation for the quark propagator.
}
\label{fig: exact SD}
\end{center}
\end{figure}
Setting $J^{a\mu} = \zeta = \overline\zeta = 0$ after differentiating
by $\zeta(y)$, we have the exact formula for the SD equation,
which is shown in Fig.\ref{fig: exact SD}.
When we solve the exact equation for the quark propagator, we have to
know the exact functional forms of the gluon propagator and the
quark-gluon vertex function.
In practical sense we need some approximation.

The ladder (or rainbow) approximation is to replace the quark-gluon
vertex function and the gluon propagator with the tree vertex and
the free propagator, respectively.
The coupling is fixed constant.
However, the approximated equation does not reproduce the property
required by QCD.
We regard the equation for QED rather than QCD.

In order to make the equation describe QCD theory, we incorporate
the property of the asymptotic freedom into it by replacing the fixed
coupling with the running coupling.
The running coupling is given by the one-loop renormalization group
equation usually.
The renormalization point of the running coupling is essentially the
gluon momentum squared, but as we will explain later several
possibilities are allowed.
The resultant approximations are called improved ladder (or rainbow)
approximations.

Then, the Schwinger-Dyson equation in the improved ladder
approximation is
\be
iS_F^{-1}(p) = \slp - \intdk\:\frac{C_2g^2(p,k)}{-l^2}
\left( g_{\mu\nu} - \eta(-l^2)\frac{l_\mu l_\nu}{l^2} \right)
\gmu iS_F(k) \gnu ~, \label{eq: SD}
\ee
where $l_\mu = (p-k)_\mu$, $C_2 = (N_c^2-1)/(2N_c)$ is the second
Casimir invariant, and $\eta(-l^2)$ is the gauge fixing
function.\cite{MN}
For later convenience we have made the gauge parameter a function of
$-l^2$, because it is possible to introduce a nonlocal gauge fixing
function in gauge theories.

In general the quark propagator is expanded by two scalar functions
\be
iS_F^{-1}(p) = A(-p^2)\slp - B(-p^2) ~. \label{eq: SF}
\ee
Substituting the eq.(\ref{eq: SF}) into the Schwinger-Dyson equation
(\ref{eq: SD}), we have
\bea
(A(-p^2)-1)p^2 &=& \intdk \frac{C_2g^2(p,k)}{-l^2} \:
\Bigg[ \left(2+\eta(-l^2)\right)p\cdot k \nn
&& + 2\eta(-l^2) \frac{p^2k^2-(p\cdot k)^2}{-l^2} \Bigg]
\frac{A(-k^2)}{B(-k^2)^2 - A(-k^2)^2k^2} ~,\nn
B(-p^2) &=& \intdk \frac{(4-\eta(-l^2))C_2g^2(p,k)}{-l^2}\nn
&&\times \frac{B(-k^2)}{B(-k^2)^2 - A(-k^2)^2k^2} ~. \label{eq: AB}
\eea

There are many possibilities for the argument of the running coupling
constant $g^2(p,k)$.
It is plausible to chose the argument such that the mass function
\[
\Sigma(x) \equiv B(x)/A(x) ~,
\]
be consistent with the result by the operator product expansion (OPE)
at the high energy\cite{Politzer}
\be
\Sigma(x) \sim \frac{g^2(x)}{x}
\left(\ln\frac{x}{\mu^2}\right)^{1+\gamma_{_\Sigma}} \qquad
\mbox{ for }~~ x \sim \infty ~, \label{eq: MF OPE}
\ee
where
\[
1+\gamma_{_\Sigma} = \frac{9C_2}{11N_c-2N_f} = \frac{4}{9}~.
\]
When we discuss the dynamical symmetry breaking, the Lagrangian do
never possess bare mass term.
Starting from the chiral limit, however, non-perturbative
approximation may violate chiral symmetry non-trivially.
The existence of the bare mass $m_0$ modifies the high energy behavior
of the system, especially $\Sigma(x)$ as $\sim m_0$ up to log
corrections.
So, in order to forbid bare mass to appear we require that the high
energy behavior of the mass function be exactly the same as that in
the case of vanishing bare mass.

Typical choices of the running coupling are
\be
g^2(p,k) = \left\{
\begin{array}{lc}
g^2(\max(-p^2,-k^2)) & (\mbox{I}) \\
g^2(-p^2-k^2) & (\mbox{II}) \\
g^2(-(p-k)^2) & (\mbox{III})
\end{array}\right. ~, \label{eq: coupling choice}
\ee
and all choices (I,II and III) reproduce the OPE result.
The type (I) is first used by Higashijima and
Miransky\cite{Higashijima,Miransky}.
In types (I) and (II) one can trivially carry out the angle
integration of the Schwinger-Dyson equation (\ref{eq: AB}) and reduce
the equation to a simple one-dimensional integral equation.
Moreover the quark propagator receives no wave function
renormalization, $A(-p^2) = 1$, in the Landau gauge
$\eta(-l^2) = 1$.\cite{ABKMN}
Thus performing the Wick rotation, we have
\be
\Sigma(x) = \int_0^\infty\!\!ydy \: \frac{\lambda(x,y)}{\max(x,y)}
\: \frac{\Sigma(y)}{y+\Sigma(y)^2}~, \label{eq: SD I-II}
\ee
where
\be
\lambda(x,y) = \frac{3C_2}{16\pi^2}\times \left\{
\begin{array}{ll}
g^2(\max(x,y)) & \quad\mbox{for type (I)}\\
g^2(x+y)       & \quad\mbox{for type (II)}
\end{array}\right.
\ee
Especially in type (I) the Schwinger-Dyson equation \num{SD I-II}
becomes a differential equation
\be
\Bigg[\frac{\Sigma'(x)}{(\lambda(x)/x)'}\Bigg]' =
\frac{x\Sigma(x)}{x+\Sigma(x)^2} ~,
\ee
with appropriate boundary conditions.\cite{Higashijima,Miransky,ABKMN}

Next, we consider the type (III) which is the most natural since the
chiral Ward-Takahashi identities hold in this case as explained later.
We can always keep $A(-p^2) = 1$ using the freedom of the gauge
function $\eta(-l^2)$ in the type (III).\cite{a:Kugo-Mitchard}
Let us explain this reason.
{}From eq.(\ref{eq: AB})
\beann
(A-1)p^2
&=& \intdk\: \frac{C_2g^2}{-l^2} \\
& & \Bigg[(2+\eta)p\cdot k + 2\eta
\frac{p^2k^2-(p\cdot k)^2}{-l^2}\Bigg]\:
\frac{A}{B^2-A^2k^2} ~,
\eeann
performing the Wick rotation and
defining $x=-p^2$, $y = -k^2$, $z = -l^2 = x+y-2\cos\theta\sqrt{xy}$,
and $p\cdot k = -\cos\theta\sqrt{xy}$,
\beann
(A-1)p^2
&=& - \frac{C_2}{12\pi^3}\int_0^\infty\!\!ydy \int_0^\pi\!\!d\theta
\frac{\sin^4\theta}{z^2}\bigg[ z\varphi'(z) - 2g^2(z)(\eta(z)-1)\bigg]
\frac{A}{B^2-A^2k^2} ~,
\eeann
where $\varphi(z) = g^2(z)(2+\eta(z))$ and the prime denotes
differentiation.
Therefore we can keep $A(x) \equiv 1$ as far as the gauge function
$\eta(x)$ satisfies the differential equation
\be
x\varphi'(x) - 2g^2(x)(\eta(x)-1) = 0 ~.\label{eq: eta diff}
\ee
Assuming that $g^2(x)$ is differentiable and finite, the differential
equation (\ref{eq: eta diff}) at $x = 0$ gives the initial condition
for itself:
\[
0 = 0 \times \varphi'(0) - 2g^2(0)(\eta(0)-1) ~,
\]
thus
\be
\eta(0) = 1 ~.
\ee
The differential equation (\ref{eq: eta diff}) means that the gauge
function $\eta(x)$ can be determined without knowing the mass function
$\Sigma(x)$.

Now, $B(x)$ becomes the mass function $\Sigma(x)$ itself, and from
eq.(\ref{eq: AB}) it is calculated by
\be
\Sigma(x) = \int_0^\infty\!\!ydy \: K_\Sigma(x,y)\:
\frac{\Sigma(y)}{y+\Sigma(y)^2} ~, \label{eq: SD equation}
\ee
with
\be
K_\Sigma(x,y) = \int_0^\pi\!\!d\theta\sin^2\theta\:
\frac{C_2g^2(z)(4-\eta(z))}{8\pi^3z} ~. \label{eq: SD kernel}
\ee

\subsection{Vacuum Expectation Value of $\:\overline q q$}
%\reseteqnum

The vacuum expectation value (VEV) of the quark bilinear
$\overline qq$ ($q = u, d$) is the simplest order parameter of the
chiral symmetry;
\be
\bigg[ iQ_5^a, \overline\psi i\gf\tb\psi \bigg] =
\delta^{ab} \frac{1}{2}\overline \psi\psi ~,
\ee
where $Q_5^a$ is the generator of the chiral rotation, $\tau^b$ is the
Pauli matrix and $\psi = (u, d)^T$.
Since $\overline q_{_L} q_{_R}$ transforms like $(2^*,2)$ under
$SU(2)_L\times SU(2)_R$, the VEV $\vevqq$ vanishes if the
system is chiral invariant in the representation level.
The symmetry in the representation level means the invariance which
the vacuum has.
When the system develops the non-vanishing VEV
$\vev{\overline\psi\psi}\ne 0$, the symmetry generated by the charge
$Q_5^a$ breaks; $Q_5^a\ket{0} \ne 0$.
$\vevqq \ne 0$ implies that the $\overline qq$ pair condensates in
the vacuum.
The symmetry of the vacuum under $SU(2)_V$ means that the VEVs of
$\overline uu$ and $\overline dd$ are identical
\be
\vev{\overline uu} = \vev{\overline dd} ~.
\ee

Even when the system possesses the chiral symmetry in the Lagrangian
level, there is no reason why the system keeps the symmetry in the
representation level.
Actually the chiral symmetry breaks spontaneously in
QCD.\cite{Nambu,NJL}

The VEV is calculated from the quark propagator
\bea
\vevqq_\Lambda &=& -N_c \tr S_F(x = 0) \nn
&=& - \frac{N_c}{4\pi^2} \int_0^{\Lambda^2}\!\!xdx
\frac{\Sigma(x)}{x+\Sigma(x)^2} ~,
\eea
where $\Lambda$ is the Euclidean momentum cutoff.
The VEV has ultra-violet divergence, and depends on the cutoff
$\Lambda$.
Instead, the renormalized VEV depends on the renormalization point
$\mu^2$.
Using the high energy behavior of the mass function
(\ref{eq: MF OPE}), we identify the renormalized VEV $\vevqq_\mu$ by
\be
\vevqq_\mu = \vevqq_\Lambda \:
\left( \frac{g^2(\Lambda^2)}{g^2(\mu^2)}
\right)^{1+\gamma_\Sigma} ~.
\ee
We note that the quantity
$\vevqq_\Lambda(g^2(\Lambda^2))^{1+\gamma_\Sigma}$ is independent of
the cutoff $\Lambda$ for sufficiently large $\Lambda$.

The vacuum expectation value of the quark bilinear is certainly the
simplest order parameter of the chiral symmetry, but needs
renormalization of the ultraviolet divergence.
The VEV is a notion which depends on a renormalization point.
The most suitable order parameter of the chiral symmetry is,
so-called, $L_{10}$ introduced by Gasser-Leutwyler\cite{b:GL}.
If $L_{10}$ develops a non-zero value, the chiral symmetry breaks.
This is a coefficient of the effective Lagrangian and suffers no
ultraviolet divergence.
Thus, the low-energy parameter $L_{10}$ does not depend on any
renormalization point, and is well-defined.
This low-energy parameter is evaluated in the improved ladder exact
approximation in Ref.~\cite{HaYo}.

Using the current algebra technique, the VEV is expressed by the
observable of the pion mass $m_\pi$, the pion decay constant
$f_\pi$ and the current quark mass $\wh m(\mu) \equiv
(m_u(\mu)+m_d(\mu))/2$\cite{GOR}
in the lowest order perturbation of the quark masses
\be
f_\pi^2 m_\pi^2 = - 2\wh m(\mu) \vevqq_\mu ~,
\ee
where the VEV is defined in the chiral limit.
With the measured value of $m_\pi$ and $f_\pi$, and with the value
$\wh m(\mbox{1 GeV}) = 7 \pm 2 \mbox{ MeV}$ from the QCD sum rules
for the axial divergence, we have an ``experimental'' value\cite{a:GL}
\be
\vevqq_{_{\rm 1 GeV}} = - (225 \pm 25 \mbox{ MeV})^3
 ~.\label{eq: GL vev}
\ee
The result in the improved ladder approximation is in good agreement
with this result \num{GL vev}.

Before closing this subsection, we comment on another order parameter.
One of the chiral coefficients, $L_{10}$, of so-called the S parameter
is the finite order parameter, contrary to the VEV $\vevqq$:
This is ultraviolet finite\cite{b:GL,PT,ILY} and renormalization group
invariant.
We can measure the chiral symmetry or asymmetry without any ambiguity
using this order parameter.
However, the calculation of it in the improved ladder approximation is
rather tedious as we will show later.

\subsection{Chiral Ward-Takahashi Identities}
\label{The Chiral Ward-Takahashi Identity}
%\reseteqnum

We are considering the QCD with two massless quarks $u$ and $d$.
Massless means vanishing bare (or current) mass here.
This system possesses the $SU(2)_L \times SU(2)_R$  chiral symmetry.
There are two Noether currents (vector and axial-vector) of this
system
\be
\partial_\mu J_5^{a\mu} = \partial_\mu J^{a\mu} = 0 ~,
\ee
where
\be
J_5^{a\mu} = \overline\psi\gmu\gf\ta\psi ~, \qquad
J^{a\mu} = \overline\psi\gmu\ta\psi ~.
\ee
By virtue of the current conservations, the chiral Ward-Takahashi (WT)
identities hold formally.
Of course one can check these WT identities hold in perturbation
expansion.
How is it in the case of the improved ladder approximation?
The axial WT identity was first studied by Maskawa and
Nakajima\cite{MN} in the ladder approximation with fixed coupling.
The more detailed and complete analyses were made by Kugo and
Mitchard\cite{a:Kugo-Mitchard,b:Kugo-Mitchard}.
The improved ladder approximations in the type (I) and (II) ( see
eq.(\ref{eq: coupling choice})) slightly violate the identities,
although the pion remains massless.
This is because the axial Ward-Takahashi identity become hold in the
soft pion limit $q_\mu\rightarrow0$ with any type of the coupling
choice (\ref{eq: coupling choice})).
We review the analyses in Refs.~\cite{a:Kugo-Mitchard,b:Kugo-Mitchard}.

The axial and vector Ward-Takahashi identities are
\bea
q^\mu\GAp &=& \ta \left[ iS_F^{-1}(p_+) \gf
+ \gf iS_F^{-1}(p_-) \right] ~, \nn
q^\mu\GVp &=& \ta  \left[ iS_F^{-1}(p_+) - iS_F^{-1}(p_-)
\right] ~, \label{eq: WTid}
\eea
where $p_\pm = p \pm q/2$.
The basic ingredients in the chiral Ward-Takahashi identities are the
quark propagator $S_F(p)$ and the three-point vertex functions of the
vector and the axial currents, $\GAp$ and $\GVp$ respectively.
We studied about the quark propagators in the improved ladder
approximation, but we do not about the three-point functions here.
We only quote the inhomogeneous BS equations of them in the improved
ladder approximation and postpone the derivation and discussions.

The Schwinger-Dyson equation reads
\be
M(p) = \intdk K(p,k) iS_F(k) ~, \label{eq: SD 2}
\ee
with
\be
M(p) \equiv \slp - iS_F^{-1}(p) ~,
\ee
and $K$ is the BS kernel
\bea
K(p,k) &=& C_2g^2(p,k)D_{\mu\nu}(p-k)~ \gmu \otimes \gnu
{}~, \nn
D_{\mu\nu}(l) &=& \frac{1}{-l^2} \left(g_{\mu\nu}-\eta(-l^2)\frac{l_\mu
l_\nu}{l^2} \right) ~. \label{eq: BS kernel}
\eea
We have used the tensor product notation:
\be
(A \otimes B) \chi \equiv A \chi B ~. \label{eq: tp notation}
\ee
If the wave function renormalization factor is unity, $A(-p^2) = 1$,
we have $M(p) = \Sigma(-p^2)$.
But we discuss the generic case $A(-p^2) \ne 1$.

The inhomogeneous BS equation for the axial-vector vertex function
$\GAp$ in the improved ladder approximation is
\be
\GAp = \gmd\gf\ta - \intdk K(p,k) \left( S_F(p_+)
\otimes S_F(p_-) \right) \GAk ~, \label{eq: iBS axial}
\ee
and the similar inhomogeneous BS equation for the vector vertex
function $\GVp$ is hold.
The vertex functions are defined by
\bea
\lefteqn{
\int\!\!dz ~e^{iqz}~ \bra{0} T \psi(x) \overline\psi(y) J_{5\mu}^a(z)
\ket{0} = } \nn
&& e^{iq\frac{x+y}{2}} \intdp ~e^{-ip(x-y)}~ S_F(p_+)
\;\GAp\; S_F(p_-) ~, \label{eq: def axial vertex}
\eea
and similarly for $\GVp$.

First we consider the axial WT identity.
The formal solution of the inhomogeneous BS equation for the axial
vertex function (\ref{eq: iBS axial}) is
\be
\GAp = \frac{1}{1-L} \gmd\gf\ta ~, \label{eq: WTid 1}
\ee
where
\be
L(p,k;q) = K(p,k)\; S_F(p_+) \otimes S_F(p_-) ~,
\ee
and we have used the ``inner product'' rule:
\be
L\chi(p;q) \equiv \intdk ~L(p,k;q)~ \chi(k;q) ~. \label{eq: ip rule}
\ee
Multiplying the eq.(\ref{eq: WTid 1}) by $q^\mu$, we have
\be
0 = \frac{1}{1-L}\slq\gf\ta - q^\mu\GAp ~. \label{eq: WTid 2}
\ee
Using the axial WT identity (\ref{eq: WTid}) and the following trivial
identities
\[
\slq\gf \equiv iS_F^{-1}(p_+)\gf + \gf iS_F^{-1}(p_-)
+ M(p_+)\gf + \gf M(p_-) ~,
\]
\[
\frac{1}{1-L} - 1 \equiv \frac{1}{1-L}L ~,
\]
we find
\be
0 = L\left(\SI(p_+)\gf + \gf\SI(p_-)\right) +
\left(M(p_+)\gf + \gf M(p_+)\right) ~, \label{eq: WTid 3}
\ee
where we remove the overall factor
\[
\left(\ta\right)\frac{1}{1-L} ~.
\]
The first term of \eq{WTid 3} is rewritten as the
following:
\beann
L\left(\SI(p_+)\gf\right)
&=& K \left( S_F(p_+)\: \SI(p_+) \gf \: S_F(p_-) \right) \nn
&=& K \left( i\gf S_F(p_-) \right) \nn
&=& - \gf \left( KiS_F(p_-) \right) ~,
\eeann
and similarly
\[
L\left( \gf\SI(p_-) \right) = - \left( KiS_F(p_+) \right) \gf ~.
\]
So, we can rewrite \eq{WTid 3} into
\be
0 = \left[\: M(p_+) - KiS_F(p_+) \:\right]\gf +
\gf\left[\: M(p_-) - KiS_F(p_-) \:\right] ~. \label{eq: id axial}
\ee
By tracing the same derivation for the axial vertex case, we
obtain the similar identity like eq.(\ref{eq: id axial}) in the vector
vertex case:
\be
0 = \left[\: M(p_+) - KiS_F(p_+) \:\right] -
\left[\: M(p_-) - KiS_F(p_-) \:\right] ~. \label{eq: id vector}
\ee
We can easily find that the only consistent manner is to require
\bea
M(p_\mp) &=& KiS_F(p_\mp) \nn
&\equiv& \intdk~ K(p,k)~iS_F(k_\mp) ~, \label{eq: id SF}
\eea
of the quark propagator.
This equality reminds us with the Schwinger-Dyson equation
(\ref{eq: SD 2}), but there is a slight difference in the momentum
flow.
When we shift the momenta $p \rightarrow p \pm q/2$ and
$k \rightarrow k \pm q/2$, the eq.(\ref{eq: id SF}) becomes
\be
M(p) = \intdk ~ C_2g^2(p_\pm,k_\pm)D_{\mu\nu}(p-k) \gmu iS_F(k) \gnu ~,
\label{eq: id SF 2}
\ee
where we consider only the case when the $k$-integration converges.
Take the difference between this equation and the Schwinger-Dyson
equation itself, we finally obtain
\be
0 = \intdk ~ C_2\bigg(~ g^2(p_\pm,k_\pm) - g^2(p,k) ~\bigg)
D_{\mu\nu}(p-k) \gmu iS_F(k) \gnu ~.
\ee
Thus, in order for the improved ladder approximation to be consistent
with the chiral WT identities, we have to chose the argument of the
running coupling such that
\be
g^2(p_\pm,k_\pm) = g^2(p,k) ~, \label{eq: consistent coupling}
\ee
holds.
The type (III) is the only choice among the three candidates
(\ref{eq: coupling choice}).
The improved ladder approximation with the type (III) coupling choice
is called the consistent ladder approximation.
We should notice that any choices (\ref{eq: coupling choice}) satisfy
(\ref{eq: consistent coupling}) in the soft moemntum limit
$q_\mu\rightarrow 0$.

Here we comment on the validity of momentum shift of the integration
variable $k_\mu$ in \eq{id SF}.
When we solve the Schwinger-Dyson equation we perform the Wick
rotation and use the momentum cutoff regularization usually.
The obtained functions $A(-p^2)$ and $B(-p^2)$ which satisfy the SD
equation identically make the integration in the \eq{id SF 2}
convergent enough.
As for the high energy momentum of the $k_\mu$, we are allowed to
ignore the momentum $q_\mu$ in the integrand of \eq{id SF}.
Even in the fixed coupling case the momentum integration of the
Schwinger-Dyson equation is convergent one, because we calculate the
function $A$ and $B$ by the following way.
The fixed coupling is regarded as the function in the cutoff.
We tune the coupling to the critical value as the cutoff goes to
infinity so as for the function $A(-p^2)$ and $B(-p^2)$ become finite.
Thus, the high energy behavior of the integrand in the RHS of
the eq.(\ref{eq: id SF}) or eq.(\ref{eq: id SF 2}) is
\be
\sim C_2g^2(p,k)D_{\mu\nu}(p-k) \gmu iS_F(k) \gnu \qquad \mbox{ as }
-\!k^2 \longrightarrow \infty ~.
\ee
This behavior is identically the same as that for the Schwinger-Dyson
equation itself which has convergent integral by construction.
We conclude that the violation of the chiral WT identities due to the
momentum shift vanishes.

\subsection{Pagels-Stokar Formula}
\label{The Pagels-Stokar Formula}
%\reseteqnum

Within the knowledge of the mass function $\Sigma(x)$, we can estimate
the pion decay constant $f_\pi$ using the Pagels-Stokar
formula\cite{PS}
\be
f_\pi^2 = \frac{N_c}{4\pi^2}\int_0^\infty\!\!xdx\:
\frac{\Sigma(x)(\Sigma(x)-\frac{x}{2}\Sigma'(x))}{(x+\Sigma(x)^2)^2}
{}~. \label{eq: PS}
\ee
If one use $f_\pi$ as an input parameter of the mass dimension, the
output is $\lqcd$.

Following Pagels and Stokar, we derive the formula.
The pion decay constant $f_\pi$ is defined by
\be
\Bra{0} \overline\psi \gmd\gf\ta
\psi(0)\Ket{\pi^b(q)} = \delta^{ab}iq_\mu \: f_\pi ~.
\ee
Using the definition of the pion BS amplitude \num{def BS amp}
we rewrite into
\be
f_\pi\: q_\mu = - \frac{N_c}{2}\intdp~ \tr\left[\gmd\gf
\chi(p;q) \right] ~. \label{eq: fpi chi}
\ee
The factor $1/2$ comes from $\tr[(\tau^a/2)(\tau^b/2)] =
\delta^{ab}/2$.
In order to calculate $f_\pi$ we need the BS amplitude only in the
order $O(q_\mu)$ since $q^2 = 0$.
In the soft momentum limit, the BS amplitude is indeed expressed in
terms of the mass function by virtue of the axial Ward-Takahashi
identity.

In any choices of the running coupling (\ref{eq: coupling choice})
the axial Ward-Takahashi identity is satisfied%
\footnote{
In a strict sense, it is enough for this identity to satisfy in the
soft momentum limit $q_\mu\rightarrow 0$.
(see (\ref{eq: consistent coupling}).)
}
\be
q^\mu \GAp = iS_F^{-1}(p+\oh q)
\gf \ta + \ta \gf
iS_F^{-1}(p-\oh q) ~, \label{eq: aWTid}
\ee
where $\GAp$ is the axial vertex function defined by eq.(\ref{eq: def
axial vertex}).
Substituting the quark propagator in the improved ladder approximation
$iS_F^{-1}(p) = \slp - \Sigma(-p^2)$ into the axial WT identity
(\ref{eq: aWTid}), we obtain, in the soft momentum limit
$q_\mu \rightarrow 0$,
\be
\lim_{q_\mu \rightarrow 0} q^\mu \GAp =
- \tau^a \gf \Sigma(-p^2) ~. \label{eq: LE theorem}
\ee
The eq.(\ref{eq: LE theorem}) shows that the soft pion contribution to
the axial vertex function is completely determined by the mass
function.
Namely, the axial WT identity fixes the contribution of the soft pion
to the axial vertex function.

In general the axial vertex function is decomposed into two parts;
the singular part due to the massless pion pole and the regular part
\be
\GAp \sim - \left(\ta\right)
\wh\chi(p;q)\:\frac{f_\pi q_\mu}{q^2}
+ \mbox{ regular at $q^2 \rightarrow 0$} ~, \label{eq: pion pole}
\ee
where $\wh\chi(p;q)$ is the amputated BS amplitude.
{}From eqs.(\ref{eq: LE theorem}) and (\ref{eq: pion pole}) we have
\be
\wh\chi(p;q \rightarrow 0) = - \gf \frac{2\Sigma(-p^2)}{f_\pi}
{}~. \label{eq: PS hchi}
\ee

Pagels and Stokar made a glorious approximation that the amputated BS
amplitude up to the order $O(q_\mu)$ be given by \eq{PS hchi}.
Thus, we find
\bea
\chi(p;q) &=& S_F(p+{\scriptstyle\frac{1}{2}}q) \wh\chi(p;q)
S_F(p-{\scriptstyle\frac{1}{2}}q) \nn
&\cong&  S_F(p+{\scriptstyle\frac{1}{2}}q)
\left( - \gf \frac{2\Sigma(-p^2)}{f_\pi} \right)
S_F(p-{\scriptstyle\frac{1}{2}}q) ~. \label{eq: PS chi}
\eea
Substituting the BS amplitude (\ref{eq: PS chi}) into the definition
of $f_\pi$ (\ref{eq: fpi chi}), we obtain the desired formula
(\ref{eq: PS}).

\subsection{Bethe-Salpeter Equation for the Pion}
\label{The Bethe-Salpeter Equation for the Pion}
%\reseteqnum

In this section we study the Bethe-Salpeter equation for the pion
with improved ladder approximation.
In order to evaluate the pion decay constant $f_\pi$, the BS amplitude
of the pion is calculated.
This calculation has been done by two groups\cite{ABKMN,a:JM} first.
The pion appears as the massless particle in the Bethe-Salpeter
equation.
The more detailed analyses were made in
Refs.~\cite{b:Kugo-Mitchard,b:JM} followingly.
The improved ladder approximation which is consistent with the chiral
Ward-Takahashi identities is found in
Refs.\cite{a:Kugo-Mitchard,b:Kugo-Mitchard}.

The Nambu-Goldstone's theorem\cite{Nambu,Goldstone} guarantees that
the pion appears as an massless particle, where the system possesses
the chiral symmetry.
However the above studies tell us that massless-ness of the pion is
guaranteed by what one use the same BS kernel in the SD and BS
equations simultaneously.
To use the same BS kernel is, of course, leads to the WT identity of
the chiral symmetry in the soft momentum limit $q_\mu\rightarrow0$
on account of (\ref{eq: consistent coupling}).
This point is discussed in
subsection~\ref{The Chiral Ward-Takahashi Identity}.
If the pion could not appear as a massless particle, the BS equation
would have no solution at zero momentum-squared $q^2=0$.
In what follows we show the equation indeed has the solution at
$q^2=0$.

The BS amplitude $\chi$ and the truncated BS amplitude $\wh\chi$ for
the pion are defined by
\bea
\Bra{0} T \psi_i(\frac{-r}{2}) \overline\psi^j(\frac{r}{2})
\Ket{\pi^a(q)} &=&
\N \delta_i^j \ta ~\intdp~ e^{-ipr}~\chi(p;q) ~,\nn
\wh\chi(p;q) &=& S_F^{-1}(p_+)~\chi(p;q)~S_F^{-1}(p_-) ~,
\label{eq: def BS amp}
\eea
where $p_\pm = p \pm q/2$, $i, j$ are color indices and $\N$ is a
constant introduced for later convenience.
When we put $\N=1$, these equations give the usual definitions for the
BS amplitude and its truncated one.
The Kronecker delta $\delta_i^j$ reflects the fact that mesons are
color singlet.
Taking into account of the spinor structure and the quantum numbers
$J^{PC} = 0^{-+}$, the (truncated) BS amplitude is expanded by four
invariant amplitudes, i.e., $S, P, \cdots, \wh Q, \wh R$, as
\bea
\chi(p;q)\gf &=& S(p;q) + (p\cdot q)\slp P(p;q) + \slq Q(p;q) + \oh
[\slp,\slq] R(p;q)~, \nn
\wh\chi(p;q)\gf &=& \wh S(p;q) + (p\cdot q)\slp \wh P(p;q) + \slq
\wh Q(p;q) + \oh [\slp,\slq] \wh R(p;q) ~.
\label{eq: inv BS amp}
\eea
Using the charge conjugation, it is easy to see that the every
invariant amplitude $X = S$, $P$, $\cdots$, $\wh Q$, $\wh R$ is an
even function in $(p\!\cdot\! q)$.

Our aim is to calculate the pion decay constant $f_\pi$ from the BS
amplitude.
The definition of $f_\pi$ is
\bea
f_\pi~ iq_\mu\delta^{ab} &\equiv& \bra{0} \overline\psi\gmd\gf\ta
\psi(0) \ket{\pi^b(q)} \nn
&=& - \: \N \frac{\delta^{ab}}{2}
   N_c \intdp~\tr \left[ \gmd\gf\chi(p;q)\right]~.
\label{eq: def fpi}
\eea
The second equality in eq.(\ref{eq: def fpi}) is obtained by taking
the trace on the definition (\ref{eq: def BS amp}) of $\chi$ with
$\gmd\gf\tau^b/2$ multiplied.
It is easy to see from eq.(\ref{eq: def fpi}) that we are enough to
calculate the BS amplitude up to $O(q_\mu)$ for the purpose of
calculating the decay constant $f_\pi$.
So, we expand the BS amplitude in powers of $q_\mu$ and keep them up
to $O(q_\mu)$:
\be
\chi(p;q) = \chi^0(p) + q^\mu\chi_\mu(p) + O(p\cdot q)^2 ~,
\ee
with
\bea
\chi^0(p) &=& \gf\; S(-p^2) ~, \nn
\chi_\mu(p) &=& p_\mu\slp\gf\; P(-p^2) + \gmd\gf\; Q(-p^2) +
\oh [\slp,\gmd]\gf\; R(-p^2) ~,
\eea
and similarly for $\wh\chi(p;q)$.
The invariant amplitude $X = S$, $P$, $\cdots$, $\wh Q$, $\wh R$ is
also expanded as
\be
X(p;q) = X(-p^2) + O(p\cdot q)^2 ~.
\ee

Next we show how the BS amplitude is calculated from the BS equation.
The homogeneous Bethe-Salpeter equation for the pion reads
\be
(T-K)\:\chi(p;q) = 0~, \qquad ( q^2 = 0)~, \label{eq: pion HBS}
\ee
where $K$ is the BS kernel defined previously by
eq.(\ref{eq: BS kernel}) which represents one-gluon exchange and
$T$ is the kinetic part defined by
\be
T(p;q) = S_F^{-1}(p_+) \otimes S_F^{-1}(p_-) ~,
\ee
where $p_\pm = p \pm q/2$.
We use the tensor product notation (\ref{eq: tp notation})
and the inner product rule (\ref{eq: ip rule}).
We notice that the homogeneous equation (\ref{eq: pion HBS}) tell us
nothing about the normalization of the BS amplitude itself, and we
need the normalization condition.

We solve the BS equation (\ref{eq: pion HBS}) order by order in
$q_\mu$.
The BS equation (\ref{eq: pion HBS}) is expanded in powers of $q_\mu$,
and the equality is hold at each order in $q_\mu$.
Namely,
\[
(T^0 + q^\mu T_\mu - K)\:( \chi^0 + q^\mu\chi_\mu )
 = 0 + O(p\cdot q)^2 ~.
\]
\[ \Downarrow \]
\bea
O(1) : & (T^0-K)\:\chi^0 = 0 \label{eq: pion O(1)BS} ~,\\
O(q_\mu) : & (T^0-K)\:\chi_\mu = - T_\mu \chi^0 ~.
\label{eq: pion O(q)BS}
\eea
The kinetic part $T$ is expanded as
\be
T(p;q) = T^0(p) + q^\mu T_\mu(p) + O(p\cdot q)^2 ~,
\ee
with
\bea
T^0(p) &\equiv& T(p;q)\bigg\vert_{\displaystyle q_{_\lambda}=0} \nn
&=& - (\slp - \Sigma(-p^2)) \otimes (\slp - \Sigma(-p^2)) ~,\nn
T_\mu(p) &\equiv& \frac{\partial}{\partial q^\mu}T(p;q)
\bigg\vert_{\displaystyle q_{_\lambda} = 0} ~,\nn
&=& (\slp - \Sigma(-p^2)) \otimes ({\scriptstyle\frac{1}{2}}\gmd
 + p_\mu\Sigma'(-p^2)) \nn
&& - ({\scriptstyle\frac{1}{2}}\gmd + p_\mu\Sigma'(-p^2))
 \otimes (\slp - \Sigma(-p^2)) ~,
\eea
where $\Sigma'(x) = d\Sigma(x)/dx$.

First we solve the $O(1)$ BS equation (\ref{eq: pion O(1)BS}).
This equation means that the operator $T^0-K$ has zero eigenvalue
on the subspace expanded by $S(-p^2)$.
We find the solution of this equation if we put $\wh S(x) \propto
\Sigma(x)$.
We show this point.
It is easy to see that after performing the Wick rotation we have
\bea
T^0\chi^0 &=& - \gf (x+\Sigma(x)^2)\: S(x) ~, \nn
K\chi^0 &=& - \gf \int_0^\infty\!\!ydy~ K_\Sigma(x,y) S(y) ~,
\label{eq: T0K0}
\eea
where $x = -p^2$ and $K_\Sigma$ is defined by eq.(\ref{eq: SD kernel}).
The relation between $\chi^0$ and $\wh\chi^0$ gives
(see also the appendix
\ref{Matrix Representations for the Pion BS Equation})
\be
- \wh S(x) = (x+\Sigma(x)^2)S(x) \qquad\longleftrightarrow\qquad
 S(x) = - \frac{\wh S(x)}{x+\Sigma(x)^2}~.  \label{eq: ShS}
\ee
Substituting \eqs{T0K0} and \num{ShS} into the $O(1)$ BS equation
\num{pion O(1)BS}, we obtain the linear equation in terms of $\wh S$
\be
\wh S(x) = \int_0^\infty\!\!\!ydy~ K_\Sigma(x,y)
 \frac{\wh S(y)}{y+\Sigma(y)^2} ~.
\ee
This equation reminds us with the Schwinger-Dyson equation
(\ref{eq: SD equation}).
Clearly $\wh S(x) \propto \Sigma(x)$ is a solution and in fact this
is the unique solution from the consequence of the axial
Ward-Takahashi identity in the soft momentum limit
(\ref{eq: LE theorem}).
To say more detail, the axial Ward-Takahashi identity \num{WTid}
uniquely determine the solution as well as the normalization constant
of the BS amplitude%
\footnote{
Usually the BS amplitude is normalized by the Mandelstam's
normalization condition.
One can show that the axial Ward-Takahashi identity and the
normalization condition for the BS amplitude give the same value of
the normalization constant.
This is the consequence of the current conservation.
}
in the soft momentum limit $q_\mu \rightarrow 0$.
We already derive the normalization constant in \eq{PS hchi}.
Taking into account the proportional constant $\N$
in eq.(\ref{eq: PS hchi}), when we put $\N = - 2/f_\pi$ we have a
important relation
\be
\wh S(x) = \Sigma(x) ~. \label{eq: Sh=Sigma}
\ee

Second we solve the $O(q_\mu)$ BS equation (\ref{eq: pion O(q)BS}).
Once we obtain $\chi^0$, i.e., eq.(\ref{eq: Sh=Sigma}), the $O(q_\mu)$
BS equation is regarded as a inhomogeneous equation for $\chi_\mu$
with the inhomogeneous term $-T_\mu\chi^0$.
The formal solution of \eq{pion O(q)BS} is
\be
\chi_\mu = \frac{-1}{T^0-K}\:T_\mu \chi^0~.\label{eq: formal sol}
\ee

Let us define the formal solution (\ref{eq: formal sol}) in the
constructive way.
It is convenient to rewrite the invariant amplitudes and the spinor
base of eq.(\ref{eq: inv BS amp}) in the compact forms:
\be
\begin{array}{rcl}
\chi^i &\equiv& (P, Q, R) ~, \\
\Gamma_{i\mu} &\equiv& (\; p_\mu\slp\gf\; ,~ \gmd\gf\; ,~
\oh [\slp,\gmd]\gf\; ) ~,
\end{array}
\qquad (i = 1, 2, 3)
\ee
where we find
\be
\chi_\mu(p) = \Gamma_{i\mu}(p) \chi^i(-p^2) ~.
\ee
We introduce the matrix representations for $T^0$, $T_\mu$ and $K$:
\bea
T^0_{ij}(x) &\equiv& \Sp{ \overline\Gamma_i^\mu(p)
T^0(p) \Gamma_{j\mu}(p) } ~, \nn
\lambda_i(x) &\equiv& \Sp{ \overline\Gamma_i^\mu(p)
T_\mu(p) \gf } ~, \nn
K_{ij}(x,y) &\equiv& \frac{1}{8\pi^3}\intdt~
\Sp{ \overline\Gamma_i^\mu(p) K(p,k) \Gamma_{j\mu}(k) } ~.
\label{eq: matrix repr}
\eea
The quantity $\overline\Gamma$ is the Dirac conjugate of $\Gamma$
before performing the Wick rotation; i.e., in general
\be
\overline \chi(p;q) \equiv \gamma^0 \chi(p^*;q)^\dagger \gamma^0
\qquad\mbox{for}\quad {}^\forall \chi ~.
\ee
The detailed expressions of $T^0_{ij}$, $\lambda_i$ and $K_{ij}$ are
summarized in the appendix
\ref{Matrix Representations for the Pion BS Equation}.
With the matrix representations eq.(\ref{eq: matrix repr}) the
$O(q_\mu)$ BS equation is rewritten into a component form.
Multiplied  by $(1/4)\tr[\overline\Gamma_i^\mu \times ~]$ from left,
the $O(q_\mu)$ BS equation (\ref{eq: pion O(q)BS}) reads
\[
\Sp{ \overline\Gamma_i^\mu\;(T^0-K)\;\Gamma_{j\mu} }\:\chi^j =
- \Sp{ \overline\Gamma_i^\mu\;T_\mu\;\gf }S
\]
\[ \Downarrow \]
\be
T^0_{ij}(x)\;\chi^j(x) - \int_0^\infty\!\!ydy~K_{ij}(x,y)\;\chi^j(y) =
\frac{\Sigma(x)}{x + \Sigma(x)^2}\;\lambda_i(x) ~.
\label{eq: constructive BS}
\ee
The notable property is that the matrices $T^0$, $K$ are self
conjugate and real definite
\be
\begin{array}{ccccc}
T^0(x){}^\dagger &=& T^0(x){}^T &=& T^0(x)~,\\
K(x,y)^\dagger &=& K(x,y)^T &=& K(y,x) ~.
\end{array} \label{eq: cc property}
\ee
This is the direct consequence of the charge conjugation properties of
the base $\Gamma_{i\mu}(p)$.
We see from the property (\ref{eq: cc property}) that we  obtain the
real definite BS amplitude $\chi^i(x)$.
It is not difficult to solve this inhomogeneous linear equation
numerically by iteration method.

Finally we make a comment on the inverse operator of $T^0-K$
(cf. eq.(\ref{eq: formal sol})) which we need for solving the BS
equation \num{constructive BS}.
One might be afraid that the operator $T^0-K$ had a zero eigenvalue
suggested by eq.(\ref{eq: pion O(1)BS}), and its inverse did not
exist.
However, one can understand that it is not the case because of the
following reason.

Let us define the four Hilbert spaces $\SS$, $\PP$, $\QQ$, $\RR$ as
\be
\begin{array}{ll}
\SS = \{~ \gf \: S(x) ~\} ~, &
\PP = \{~ \Gamma_{1\mu} \: P(x) ~\} ~,\\
\QQ = \{~ \Gamma_{2\mu} \: Q(x) ~\} ~, &
\RR = \{~ \Gamma_{3\mu} \: R(x) ~\} ~,
\end{array}
\ee
which are spanned by any functions $S(x)$, $P(x)$, $Q(x)$ and $R(x)$.
Then
\be
\begin{array}{llll}
\chi^0 \in \SS~, &
\chi_{1\mu} \in \PP~, &
\chi_{2\mu} \in \QQ~, &
\chi_{3\mu} \in \RR~, \\
\chi_\mu \in \PP\QQ\RR~, &
\chi \in \PP\QQ\RR\SS~, & &
\end{array}
\ee
where $\PP\QQ\RR \equiv \PP \oplus \QQ \oplus \RR$ and so on.
The spaces $\SS$, $\PP\QQ$ and $\RR$ are orthogonal with each other in
a sense that
\be
\begin{array}{cccc}
\tr\left[\overline\gf \Gamma_{i\mu}\right] = 0 &
(i = 1,2,3) &
\longrightarrow & \SS \perp \PP\QQ~, \RR \\
\tr\left[\overline\Gamma_3^\mu\Gamma_{i\mu}\right] = 0 &
(i = 1,2) &
\longrightarrow & \RR \perp \PP\QQ ~.
\end{array}
\ee
It is not difficult to check (cf. matrix forms in appendix
\ref{Matrix Representations for the Pion BS Equation})
that these orthogonalities conclude that the multiplications of the
subspaces $\SS$, $\PP$, $\QQ$, $\RR$ by the operators $T^0$, $T_\mu$
generate transitions between those spaces as
\be
\begin{array}{ll}
T^0 \hookrightarrow \SS~, &
T^0 \hookrightarrow \PP\QQ\RR~, \\
T_\mu : \SS \rightarrow S \cup \PP\QQ\RR ~, &
T_\mu \hookrightarrow \PP\QQ\RR ~,
\end{array}
\ee
where $T \hookrightarrow X$ means $T : X \rightarrow X$.
To say more concretely, for example, $T^0\chi^0$ is a element in the
subspace $\SS$ and so on.
Thus, $\PP\QQ\RR$ is a subspace; i.e., it is closed under the
multiplications by $T^0$ and $T_\mu$, and $\SS$ is a quotient space.
We note that the multiplication of $\SS$ by $T_\mu$ generate
inhomogeneous part in $\PP\QQ\RR$.
Similarly, it is easy to see
(cf. the appendix
\ref{Matrix Representations for the Pion BS Equation}) that
\be
\begin{array}{ccc}
K\hookrightarrow \SS~, &
K\hookrightarrow \PP\QQ~, &
K\hookrightarrow \RR~.
\end{array}
\ee
$\SS$, $\PP\QQ$ and $\RR$ are closed under the multiplication
by $K$.
Namely,
$\PP\QQ\RR$ is closed under the multiplications by $T^0$,
$T_\mu$ and $K$, and especially $\SS$ is also closed by means of
quotient space.
Therefore we conclude the eq.(\ref{eq: pion O(q)BS}) can be solved on
the subspace $\PP\QQ\RR$ and the zeros of the operator $T^0-K$ in
$\SS$ does not bring about that in $\PP\QQ\RR$.

\section{The Low Energy Parameter $L_{10}$ in QCD and S Parameter}
\reseteqnum

Strong interactions of the light degrees of freedom respect chiral
symmetry.
The symmetrical aspects of the strong interaction are expressed in
terms of the chiral Lagrangian.
There are ten low-energy parameters $L_1$, $\cdots$, $L_{10}$ which
are introduced by Gasser-Leutwyler\cite{b:GL} besides the pion
decay constant $f_\pi$.
The S parameter in QCD is related to one of the low-energy parameters
$L_{10}$ by the ratio $S=-16\pi L_{10}$.
The S parameter is a finite order parameter of the chiral
symmetry\cite{ILY} while the vacuum expectation value of the quark
bilinear suffers ultraviolet divergence and is to be regularized.

In this section we calculate the low-energy parameter $L_{10}$, or
so-called S parameter $S$ in QCD, and the pion decay constant $f_\pi$,
using the inhomogeneous BS equation in the improved ladder
approximation.
In order to hold the chiral symmetry, we keep the same approximation
to the SD equation which determines the quark mass function.
The obtained mass function is used in the inhomogeneous BS equation.
The QCD S parameter $S$ and the pion decay constant $f_\pi$ are
evaluated, in the chiral limit, from the current two-point function of
`V-A' type $\vma$ in the space-like region.

\subsection{QCD S Parameter and Two-Point Function}

In order to calculate the `V-A' tow-point function $\vma$ we
consider the vector and axial-vector currents $V^a_\mu(x)$ and
$A^a_\mu(x)$ defined by
\bea
V^a_\mu &=& \overline\psi\ta\gmd\psi ~, \nn
A^a_\mu &=& \overline\psi\ta\gmd\gf\psi ~,
\eea
where $\psi=(u,d)^T$ and $\tau^a$ ($a\!=\!1,2,3$) is Pauli matrix.
The two-point functions $\Pi_{JJ}$ of the current $J^a_\mu$
($=V^a_\mu, A^a_\mu$) is given by
\be
\delta^{ab}\Pi_{JJ}(q^2) = \epsilon^\mu \epsilon^\nu i\int d^4x
e^{iqx} \bra{0} T J^a_\mu(x) J^b_\nu(0) \ket{0} ~,\label{eq: pijj}
\ee
where $\epsilon^\mu$ is the polarization vector defined by
$\epsilon\!\cdot\! q=0$, $\epsilon^2=-1$.
The QCD S parameter $S$ and pion decay constant $f_\pi$ are calculated
using the two-point function as
\bea
S &=& 4\pi \frac{d}{dq^2}\Big[ \vma \Big]\bigg\vert_{q^2=0} ~,\nn
f_\pi^2 &=& \Pi_{VV}(0) - \Pi_{AA}(0) ~.\label{eq: s def}
\eea
First one of \eq{s def} is the definition of the QCD S parameter, and
second one is derived from the definition of $f_\pi$ using QCD sum
rule.
In what follows, the isospin indices $a, b, \cdots$ are suppressed for
simplicity.

Let us derive the sum rules of the quantities $S$ and $f_\pi$.
We expand the two-point Wightman function $W_{\mu\nu}(x)$ of the
current $J_\mu$ by using the physical complete set $1=\sum_n
\ket{n}\bra{n}$, and we insert the identity $1=\int\!
d^4q~\delta^4(q-q_n)$ in the summand.
Thus,
\bea
W_{\mu\nu}(x) &\equiv& \bra{0}J_\mu(x)J_\nu(0)\ket{0} \nn
&=& \sum_n \bra{0}J_\mu(0)\ket{n}
 e^{-iq_nx} \bra{n}J_\nu(0)\ket{0} \nn
&=& \int d^4q e^{-iqx} \sum_n \delta^4(q-q_n)
\bra{0}J_\mu\ket{n}\bra{n}J_\nu\ket{0} ~,\label{eq: wightman}
\eea
where $q_n^\mu$ is the momentum of the state $\ket{n}$ and we use the
short hand notation $\bra{0}J_\mu(0)\ket{n}=\bra{0}J_\mu\ket{n}$.
When we consider in the rest frame ($\vec q_n=0$) of the massive state
$\ket{n}$, we find
\be
\bra{0}J_\mu\ket{n} = 0 ~,
\ee
otherwise $\mu\!=\!0$ and $\ket{n}$ is scalar state, or
$\mu\!\ne\!0$ and $\ket{n}$ is vector state.
This follows from the fact that the quantity $\bra{0}J_\mu\ket{n}$
has to be a scalar.
Here, scalar and vector are the notions of the three-dimensional
continuous rotation in the rest frame of $\ket{n}$.
The states which appears in the summand of \eq{wightman} survive if
and only if those are spin-$0$ or spin-$1$ states.
The Wightman function $W_{\mu\nu}(x)$ consists of spin-$0$ part
$W_{\mu\nu}^\ze(x)$ and spin-$1$ part $W_{\mu\nu}^\on(x)$ as
\be
W_{\mu\nu}(x) = W_{\mu\nu}^\ze(x) +  W_{\mu\nu}^\on(x) ~.
\ee
We define the Fourier transform of $W_{\mu\nu}(x)$ as
\be
W_{\mu\nu}(q) = \int\!\frac{d^4x}{2\pi}~ e^{iqx} ~ W_{\mu\nu}(x) ~,
\ee
and similarly for $W_{\mu\nu}^\ze(x)$ and $W_{\mu\nu}^\on(x)$.

Massive spin-$0$ particles do not couple to conserved currents because
$0$ = $q^\mu\bra{0}J_\mu\ket{n}$ = $q^0\bra{0}J_0\ket{n}$ in the rest
frame.
Massless spin-$0$ particles can couple to conserved currents, which
defines the decay constant
\be
\bra{0}J_\mu\ket{n} = iq_\mu \, f_n^\ze ~.
\ee
We know only the pion as such particles, thus
\be
\bra{0}A_\mu\ket{\pi} = iq_\mu \, f_\pi ~,
\ee
and $f_n^\ze=0$ for others.
Then, the spin-$0$ part of the Wightman function $W^\ze(x)$ is
given as
\bea
W_{\mu\nu}^\ze(q) &=& (2\pi)^3 \sum_n
 \delta^4(q-q_n)
\bra{0}A_\mu(x)\ket{n}\bra{n}A_\nu(0)\ket{0} \nn
&=& (2\pi)^3\int\frac{d^4q'}{(2\pi)^3}\theta(q'{}^0)\delta(q'{}^2)
\delta^4(q-q') \Big( -f_\pi^2 q_\mu q_\nu \Big) \nn
&=& - f_\pi^2~ \theta(q^0)\delta(q^2) \; q_\mu q_\nu ~,
\eea
where $n$ runs over all spin 0 states.

The decay constants of massive spin-$1$ particles are defined by
\be
\bra{0} J_\mu \ket{n,\epsilon} = \epsilon_\mu ~ f_n^\on m_n ~,
\ee
where $m_n$ is the mass and $\epsilon^\mu$ is the polarization
vector of the state $\ket{n}$
which satisfies $\epsilon\c q_n=0$ and $\epsilon^2\!=\!-1$.
Of course, continuum states also have couplings $f_n^\on$.
Summing over the polarizations, we have
\bea
\sum_\epsilon \bra{0}J_\mu\ket{n,\epsilon}
\bra{n,\epsilon}J_\nu\ket{0}
&=& \sum_\epsilon \epsilon_\mu \epsilon_\nu
 \Big\vert f_n^\on m_n \Big\vert^2 \nn
&=& \left(-g_{\mu\nu} + \frac{q_{n\mu} q_{n\nu}}{q_n^2}\right)
 \Big\vert f_n^\on m_n \Big\vert^2 ~.
\eea
Then, we obtain
\bea
W_{\mu\nu}^\on(q) &=& (2\pi)^3 \sum_n
\delta^4(q-q_n)
\bra{0}J_\mu(x)\ket{n,\epsilon}\bra{n,\epsilon}J_\nu(0)\ket{0} \nn
&=& (2\pi)^3 \sum_n \int\frac{d^4q_n}{(2\pi)^3}
 \theta(q_n^0)\delta(q_n^2-m_n^2) \times
 \delta^4(q-q_n) \left(-g_{\mu\nu} + \frac{q_\mu q_\nu}{q^2}\right)
 \Big\vert f_n^\on m_n \Big\vert^2 ~\nn
&=& \sum_n \Big\vert f_n^\on m_n \Big\vert^2
 \left(-g_{\mu\nu} + \frac{q_\mu q_\nu}{q^2}\right)
 \theta(q^0)\delta(q^2-m_n^2) ~,
\eea
where $n$ runs over all spin 1 states.
Inserting the identity
\[
1 = \int_0^\infty\! ds ~ \delta(s-m_n^2)
\]
and putting
\be
\rho_{_J}(s) = \sum_n \Big\vert f_n^\on m_n \Big\vert^2
\delta(s-m_n^2) ~,
\ee
we finally find
\be
W_{\mu\nu}^\on(q) = \int_0^\infty ds ~ \rho_{_J}(s)
 \left(-g_{\mu\nu} + \frac{q_\mu q_\nu}{s}\right)
 \theta(q^0)\delta(q^2-s) ~.
\ee
The quantity $\rho_{_J}(s)$ is called the spectral function of the
current $J_\mu$ and represents the density of decay constants squared.

It is more convenient to rewrite the Wightman function in the form of
a commutator or $T$-product;
we easily obtain
\bea
\bra{0}[J_\mu(x), J_\nu(0)]\ket{0} &=& \delta_{_J,_A}\,f_\pi^2
{}~\partial_\mu \partial_\nu i\Delta(x;0)
- \int_0^\infty\!ds ~\rho_{_J}(s)
 \left(g_{\mu\nu} + \frac{\partial_\mu\partial_\nu}{s}\right)
i\Delta(x;s) ~,\nn
\bra{0}TJ_\mu(x)J_\nu(0)\ket{0} &=& \delta_{_J,_A}\,f_\pi^2
{}~\partial_\mu \partial_\nu\Delta_F(x;0)
- \int_0^\infty\!ds ~\rho_{_J}(s)
 \left(g_{\mu\nu} + \frac{\partial_\mu\partial_\nu}{s}\right)
\Delta_F(x;s) ~,\nn\label{eq: sum rule}
\eea
where $\Delta(x;s)$ and $\Delta_F(x;s)$ are the invariant delta
function and Feynman propagator respectively.
Using \eq{sum rule}, we have a spectral representation given by
\bea
\lefteqn{
i\int\!d^4x\;e^{iqx} ~ \bra{0}T\Big[ V_\mu(x)V_\nu(0)
- A_\mu(x)A_\nu(0) \Big]\ket{0} }\nn
&&\qquad = f_\pi^2~\frac{q_\mu q_\nu}{q^2 +i\epsilon}
- \int_0^\infty\!\frac{ds}{s}\;
\frac{\rho_{_V}(s)-\rho_{_A}(s)}{s-q^2-i\epsilon}
\Big( sg_{\mu\nu}-q_\mu q_\nu \Big) ~.\label{eq: sum rule currents}
\eea

Multiplying \eq{sum rule currents} by $q^\mu$ and using the current
conservations, we obtain the first Winberg sum rule\cite{Weinberg}
\be
\int_0^\infty\!\frac{ds}{s}\Big(\rho_{_V}(s)-\rho_{_A}(s)\Big)
= f_\pi^2 ~,
\ee
which means that the decay constants of the spin-$1$ states are
related to the pseudoscalar decay constant $f_\pi$.

Multiplying \eq{sum rule currents} by the polarization vectors
$\epsilon^\mu\epsilon^\nu$, we obtain the spectral representation of
the `V-A' two-point function
\be
\vma = \int_0^\infty\!\frac{ds}{s}~
\frac{\rho_{_V}(s)-\rho_{_A}(s)}{s-q^2-i\epsilon}
{}~.\label{eq: sp repr vma}
\ee
Thus, we find the sum rule of the QCD S parameter
\be
S = 4\pi\int_0^\infty\!\frac{ds}{s^2}~
\Big(\rho_{_V}(s)-\rho_{_A}(s)\Big) ~,
\ee
which is called Das-Mathur-Okubo sum rule\cite{DMO}.

\subsection{Finiteness of the `V-A' Two-Point Function}

Let us consider the following current-inserted three-point function
\bea
\lefteqn{ \delta_i^j~\frac{\tau^a}{2}
\int \frac{d^4p}{(2\pi)^4}~e^{-ipr}~
\chi_{\alpha\beta}(p;q,\epsilon) ~~ = }\nonumber\\
&&~~~~~~~~ \epsilon^\mu \int d^4x~e^{iqx}~
\Bigl\langle{0}\Bigr| {\rm T}\,\psi_{\alpha i}(r/2)\,
   \overline\psi_\beta^j(-r/2)\,J^a_\mu(x)
   \Bigl|{0}\Bigr\rangle~,
\eea
where $\chi$ ($=\chi_{_V}, \chi_{_A}$) is bispinor, which we call the
inhomogeneous BS amplitude.
The inhomogeneous BS amplitude $\chi$ has definite spin, parity and
charge conjugation quantum numbers; i.e., $J^{PC}=1^{--}$ for the
vector case and $J^{PC}=1^{++}$ for the axial-vector case.
We calculate these quantities $\chi_{_V}$, $\chi_{_A}$ using the
inhomogeneous BS equations in the improved ladder approximation.
Thanks to the current conservations, the amplitude $\chi$ has no
ultraviolet divergence.
Strictly speaking, the chiral WT identities may be violated by the
choices of running coupling (type (I) and (II)) in the improved
ladder approximation.
However, the violations occur in the order of $O(q_\mu)$, if any.
Thus, we find no ultraviolet divergence in the quantity $\chi$.
This fact is easily checked by explicitly writting down the Feynman
diagrams.

Closing the fermion legs of the three-point function, we find that the
two-point function $\Pi_{JJ}(q^2)$ is expressed in terms of the
inhomogeneous BS amplitude $\chi$ by
\bea
\Pi_{JJ}(q^2) &=&
   -\frac{1}{3}\sum_\epsilon \int\frac{d^4p}{i(2\pi)^4} \;
   \frac{N_c}{2}~{\rm tr}\Bigl[ (\epsilon \cdot G)\,
 \chi(p;q,\epsilon)\Bigr]~,\nonumber\\
G_\mu &=& \left\{ \begin{array}{ll}
   \gamma_\mu         & \mbox{ for vector vertex }\\
   \gamma_\mu\gamma_5 & \mbox{ for axial-vector vertex }
   \end{array}\right. ~.
\eea
We average over the polarization vectors $\epsilon_\mu$ for
convenience, becase $\Pi_{JJ}(q^2)$ defined by \eq{pijj} does not
depend on $\epsilon_\mu$.
This reason is also understood from \eq{sp repr vma}.
Although the vector and axial-vector two-point function themselves are
logarighmically divergent, the chiral symmetry guarantees the
finiteness of the `V-A' two-point function $\vma$;
The divergences appearing in each two-point function cancel, because
the structures of the divergences are exactly the same.\cite{ILY}
\begin{figure}[htbp]
\begin{center}
\ \epsfbox{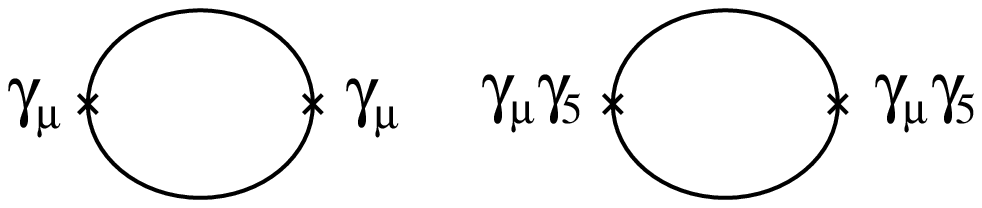}
\vspace{-15pt}
\caption[]{
The divergent Feynman diagrams contributing the two-point function
$\vma$.
Fermion lines represent full propagators of the quark in the
improved ladder approximation.
}
\label{fig: divergences}
\end{center}
\end{figure}
In the improved ladder approximation the divergent Feynman diagrams
contributing the two-point function $\vma$ are the `one-loop'
diagrams as shown in Fig.~\ref{fig: divergences} only.
It is easily checked that the ultraviolet divergences of these two
diagrams exactly cancel out.
As far as we hold the chiral symmetry, we find the `V-A' two-point
function $\vma$ as finite quantity.
In the numerical calculation we use the `regularization' by
introducing the Euclidean momentum cutoff and discretizing the
relative momenta $p$, $k$.
We should emphasize that calculations of finite quantities do not
depend on any regularization.
Then, in order to obtain the `V-A' two-point function, we carry out
the momentum integration after taking the difference; i.e.,
\be
\Pi_{VV}(q^2) - \Pi_{AA}(q^2) =
   -\frac{1}{3}\sum_\epsilon \int\frac{d^4p}{i(2\pi)^4} \;
   \frac{N_c}{2}~
   {\rm tr}\Big[\; \sle\chi_{_V}(p;q,\epsilon)
 - \sle\gf\chi_{_A}(p;q,\epsilon) \; \Big]~.
\label{eq: two-point function}
\ee
The numerical momentum integration of \eq{two-point function}
converges.

\subsection{Inhomogeneous BS Equation}

In this subsection we show the basic formulations for solving the
vector and axial-vector inhomogeneous BS equations in the improved
ladder approximation.
These equations are solved in the space-like region for a total
momentum $q_\mu$, $q_{_E}^2\equiv -q^2>0$.
SD equation is consistently solved in the same approximation.

In the improved ladder approximation with Landau gauge, the BS kernel
$K$ is given by
\be
K(p,k) = C_2g^2(p,k)\frac{1}{-l^2}\left( g_{\mu\nu}-\frac{l_\mu
l_\nu}{l^2} \right) \gmu \otimes \gnu ~,
\ee
where $l_\mu = (p-k)_\mu$.
We adopt the type (I) approximation for the running coupling;
$g^2(p,k)=g^2(\max(-p^2,-k^2))$.
The actual form of the running coupling $g^2(\mu^2)$ is found later.

When we solve the inhomogeneous BS equations, we need the full quark
propagator $S_F(p)$, which is calculated from the SD equation with
the same approximation as that for the BS equation, and takes the form
\be
S_F(p) = \frac{i}{\slp - \Sigma(-p^2)} ~.
\ee
The mass function is calculated from
\be
i\Sigma(-p^2) = K S_F(p) ~.
\ee

Now, the inhomogeneous BS equation for $\chi(p;q,\epsilon)$ is
\be
(T-K)\chi = (\epsilon\!\cdot\!G) ~,\label{eq: IBS}
\ee
where $T$ is the kinetic part defined by
\be
T = T(p;q) \equiv S_F^{-1}(p+\frac{q}{2})
   \otimes S_F^{-1}(p-\frac{q}{2}) ~.
\ee
The formal solution is given by
\be
\chi = \frac{1}{T - K} \; \slG ~.
\ee

Let us rewrite \eq{IBS} in a component form for the numerical
calculation.
The inhomogeneous BS amplitudes $\chi_{_V}$ and $\chi_{_A}$ are
expanded into eight invariant amplitudes $\chi_{_J}^1$, $\cdots$,
$\chi_{_J}^8$ as
\be
\chi_{{}_J}(p;q,\epsilon) =
\displaystyle\sum_{i = 1}^8 \Gamma^{(J)}_i(p;q,\epsilon)~
   \chi^i_{{}_J}(p;q)~,\qquad (~J = V, A~)
\ee
where $\chi^i_{{}_J} \: (i = 1,\cdots, 8)$ is scalar quantity.
$\Gamma^{(J)}_i$ is the vector or
axial-vector bispinor base defined by
\be
\ba{llll}
\Gamma^{(V)}_1 = \sle , &
\Gamma^{(V)}_2 = \frac{1}{2}[\sle ,\slp](p\c\whq) , &
\Gamma^{(V)}_3 = \frac{1}{2}[\sle ,\slqh] , &
\Gamma^{(V)}_4 = \frac{1}{3!}[\sle ,\slp,\slqh] ,\bigskip\\
\Gamma^{(V)}_5 = (\epsilon\c p) , &
\Gamma^{(V)}_6 = \slp(\epsilon\c p) , &
\Gamma^{(V)}_7 = \slqh(p\c\widehat{q})(\epsilon\c p) , &
\Gamma^{(V)}_8 = \frac{1}{2}[\slp,\slqh](\epsilon\c p) , \bigskip
\ea \label{eq. v base}
\ee
and
\be
\Gamma^{(A)}_i = \Gamma^{(V)}_i \gamma_5~, \label{eq. a base}
\ee
where $\widehat{q}_\mu = q_\mu/\sqrt{q_E^2}$ and $[a,b,c] \equiv
a[b,c] + b[c,a] + c[a,b]$.
The dependence on the polarization vector $\epsilon_\mu$ is
isolated in the base $\Gamma_i^{(J)}(p;q,\epsilon)$.

We introduce the real scalar variables $u$, $x$, $v$ and $y$ as
\bea
p\c \whq = -u , &~& p^2 = -u^2-x^2 , \nn
k\c \whq = -v , &~& k^2 = -v^2-y^2   ~.
\eea
Multiplying \eq{IBS} by the Dirac conjugate base
$\overline\Gamma_i$,
taking the trace and summing over the polarizations,
we convert it into the component form%
\footnote{
We evaluate the matrix elements of $T_{ij}(u,x)$ and
$K_{ij}(u,x;v,y)$ by REDUCE program.}
\be
\sum_j~( T_{ij} - K_{ij} )\, \chi^j = I_i~,\label{eq: ibs matrix}
\ee
where
\bea
I_i(u,x) &=& \Sp{ \overline\Gamma_i(p;\widehat{q},\epsilon)\, \slG }~,\nn\\
T_{ij}(u,x) &=& \Sp{ \overline\Gamma_i(p;\widehat{q},\epsilon)\,
   T(p;q)\, \Gamma_j(p;\widehat{q},\epsilon) }~,\nn\\
K_{ij}(u,x;v,y) &=& \int_{-1}^1d\cos\theta\;
   \Sp{ \overline\Gamma_i(p;\widehat{q},\epsilon)\,
   K(p,k) \,\Gamma_j(k;\widehat{q},\epsilon)}~.
\eea
Here $\theta$ is the angle between the three-vector part of $p$ and
$k$; $\cos\theta \equiv \bp\c\bkk /|\bp||\bkk|$.
The $v$ and $y$ integrations are promised when we multiply the
imvariant amplitude $\chi^j(v,y)$ by $K_{ij}(u,x;v,y)$.
We should note that the Dirac conjugation is taken to be
\be
\overline \chi(p;q,\epsilon)
   \equiv
   \gamma_0 \chi(p^*;q^*,\epsilon)^\dagger \gamma_0~.
\ee
The complex conjugate on $p$ and $q$ should be taken to preserve
Feynman causality of inhomogeneous BS amplitude when those momenta
become complex by analytic continuation.
In our choice of the bases (\ref{eq. v base}) and (\ref{eq. a base}),
the matrix $K_{ij}$ is independent of $q_{_E}^2\equiv-q^2$, so the
$q_{_E}^2$ dependence of $\chi^i$ solely comes from $T_{ij}$.

Here we notice that the invariant amplitude $\chi^i$ is an even
function in $u$:
\be
\chi^i(u,x) = \chi^i(-u,x)~.\label{eq. chi even}
\ee
This result follows from the charge conjugation properties
\bea
C \chi(-p;q,\epsilon)^T C^{-1} &=& -\chi(p;q,\epsilon)~,\nn
C \Gamma_i(-p;q,\epsilon)^T C^{-1} &=& -\Gamma_i(p;q,\epsilon)~,
\eea
where $C =  i\gamma_0\gamma_2$ is charge conjugation matrix.
Similarly from this property,
one can easily check that $T$ and $K$ are real definite and symmetric
matrices:
\bea
T_{ij}(u,x) &=& T_{ji}(u,x)~,\nn
K_{ij}(u,x;v,y) &=& K_{ji}(v,y;u,x)~.
\eea
This is an important property, from which we find that the
inhomogeneous BS amplitudes are real definite, and thus the
two-point functions are real definite.

We restrict the integral region of $v$ to be positive, subject to
replace the BS kernel in \eq{ibs matrix} as
\be
\int\!dv~ K_{ij}(u,x;v,y)\chi^j(v,y)
= \mathop{\int}_{v>0}\!dv~
 \Big[ K_{ij}(u,x;v,y) + K_{ij}(u,x;-v,y) \Big]\; \chi^j(v,y) ~.
\ee
Then, we treat all the variables $u$, $x$, $v$, $y$ as positive.
In the numerical calculation, we use the variables $U$, $X$, $V$, $Y$
defined by
\be
u = e^U ~,\quad x = e^X ~,\quad v = e^V ~,\quad y = e^Y ~,
\ee
and we discretize these variables at $N_{BS}=22$ points evenly spaced
in the intervals
\bea
   U, V &\in& \left[\;\iru\,,\,\uvu\;\right]
   = \left[\;-5.5\,,\,2.5\;\right]~, \nn
   X, Y &\in& \left[\;\irx\,,\,\uvx\;\right]
   = \left[\;-2.5\,,\,2.5\;\right]~.\label{eq: cutoff}
\eea
The momentum integration becomes the summation
\be
\mathop{\int}_{v>0}y^2dydv ~\cdots  \quad\longrightarrow\quad
DVDY\sum_{V,Y}VY^3~ \cdots ~,
\ee
where the measures $DV$, $DY$ are given by
\be
DV = \frac{\uvu-\iru}{N_{BS}-1} ~,\qquad
DV = \frac{\uvx-\irx}{N_{BS}-1} ~.
\ee
The BS kernel $K_{ij}(u,x;v,y)$ has integrable logarithmic
singularities at $(u,x)=(v,y)$.
To avoid the singularities we take the four-point average
prescription\cite{AKM} as
\bea
K_{ij}(u,x;v,y) &\longrightarrow&
\frac{1}{4}\Big[\;
K_{ij}(u,x;v_+,y_+) + K_{ij}(u,x;v_+,y_-) \nn
&&\qquad + K_{ij}(u,x;v_-,y_+) + K_{ij}(u,x;v_-,y_-) \;\Big] ~,
\eea
where
\be
v_\pm = \exp\Big( V\pm\frac{1}{4}DV \Big) ~,\qquad
y_\pm = \exp\Big( Y\pm\frac{1}{4}DY \Big) ~.
\ee
Now, we solve the inhomogeneous BS equation for $\chi_{_V}^i(u,x)$ and
$\chi_{_V}^i(u,x)$ separately.
We use {\small FORTRAN} subroutine package for these numerical
calculations.

\subsection{Results}

According to Ref.~\cite{AKM}, we adopt the following form of the
running coupling
\be
\alpha(\mu^2) \equiv \frac{g^2(\mu^2)}{4\pi} =
   \alpha_0 \times \left\{\begin{array}{ll}
\displaystyle \frac{1}{t} & \mbox{ if $t_F < t$ } \smallskip\\
\displaystyle \frac{1}{t_F} + \frac{(t_F - t_C)^2
   - (t - t_C)^2}{2t_F^2(t_F - t_C)} &\smallskip
   \mbox{ if $ t_C < t < t_F$ } \\
\displaystyle \frac{1}{t_F} + \frac{(t_F - t_C)^2}{2t_F^2(t_F - t_C)} &
   \mbox{ if $ t < t_C$ } \smallskip
   \end{array}\right.~,\label{eq. alpha}
\ee
where $\alpha_0 = 4\pi/9$, $t_C=-2$ and $t = \ln \mu^2$.

After obtaining the vector and axial-vector inhomogeneous BS
amplitudes $\chi_{_V}^i(u,x)$, $\chi_{_A}^i(u,x)$ numerically, we
calculate the `V-A' two-point function $\vma$ using
\eq{two-point function}.

We check all the dependences on the parameters; i.e., the momentum
cutoff \num{cutoff}, the lattice size $N_{BS}$ and a infrared cutoff
$t_F$ of the running coupling.
We use a suitable choice of them which affect the numerical error at
most a few percent.\cite{HaYo}
\begin{figure}[hbtp]
\begin{center}
\ \epsfbox{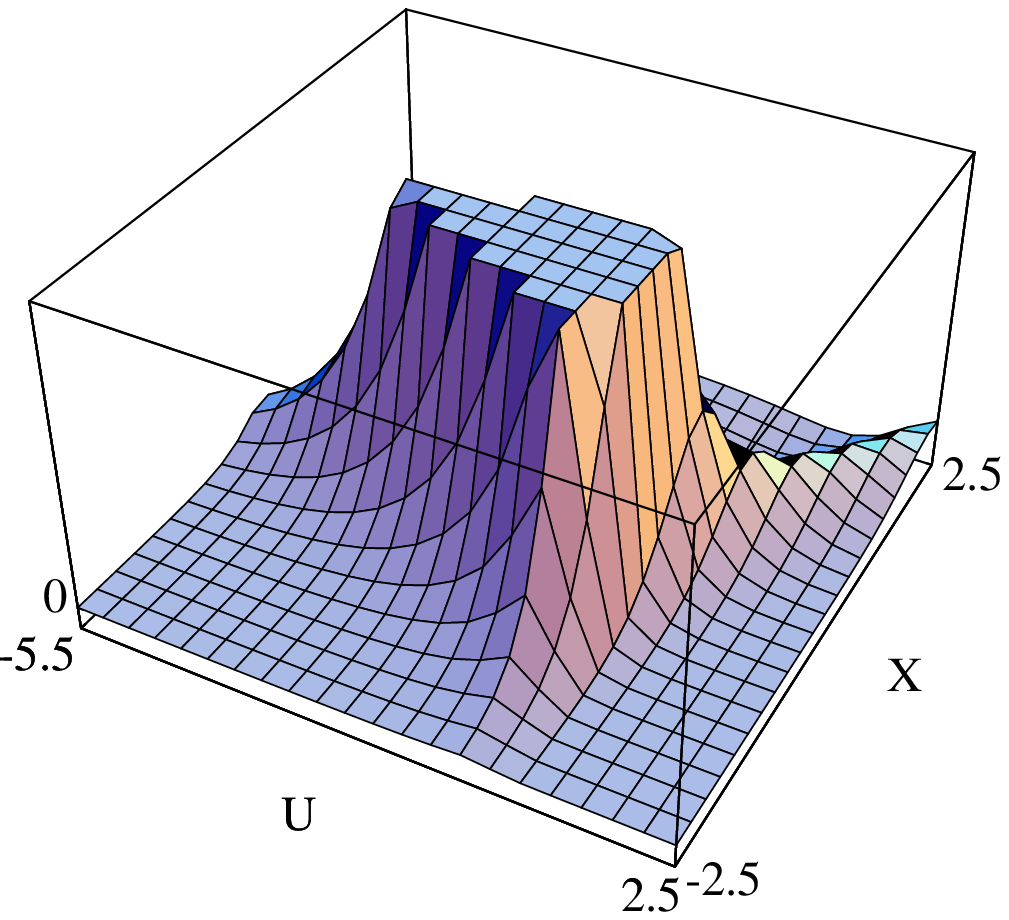}
\vspace{-15pt}
\end{center}
\caption[]{The typical form of the support of the `$V-A$' two-point
function.
The upper $9/10$ of the figure is cripped.}
\label{fig: support}
\end{figure}
The typical example of the support of the integration
\eq{two-point function} is shown in Fig.~\ref{fig: support}.
This figure explicitly shows the cancellation of the ultraviolet
divergences which results in the finiteness of the `V-A' two-point
function.

The QCD S parameter $S$ and the pion decay constant $f_\pi$ are
evaluated from formulae \num{s def}.
\begin{table}[hbtp]
\center{
\begin{tabular}[t]{|c||c|c|c|c|c|}
\hline
$t_F$
   & 0.3    & 0.5    & 0.7    & 0.9    & 1.1    \\\hline\hline
$S$
   & 0.432 & 0.464   & 0.478  & 0.481  & 0.470  \\ \hline
$(f_\pi/\Lambda_{\rm QCD})^2 \times 100$
   & 3.43  & 4.05    & 4.10   & 3.75   & 3.12   \\ \hline
$\Lambda_{\rm QCD}$ [MeV]
   & 502   & 462     & 459    & 481    & 526    \\ \hline
\end{tabular}}
\vspace{-7pt}
\caption[]{$t_F$ dependence of the value of the QCD $S$ parameter and
$f_\pi/\Lambda_{\rm QCD}$.
We also show the values of $\Lambda_{\rm QCD}$ calculated by imposing
$f_\pi = 93$ MeV.
We fix the lattice size as $N_{BS} = 22$.
}\label{tab: tF dependence}
\end{table}
We show the values of $S$ and $f_\pi$ for several values of $t_F$
in Table.~\ref{tab: tF dependence}.
We conclude that the QCD S parameter $S$ takes the value
\be
S = 0.43 \sim 0.48 ~,\label{eq: s value}
\ee
in the improved ladder exact approximation.
This value is 30\% larger than the experimental value\cite{b:GL}.
Our value \num{s value} implies that the chirl symmetry is dynamically
breaking in the improved ladder exact approximation.
In addition, our value of $\lqcd$ is consistent with the result from
the homogeneous BS equation for the pion\cite{ABKMN}.

Next, we extract the mass $m_\rho$ and decay constant $f_\rho$ of the
$\rho$ meson.
The decay constant of the $\rho$ meson is defined by
\be
\Bigl\langle{0}\Bigr| V_\mu^a(0) \Bigl|{\rho^b(q,\epsilon)}\Bigr\rangle =
   \delta^{ab}\epsilon_\mu f_\rho m_\rho~.
\ee
We use the $\chi^2$ method to extract these two values from the
functional form of our two-point function $\Pi(q_{_E}^2)\equiv\vma$.
The $\chi^2$ is defined by
\be
\chi^2 = \sum_{q_{_E}^2} \left\vert
\frac{\Pi(q_{_E}^2)-F(q_{_E}^2)}{\Pi(q_{_E}^2)} \right\vert^2 ~,
\label{eq: pole fit}
\ee
where $F(q_{_E}^2)$ is a three-resonance form given by
\be
F(q_{_E}^2) =
   \frac{f_\rho^2 m_\rho^2}{q_{_E}^2+m_\rho^2} -
   \frac{f_{_{R1}}^2 m_{_{R1}}^2}{q_{_E}^2+m_{_{R1}}^2} +
   \frac{f_{_{R2}}^2 m_{_{R2}}^2}{q_{_E}^2+m_{_{R2}}^2} ~.
\ee
\begin{table}[hbtp]
\center{
\begin{tabular}[t]{|c||c|c|}
\hline
\raisebox{-8pt}{\ }\raisebox{13pt}{\ }
& Our value & Experiment \\
\hline\hline
\raisebox{-8pt}{\ } $f_\rho$ [MeV] \raisebox{13pt}{\ }
& 133 & 144 $\pm$ 8 \\
\hline
\raisebox{-8pt}{\ } $m_\rho$ [MeV] \raisebox{13pt}{\ }
& 643 & 770 \\
\hline
\end{tabular}}
\caption[]{
The best fitted values of the mass and decay constant of $\rho$ meson.
We use our value of $\Lambda_{\rm QCD}$ ($ = 462$ [MeV]).
}\label{tab: rho}
\end{table}
Minimizing $\chi^2$ by the six parameters in $F(q_{_E}^2)$, we obtain
the best fitted values of $f_\rho$ and $m_\rho$, which is shown in
Table.~\ref{tab: rho}.
Our value of $m_\rho$ is free from the regularization of ultraviolet
divergences because of the finiteness of the `V-A' two-point function
$\Pi(q_{_E}^2)$.
This point is superiour to the result in Ref.~\cite{AKM}.
The masses and decay constants of the heavier mesons are unstable for
fitting;
we obtain several best fitting curves with different values of the
masses and decay constants of the heavier mesons,
although the best fitting curves seem to be almost the same.%
\footnote{
All best fitting curves satisfy the first and second Weinberg sum
rules.
}
On the other hand, the lowest meson mass and decay constant are
very stable.

\begin{figure}[hbtp]
\begin{center}
\ \epsfbox{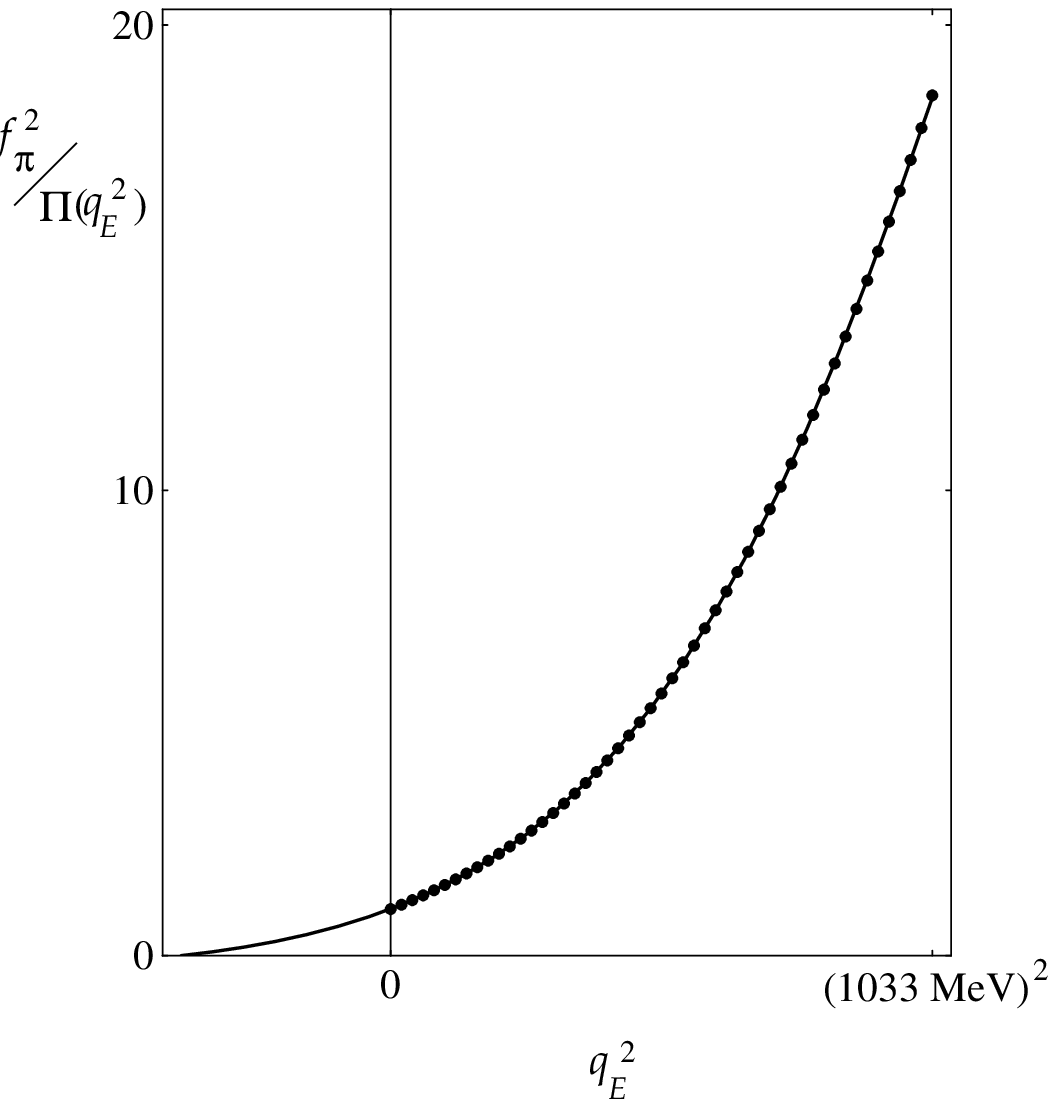}
\vspace{-7pt}
\caption[]{
Three pole fitting of our two-point function, $\Pi(q_E^2) =
\Pi_{VV}(q_E^2) - \Pi_{AA}(q_E^2)$.
The solid line is the best fitting curve, the dots denote the value of
our $\Pi(q_E^2)$.}
\label{fig: V-A}
\end{center}
\end{figure}

We find that the sum of the pole residues
$f_\rho^2m_\rho^2 - f_{R_1}^2m_{R_1}^2 + f_{R_2}^2m_{R_2}^2$
vanishes, which implies that our $\Pi(q_E^2)$ behaves as $1/q_E^4$ in
the high energy region.
This means that the spectral functions of our $\Pi(q_E^2)$ satisfies
the second Weinberg sum rule:
\be
\int_0^\infty ds \left[ \rho_V(s) - \rho_A(s) \right] = 0~.
\ee
The result from the improved ladder approximation reproduces the high
energy behavior of the `$V-A$' two-point function required by that
from Operator Product Expansion.

\newpage
\part{Heavy Quark Symmetry and Isgur-Wise Function}

\section{The Determination of $\Vcb$}
\reseteqnum

\subsection{The Basic Notion of the Heavy Quark Effective Theory}

The heavy quark effective theory (HQET) will give the most efficient
method to treat the decay constants of the heavy mesons and the
various form factors of the semi-leptonic heavy quark decay.
The form factors are given by appropriate overlap integrals between
the initial and final state wave functions of the heavy mesons.

The basic idea of the HQET is the following:
\begin{quote}
The heavy mesons such as $B$, $D$ and $D^*$ are the bound states
formed by the strong interaction.
The bound states consist of one heavy quark ($Q=b$ or $c$) and
one light antiquark ($\overline q=\overline u$ or $\overline d$).
This situation reminds us with the hydrogen atom in which the light
electron is bounded around the heavy proton.
\end{quote}

The masses of the heavy quarks are thought to be ideally infinity
$M_Q=\infty$ in the HQET.
We may expect that the $Q\overline q$ bound state by the strong
interaction has the same properties as that the hydrogen atom has.
Let us enumerate some of the important properties.
\begin{itemize}
\item[1)]
The heavy quark behaves like `free particle'.
Namely, the momentum recoil from the light antiquark, or we should say
the light degree of freedom, is negligible compared with the mass of
the heavy quark ($\lqcd / M_Q = 0$).
\item[2)]
{}From 1) the momentum of the heavy quark becomes a conserved
quantity, and we are allowed to take the rest frame of the heavy quark
as a Lorenz frame of the observer.
\item[3)]
{}From 2) the spin of the heavy quark becomes also a good quantum
number, and the heavy spin symmetry $SU(2)_{HS}$ arises.
The heavy flavor symmetry $SU(2)_{HF}$ emerges because the strong
interaction does not distinguish the flavor of the heavy quarks $b$
and $c$.
The symmetry group $SU(2)\times SU(2)$ enhances to $SU(4)$ like
non-relativistic $SU(6)$ model of light quarks.
Then the global symmetry $SU(2)_f$ which originally exists in the
total system ($u$, $d$ are doublet and $b$, $c$ are singlet) extends
to $SU(2)_f \times SU(4)_{HQ}$.
The symmetry $SU(4)_{HQ}$ is called the heavy quark spin-flavor
symmetry.
In this case the heavy mesons $D$, $D^*$, $B$ and $B^*$ form the
quartet of this symmetry.
\item[4)]
{}From 2) the BS amplitude $\chi=\bra{0}\T Q(x) \overline q(y)
\ket{X(q)}$ which is the wave function of the heavy meson
$X=D,D^*,B,B^*$ factories into the heavy quark sector and the light
quark sector:
\[
\chi = e^{-iqX} \Lp \times \left[
\ba{c}
\mbox{ the light quark sector which is described by }\\
\mbox{ only the light degrees of freedom in QCD }
\ea
\right] ~,
\]
where $X^\mu$ is the center-of-mass coordinate between $x^\mu$ and
$y^\mu$ and $\Lp=(1+\slv)/2$ with $v_\mu=q_\mu/M_B$ is the projection
operator which projects the heavy quark onto the positive energy.
\end{itemize}

Then, the form factors of the semi-leptonic $B$ decay given by overlap
integrals of the wave functions factorize into the heavy quark sector
and the sector of the light degrees of freedom.
The heavy quark sector takes a simple form which represents that the
`free' $b$ quark decays to the `free' $c$ quark though an appropriate
weak current.
The `free' heavy quark means that she propagates without any recoils
from the light degrees of freedom, and does not mean actual free
quark.
The light sector given by the overlap integral represents the
transition from the cloud around the $b$ quark formed by the light
degrees of freedom to the cloud around the $c$ quark.

It is very difficult to calculate these overlap integrals from the
first principle.
Fortunately, these overlap integrals are expressed in terms of a
single universal function, so-called the Isgur-Wise function.
When we consider the form factors of the transitions (decays) between
the multiplets of the heavy quark spin-flavor symmetry,
the marvelous theorem of Wigner-Eckert relates them to some universal
scalar quantity up to the Clebsch-Gordan coefficients.
Usually such universal scalar quantities are called reduced matrix
element.
The Isgur-Wise function is just the reduced matrix element:
\beann
\bra{D^*(\hq',\epsilon)}\overline c\gmd(1-\gf)b \ket{B(v)} &=&
\left[ (\vpv+1)\epsilon_\mu  -(\epsilon\!\cdot\!\hq)\hq'_\mu +
i\varepsilon_{\mu\nu\rho\sigma}\epsilon^\nu\hq'{}^\rho\hq^\sigma
\right]\xi(\vpv) ~, \\
\bra{D(v)}\overline c\gmd(1-\gf)b \ket{B(v)} &=&
(v'+v)_\mu \xi(\vpv) ~,
\eeann
and so on where we adopt the mass independent normalization condition
to the bound states $\iprod{B(v')}{B(v)} =
2v^0(2\pi)^3\delta^3(\bq-\bq')$.

The Isgur-Wise function is normalized absolutely at the kinematical
end point.
This fact is easily shown using the heavy quark spin-flavor symmetry
and the non-renormalization theorem of the conserved current.
The $B\rightarrow B$ matrix element of the $b$ quark current is given
in terms of the Isgur-Wise function $\bra{B(v')}\overline b\gmd
b\ket{B(v)} = (v'+v)_\mu \,\xi(\vpv)$, while the conserved current
$\overline b\gmd b$ receives no renormalization at the vanishing
momentum transfer $v'=v$.
Thus, we find $\xi(1) = 1$.

As for the leptonic decay of the heavy mesons, the corresponding
reduced matrix element is $F_B\sqrt{M_B}$ where $F_B$ is the decay
constant of the $B$ meson:
\[
\ba{ccccc}
\bra{0}\overline d \gmd\gf c \ket{D(v)} &=&
\bra{0}\overline u \gmd\gf b \ket{B(v)} &=&
  iF_B\sqrt{M_B} ~ v_\mu ~,\\
\bra{0}\overline d \gmd c \ket{D^*(v,\epsilon)} &=&
\bra{0}\overline u \gmd b \ket{B^*(v,\epsilon)} &=&
  F_B\sqrt{M_B} ~ \epsilon_\mu ~.
\ea
\]
Here we comment on the determination of the heavy meson decay
constant.
Usually the decay constant of a charged meson $M$ is determined from
the combined lepton decay $M\rightarrow l\overline\nu +
l\overline\nu\gamma$.
In the case of heavy mesons, there is another way to extract the decay
constant using the non-leptonic decays $B\rightarrow DD_s$ and
$B\rightarrow D^*D_s$.\cite{Rosner}
The decay constant $F_{D_s}$ is directly estimated from these
processes using the factorization hypothesis.
Decay constants of the other heavy mesons are calculated using the
heavy quark symmetry.

Therefore, once we know some two matrix elements, such as
$\xi(\vpv)$ and $F_B\sqrt{M_B}$,
the other channels are completely determined using the heavy
quark symmetry by virtue of the Wigner-Eckert theorem.

\subsection{Experimental Point of View}

\begin{figure}[bhtp]
\begin{center}
\ \kern-78pt\epsfbox{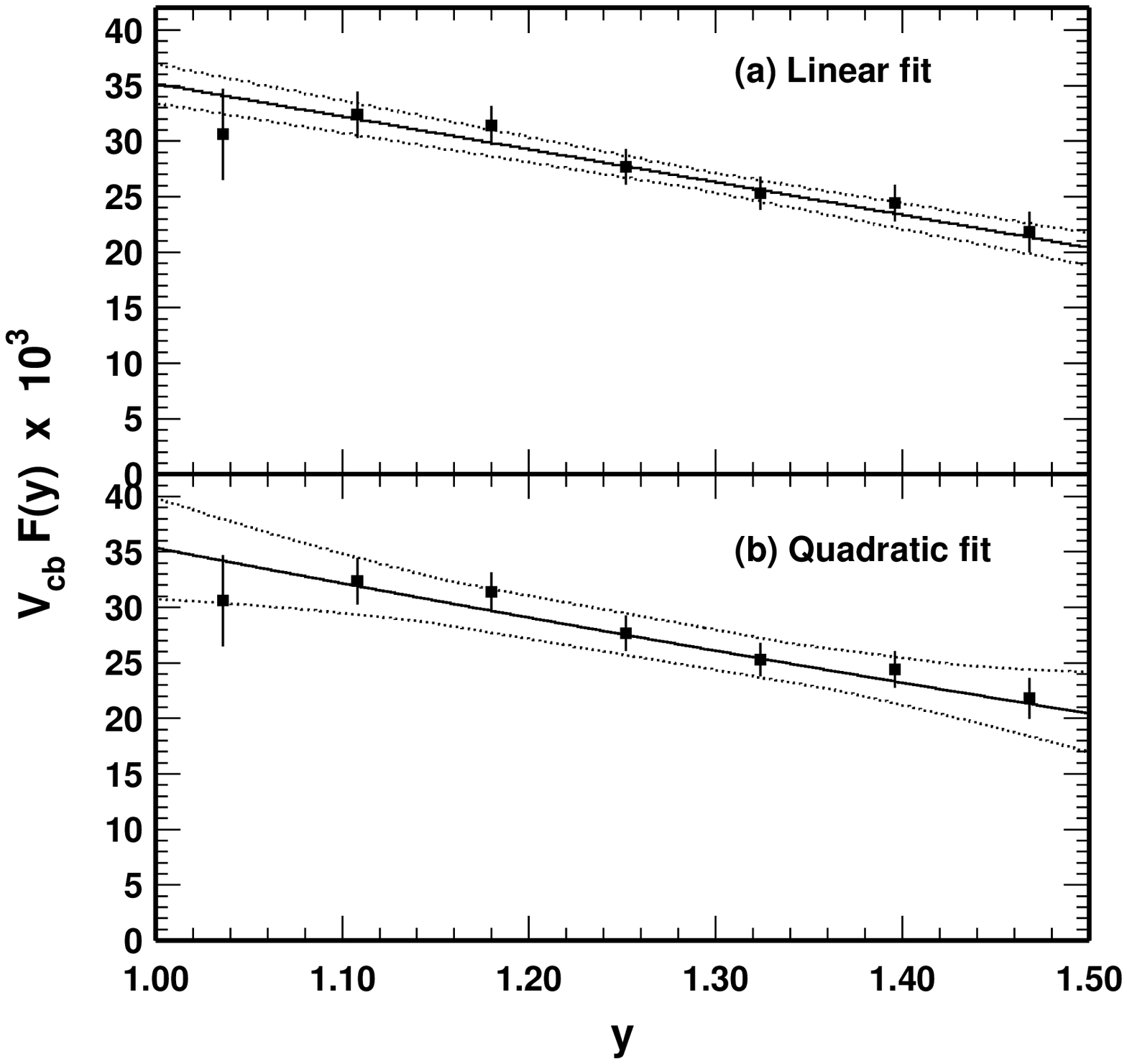}
\vspace{-5pt}
\caption[]{
The product $\Vcb\xi_{A1}(t)$ as a function of $t$ in the CLEO
data\cite{CLEO} where $y=t$ and $F=\xi_{A1}$.
$\Vcb$ is extracted from the intercept by the extrapolation to $t=1$.
The points are data for the decay rate of the decay $B\rightarrow
D^*l\overline\nu$ divided by the factors other than $\Vcb\xi_{A1}(t)$
of \eq{dif decay rates}.
The error bars are statistical only.
The dotted lines indicate the contours for $1\sigma$ variations of the
fit parameters.
They use (a) a linear function and (b) a quadratic function for the
extrapolation.
}
\label{fig: CLEO}
\end{center}
\end{figure}

The $\Upsilon(4S)$ resonance produces the $B^0\overline{B^0}$ or
$B^+B^-$ pair.
The $B$ mesons are produced almost in the laboratory frame because the
invariant mass of those pairs are almost the same as the mass of the
$\Upsilon(4S)$ resonance.
The semi-leptonic decays of a $B$ meson which is one of those are used
for the determination of $\Vcb$.
Let us define the four-velocities of the $B$ and $D^{(*)}$ mesons as
$v_\mu$ and $v_\mu{}'$ respectively.
Usually the kinematic variable is defined by $t=\vpv$ which is
essentially the momentum transfer $(q-q')^2$ to the lepton pair given
by
\be
t = \frac{M_B^2 + M_{D^{(*)}}^2 - (q-q')^2}{2M_BM_{D^{(*)}}} ~.
\ee
The momentum transfer has maximum value $(M_B-M_{D^{(*)}})^2$ at the
zero recoil point and must be positive because the invariant mass of
the lepton pair must be positive.
Thus we have the kinematically allowed region
\be
1 \le t \le \frac{M_B^2 + M_{D^{(*)}}^2}{2M_BM_{D^{(*)}}}
\cong 1.5 \sim 1.6 ~.\label{eq: t range}
\ee
The differential decay rates are given by\cite{Neubert91,Neubert94}
\bea
\frac{d\Gamma(B\!\rightarrow\! Dl\overline\nu)}{dt} &=&
\Vcb^2\frac{G_F^2}{48\pi^3} (M_B\!+\!M_D)^2M_D^3(t^2\!-\!1)^{3/2}
\left\vert\xi_+(t) \!-\!
\frac{M_B\!-\!M_D}{M_B\!+\!M_D}\xi_-(t)\right\vert^2 ,\nn
\frac{d\Gamma(B\!\rightarrow\! D^*l\overline\nu)}{dt} &=&
\Vcb^2\frac{G_F^2}{48\pi^3}
(M_B\!-\!M_{D^*})^2M_{D^*}^3(t^2\!-\!1)^{1/2}
F(t,\frac{M_B}{M_{D^*}}) \Big\vert \xi_{A1}(t)\Big\vert^2 ,\nn
\label{eq: dif decay rates}
\eea
where $\xi_\pm(t)$, $\xi_{A1}(t)$ are the form factors and $F$ is a
given function
\[
F(t,x) = (t+1)^2\left[1+\frac{4t}{t+1}\frac{x^2-2tx+1}{(x-1)^2}\right]
{}~.
\]
We will give the definitions of the form factors $\xi_\pm(t)$ and
$\xi_{A1}(t)$ later.
The charged lepton is regarded as massless without any significant
loss of accuracy.
The helicities of the $D^*$ meson are summed over.
Let us show the experimental results by the CLEO
collaborations\cite{CLEO} in fig.\ref{fig: CLEO}.
In the heavy quark effective theory the KM matrix element $\Vcb$ is
determined at the kinematical end point, where
the $B$ meson at rest decays into the $D$ or $D^*$ mesons at rest and
the lepton and neutrino pair are emitted back to back.
The kinematical end point is where the $D$ or $D^*$ meson receives no
momentum recoil.
It becomes possible in the heavy quark limit to determine the $\Vcb$
at the zero recoil point without any ambiguity stems from the
functional form of the form factors by virtue of the heavy quark
effective theory.

Moreover, the differential decay rate of $B\rightarrow
D^*l\overline\nu$, not $B\rightarrow D l\overline\nu$, is protected
from $1/M_Q$ order corrections which break the heavy quark
symmetry.\cite{Luke,BoydBrahm}
So, the determined value of $\Vcb$ in the heavy quark limit is correct
up to $1/M_Q$ order.
We notice that the form factors which are protected from $1/M_Q$
corrections at the kinematical end point are just $\xi_+(t)$ and
$\xi_{A1}(t)$, not $\xi_-(t)$.
Then, the semi-leptonic decay $B\rightarrow D^*l\overline\nu$ is the
best to extract $\Vcb$.\cite{Neubert94}

However there are large statistical error at the zero recoil point.
The zero recoil point (or kinematical end point) is actually the end
point in phase space, and phase space volume is almost zero.
The kinematical suppression factors such as $(t^2-1)^{3/2}$ and
$(t^2-1)^{1/2}$ appear in \eq{dif decay rates}.
The suppression for $B\rightarrow Dl\overline\nu$ is relatively
stronger.
We have thus relatively large statistical error because the number of
events are suppressed at the zero recoil point.

In addition, there is more serious problem of an ambiguity related to
the emission of the soft pion at the kinematical end point, which
enhances the relative systematic error at that point.
The $B$ meson decays to the standing $D$ or $D^*$ mesons.
Successively, the $D^*$ meson also decays to the $D$ meson with the
soft pion emitted:
\[
\ba{cc}
\mbox{ decay channel } & \mbox{ branching ratio } \\
\hline
D^{*\pm}(2010) \rightarrow D^0(1865) \pi^\pm(140) & 68 \% \\
D^{*\pm}(2010) \rightarrow D^\pm(1869) \pi^0(135) & 31 \%
\ea
\]
The value of the mass difference between $D^*$ and $D$ is almost the
same as that of the pion mass.
Then, the emitted pion does not acquire the kinetic energy, and does
not reach to the particle detectors.
We note that the lifetimes of the pion are $\tau_{\pi^\pm}=7.8$ m and
$\tau_{\pi^0}=25$ nm.
The decay $D^*\rightarrow D^\pm\pi^0$ is more serious than
$D^*\rightarrow D^0\pi^\pm$.
As a result we become hard to distinguish which process occur,
$B\rightarrow D^*$ or $B\rightarrow D$ at the zero recoil point.
We have thus relatively large systematic error.
The unexpected suppression of the differential decay rate in
Fig.\ref{fig: CLEO} at the kinematical end point might stem from the
above reason.
Here we have a comment.
As far as we believe the experimental data in Fig.\ref{fig: CLEO}, it
is seemed that the derivative of the form factor turns out to become
positive around $t=y=1.15$, and is positive in the kinematical end
point.
This corresponds that the slope parameter $\rho^2$ of the Isgur-Wise
function $\xi(t)$ becomes negative.
The slope parameter is defined by
\[
\rho^2 = - \left.\frac{d\xi(t)}{dt}\right\vert_{t=1} ~.
\]
However it is shown by the QCD sum rule with a rigorous
treatment\cite{Bjorken,IWc} that the slope parameter is bounded
from below $\rho^2>1/4$ in the infinite heavy quark mass limit.
Even though the actual data is not of the infinite heavy quark mass
limit, we think that a contradiction exists.

\subsection{Our Standpoint}

Taking the experimental difficulties at the kinematical end point into
account, therefore, in order to extract $\Vcb$ we are forced to
extrapolate the value of the differential decay rate at the zero
recoil point from the experimental data away from the point.
How do we, however, know the functional form of the form factor?
A linear and a quadratic functions are used in Ref.\cite{CLEO}.
The parameters in the functions such as $\rho^2$ are determined from
the experimental data by the $\chi^2$ fit.

There are several phenomenological models for the form factors.
Some legitimate forms of the Isgur-Wise function are
\bea
\xi_{_{Linear}}(t) &=& 1 - \rho^2(t-1) ~,\nn
\xi_{_{BSW}}(t) &=& \frac{2}{t+1}
\exp\left[\; -(2\rho^2-1)\frac{t-1}{t+1} \;\right] ~,\nn
\xi_{_{ISGW}}(t) &=& \exp\left[\; -\rho^2(t-1) \;\right]~,\nn
\xi_{_{KS}}(t) &=& \left(\frac{2}{t+1}\right)^{2\rho^2}
{}~,\label{eq: models}
\eea
with the slope parameter $\rho^2$ as an free parameter.
The first function is just a linear approximation.
The second one is called the BSW model\cite{BSW} and is derived from a
relativistic oscillator model\cite{NR92}.
The third one is called the ISGW model\cite{ISGW} and is calculated
from a valence quark model with the Coulomb and linear potential.
The last one is called KS model\cite{KS} or the pole ansatz model.
In Ref.\cite{KS} the exclusive semi-leptonic $B$ decays are studied in
the spectator quark model.
The helicity structure of the decays is matched to that of the
semi-leptonic $b\!\rightarrow\!c$ `free' quark decay at minimum
momentum transfer $(q-q')^2=0$.
The $(q-q')^2$ or $t$ dependence of the form factor is fixed by
nearest one meson dominance in the $\overline c\gmd(1-\gf)b$ current
channel, which produces a simple pole of the meson couples to that
current.
So, the slope parameter is given by $2\rho^2=1$ in the original
literature.

Let us carry out the $\chi^2$ fitting to extract $\Vcb$ using the
above four forms.
The experimental result of CLEO\cite{CLEO} is used.
The cited errors are only statistical, and the systematic errors are
not included.
We multiply the error bar at the kinematical end point by five just
for convenience.
\begin{table}[hbtp]
\begin{center}
\begin{tabular}{c|ccc}
model & $\Vcb$ & $\rho^2$ & $\chi^2$ \\
\hline
Linear & 0.0362 & 0.866 & 0.846 \\
BSW    & 0.0387 & 1.42  & 0.899 \\
ISGW   & 0.0376 & 1.16  & 0.806 \\
KS     & 0.0382 & 1.32  & 0.854
\end{tabular}
\caption[]{
The best fitted values of the KM matrix element $\Vcb$ and slope
parameter $\rho^2$.
These are determined from the experimental data of CLEO\cite{CLEO}.}
\label{tab: model fit}
\end{center}
\end{table}
We give the value of the KM matrix element $\Vcb$, the slope parameter
$\rho^2$ and $\chi^2$ extracted from the the experiment in
Table~\ref{tab: model fit}.
The $\chi^2$ is defined by
\be
\chi^2 = \sum_i \left\vert\frac{\Vcb\xi_{_X}(t_i) - F_i}{\sigma_i}
\right\vert^2 ~,
\ee
where $X= Linear$, $BSW$, $ISGW$, $KS$ and $t_i$ is experimental data
point, $F_i$ is experimental data of the combination $\Vcb\xi(t)$ at
$t_i$ and $\sigma_i$ is relative error at $t_i$.
It will be the best model that gives the least value of $\chi^2$ by
tuning $\Vcb$ and $\rho^2$.
The ISGW model gives the least $\chi^2$.
The values of the KM matrix element $\Vcb$ and slope parameter
$\rho^2$ deviate around 3\% and 24\% respectively.
The determined values of $\Vcb$ are rather stable over these four
models, but those of $\rho^2$ are not so stable.
Square root of the slope parameter $\rho$ represents the size of the
heavy mesons in unit of $\lqcd$.
The large value of $\rho^2$ means large size of the `cloud' around the
heavy quark in unit of the typical scale of binding energy and
vice-versa.
The size of the heavy meson becomes just input in the phenomenological
models ignoring the strong interaction dynamics of the light degrees
of freedom.

We would like to know the functional form of the form factor in more
model independent way.
For this purpose we need the detailed information of the wave function
of the heavy mesons based on the dynamics of the strong interaction.
Our main subject in Ref.\cite{KMY} is to calculate the form factors in
the heavy quark limit, i.e. the Isgur-Wise function, numerically from
the Bethe-Salpeter amplitudes of the heavy mesons.
The Isgur-Wise function is calculated as an overlap integral of the BS
amplitudes of the heavy mesons.

The BS amplitudes are calculated from the BS equation for the
pseudoscalar heavy meson with the three approximations i) constant
mass, ii) improved ladder and iii) heavy quark limit
($M_Q\rightarrow\infty$).
The first approximation is applied by replacing the mass functions of
the heavy and light quarks with appropriate constants, i.e. some kind
of a constituent mass.
This replacement is just for the simplicity of the calculations.
Ideally we should use the mass function with the same approximation
ii), but we leave it in the future work.
The typical values of the constant light quark mass is determined by
the conditions $m=\Sigma(cm^2)$ with $c=1,4$.
The mass function $\Sigma(-p^2)$ is calculated by the Schwinger-Dyson
equation with the improved ladder approximation.
As for the last one we adopt it to make good use of the basic idea
(includes the heavy quark symmetry) implemented in the heavy quark
effective theory.

\section{Isgur-Wise Function from Bethe-Salpeter Amplitude}
\reseteqnum

This section is devoted to the review of the literature in
Ref.\cite{KMY}.
We show the basic formulations for the Bethe-Salpeter equation.
We expand the BS equation in the inverse power of the heavy quark
mass and keep only the leading order terms.
The Isgur-Wise function is expressed by the overlap integral of the
leading BS amplitudes.
Finally we give the numerical results.

\subsection{BS Amplitude for the $B$ Meson}

We consider the bound state of the pseudoscalar heavy meson $\bB$.
The bound state momentum is $q_\mu$ and its mass is $M_B$, so
$q^2=M_B^2$.
The bound state consists of the heavy quark $\Psi$ and the light
anti-quark $\overline \psi$ with masses $M$ and $m$ respectively.
The BS amplitude $\chi_{\alpha\beta}(p;q)$ of the bound state $\bB$ is
defined by
\be
\bra{0}\T \Psi_{\alpha i}(x) \overline \psi_\beta^j(y)\bB
= e^{-iqX} \intdp~e^{-ipr} \delta_i^j ~\chi_{\alpha\beta}(p;q) ~,
\ee
where $\alpha, \beta$ are spinor indices and $i, j$ are color indices.
The Kronecker delta $\delta_i^j$ means that the bound state is color
singlet.
The center-of-mass coordinate $X^\mu$ and the relative coordinate
$r^\mu$ are defined by
\be
X^\mu = \zeta x^\mu + \eta y^\mu ~,\qquad r^\mu = x^\mu - y^\mu ~,
\ee
with weights
\be
\zeta = \frac{M}{M+m}~, \qquad \eta = \frac{m}{M+m}
 ~.\label{eq: weights}
\ee
Only the division with these weights \num{weights} allow us to
carry out the Wick rotation for any BS amplitudes.

The BS amplitude is decomposed into four invariant amplitudes $A$,
$B$, $C$, $D$ as
\be
\chi(p;q) = \sum_{i=1}^2\left[~
\Gamma^{(+)}_i(p;\hq)
\left(\ba{c}A(p;q)\\ B(p;q)\ea\right)^i
+ \Gamma^{(-)}_i(p;\hq)
\left(\ba{c}C(p;q)\\ D(p;q)\ea\right)^i
{}~\right] ~,\label{eq: inv hqBS amp}
\ee
where $\hq_\mu = q_\mu/M_B$ is the four-velocity of the $B$ meson.
The bispinor base $\Gamma^{(\pm)}_i(p;\hq)$ is defined by
\be
\Gamma^{(\pm)}_1(p;\hq) = \Lpm \gf ~,\qquad
\Gamma^{(\pm)}_2(p;\hq) = \Lpm \slbp \gf ~,
\ee
where $\bp^\mu \equiv p^\mu - (p\!\cdot\!\hq)\hq^\mu$ and
$\Lpm = (1\pm \slv)/2$ is the projection operator.
We note that the projection operators project an arbitrary state
of the heavy quark onto a positive energy or negative energy
amplitude and has the properties
\be
\Lpm\Lmp = 0 ~,\qquad \slv\Lpm=\pm\Lpm~, \qquad \Lpm\slbp\Lpm=0~.
\ee
The invariant amplitudes are scalar functions in $p^2$ and
$p\cdot q$.
It is convenient to introduce the real variables $u$ and $x$ by
\be
\pv = iu ~,\qquad p^2 = -u^2-x^2 ~,
\ee
when we carry out the Wick rotation on the BS amplitude.
We mean that $u$ is the time component of $p^\mu$ and $x$ is the
magnitude of the spatial component of $p^\mu$ in the rest frame for
the $B$ meson $q^\mu = (M_B,\vec 0)$.
The invariant amplitude is a scalar function in $u$ and $x$; i.e.
$X(p;q) = X(u,x)$ for $X = A,B,C,D$.

The conjugate BS amplitude $\overline\chi(p;q)$ of the bra-state $\kB$
is defined by%
\footnote{
This definition is different from the usual one by the negative sign.
}
\be
\kB \T \psi_{\beta j}(y) \overline \Psi_\alpha^i(x) \ket{0} = -
e^{iqX}\intdp~ e^{ipr} \delta^i_j ~\overline\chi_{\beta\alpha}(p;q)
\ee
The conjugate BS amplitude is also decomposed into four invariant
amplitudes $\overline A$, $\overline B$, $\overline C$, $\overline D$
as
\be
\overline \chi(p;q) = \sum_{i=1}^2\left[~
\overline\Gamma^{(+)}_i(p;\hq)
\left(\ba{c}\overline A(u,x)\\ \overline B(u,x)\ea\right)^i
+ \overline\Gamma^{(-)}_i(p;\hq)
\left(\ba{c}\overline C(u,x)\\ \overline D(u,x)\ea\right)^i
{}~\right] ~,\label{eq: inv conj BS amp}
\ee
where the conjugate bispinor base is defined by
$\overline\Gamma^{(\pm)}_i(p;\hq) \equiv
\gamma_0[\Gamma^{(\pm)}_i(p^*;\hq)]^\dagger\gamma_0$.
Once we know the BS amplitude $\chi$, its conjugate BS amplitude
$\overline\chi$ is given by the relation
\be
\overline\chi(p;q) = \gamma_0 [\chi(p^*;q)]^\dagger \gamma_0 ~,
\label{eq: conj BS amp}
\ee
where performing the Wick rotations on $\chi$ and $\overline\chi$ is
understood.
This equality is found to hold provided that the mass $M_B$ of the
bound state is real, using the BS equation for the conjugate
amplitude.
In the numerical calculation we actually encounter the complex value
of $M_B$.
This point will be explained later.
Comparing \eq{inv conj BS amp} with \eq{conj BS amp}, we have
$\overline X(u,x) = X^*(-u,x)$ for $X=A,B,C,D$.
Because imaginary number $i$ comes into $X$ solely in the
combination with $iu$, we find the relation
\be
X^*(-u,x) = X(u,x) ~,\label{eq: reality}
\ee
for $X=A,B,C,D$.

The Bethe-Salpeter equation for the $B$ meson in the improved ladder
approximation reads
\be
T \chi(p;q) = K\chi(p;q) ~.\label{eq: hqbs}
\ee
The kinetic part $T$ and the BS kernel $K$ are defined by
\bea
T(p;q) &=& S_H(p+\zeta q) \otimes S_L(p-\eta q) ~,\nn
K\chi(p;q) &=& \intdk \gmu\chi(k;q)\gnu D_{\mu\nu}(p,k)
\eea
with
\be
D_{\mu\nu}(p,k) = C_2g^2(p,k)\frac{1}{l^2}
\left(g_{\mu\nu}-\frac{l_\mu l_\nu}{l^2}\right)\; ~,\label{eq: Dmunu}
\ee
where $l_\mu=(p-k)_\mu$, $C_2=(N_c^2-1)/(2N_c)$ is the second Casimir
invariant of color gauge group.
The inverse propagators of the heavy and the light quarks are given in
the constant mass approximation by
\bea
S_H(p+\zeta q) &=& \slp + \zeta \slq - M ~,\nn
S_L(p-\eta q) &=& \slp - \eta \slq - m ~.
\eea

As we show later the BS equation becomes an eigenvalue equation of the
binding energy $E$ of the bound state and the BS amplitude in the
heavy quark limit.
The binding energy $E$ of the bound state is defined by
\be
M_B = M+m-E = \zeta^{-1}(M-\zeta E) ~.
\ee

For the development of the notation we introduce bra-ket notations
\be
\chi(p;q) \llra \kket{\chi} ~,\quad
\overline\chi(p;q) \llra \bbra{\chi} ~,
\ee
\be
A\chi(p;q) B  \llra A\otimes B \kket{\chi}~, \quad
\ds \intdk~K(p,k)\chi(k;q) \llra K\kket{\chi} ~,
\ee
It is convenient to introduce a inner product:
\be
\iiprod{\psi}{\chi} = N_c \intdp
 \tr\big[\,\overline\psi(p;q)\chi(p;q)\,\big] ~.
\ee
Then, the BS equation (\ref{eq: hqbs}) is rewritten by the simple form:
\be
T\kket{\chi} = K\kket{\chi} ~.\label{eq: bra-ket hqbs}
\ee

In the heavy quark system it is useful to adopt the following
normalization to the bound state:
\be
\iprod{B(q)}{B(q')} = 2\hq^0(2\pi)^3 \delta^3(\bq-\bq') ~,
\label{eq: hq norm}
\ee
where $\hq_\mu=q_\mu/M_B$.
We call this the mass independent normalization.
It is convenient to introduce the quantity
\be
\rho \equiv \hq^\mu \frac{\partial T}{\partial q^\mu} =
\zeta \slv\otimes S_L - \eta S_H\otimes\slv
{}~.\label{eq: def rho}
\ee
The normalization condition for the BS amplitude is given by the
Mandelstam formula, which reads
\be
\bbra{\chi}\rho\kket{\chi} = 2 ~.\label{eq: norm cond}
\ee

When we obtain the normalized BS amplitude, we can immediately
calculate the decay constant of the $B$ meson.
The decay constant of the $B$ meson is defined by
\be
i\hq_\mu\,F_B\sqrt{M_B}
 = \bra{0}\overline \Psi\gmd\gf \psi(0)\ket{B(q)} ~,\label{eq: FB def}
\ee
where the factor $\sqrt{M_B}$ comes from our definition of the state
normalization (\ref{eq: hq norm}).
Multiplying \eq{FB def} by $v^\mu$ and substituting the BS amplitude
$\chi$ into \eq{FB def}, we have
\be
F_B\sqrt{M_B}
 = -N_c \intdp \tr\Big[\slv\gf \chi(p;q)\Big] ~.
\ee
For the numerical calculation it is rewritten in terms of the invariant
amplitudes as
\be
F_B\sqrt{M_B} = N_c\int\frac{x^2dxdu}{2\pi^3}[ A(u,x) - C(u,x) ]
 ~.\label{eq: FB}
\ee

The decay constant of the $B$ meson is measured from the leptonic
decay induced by the non-conserving weak current $\overline u\gmd b$
(with ambiguous CKM mixing matrix element $\KM{ub}$).
The conservation of this current is violated only by the mass of the
heavy quark, so there is no ultraviolet divergence to cause the
renormalization of the current.
The current receives only a finite renormalization because it is a
dimensionful parameter that violates current conservation.
The decay constant is calculated using a ultraviolet cutoff in our
calculation.
Then, the decay constant of the $B$ meson is regarded as the
quantity in such cutoff scale, although it has no ultraviolet
divergence.
However, the resulting decay constant is directly connected with the
real observable because it has no dependence of renormalization point,
and is the on-shell quantity (of the $B$ meson).

In the full theory the renormalization constant of the current is
indeed finite and depends on the heavy quark masses by a logarithmic
correction.
When we take the heavy quark limit such `finite' renormalization
constant may diverges.
In the heavy quark effective theory where the heavy quark limit is
taken for the first time, the current needs renormalization
of the ultraviolet divergence because the theory is defined in the low
energy limit much lower than the scale of the heavy quark masses, and
the current depends on the renormalization
point.\cite{PW,FGGW,Neubert92}
Some of the two-point functions, say $\bra{0}T\overline u\gmd b(x)
\overline u\gmd b(0)\ket{0}$ has divergence because it is calculated
on the vacuum using the inhomogeneous BS equation.
This divergence is not of the current itself.

In the homogeneous BS formulation (with ladder diagrams), however, it
needs no renormalization at all likewise the full theory defined by
some kind of mass-dependent renormalization scheme\cite{NY} where the
threshold effect is naturally built in.
Any quantities, as well as the current $\overline u\gmd b$, which are
defined on the non-perturbative bound states, not the vacuum, have no
divergences.
We introduce a ultraviolet cutoff to define the system for
calculations, but it is just a convention for the calculations.

\subsection{Heavy Quark Mass Expansion}

Now, we expand all quantities in terms of the parameter $m/M$.
The light quark mass $m$ appearing in the expansion parameter is just
for the convenience to make dimension-less parameter.
In this sense we are allowed to use $\lqcd$ in stead of $m$.
The important point is that we should expand in the inverse power of
the heavy quark mass $M$.

The binding energy and BS amplitude are expanded as
\bea
\zeta E &=& E_0 + \mom E_1 + \mom^2 E_2 + \cdots ~,\nn
\chi &=& \chi_0 + \mom \chi_1
+ \mom^2 \chi_2 + \cdots ~.
%\kket{\chi} &=& \kket{\chi_0} + \mom \kket{\chi_1}
%+ \mom^2 \kket{\chi_2} + \cdots ~.
\eea
The expansion of the BS amplitude begins with the zeroth order in
$m/M$ because of the mass independent normalization \eq{norm cond}.
Even when we do not use such normalization condition, the difference
of the two cases is absorbed in the overall constant of amplitudes.
The expansion of the binding energy also begins with the zeroth order.
We think that the binding energy of the first few lowest-lying states
are of the order of the typical scale in the strong interaction.
Then, we regard it as the order one $O(1)$ quantity.
Although the binding energy depends on the heavy quark mass in
general, the dependence is of the sub-leading order in $O(1)$.

Expanding the quark inverse propagators, we find
\be
\ba{ccl}
S_H &=&\ds \mom^{-1} S_{H-} + S_{H0} + \mom S_{H1} + \cdots ~,\mm
S_{H-} &\equiv& -m(1-\slv) = -2m\Lm~,\mm
S_{H0} &\equiv& \slp - E_0 \slv ~,\mm
S_{H1} &\equiv& - E_1 \slv ~,\mm
\vdots &&\qquad\vdots
\ea
\ee
and
\be
\ba{ccl}
S_L &=&\ds S_{L0} + \mom S_{L1} + \mom^2 S_{L2} + \cdots ~,\mm
S_{L0} &\equiv& \slp - m(1+\slv) = \slp - 2m\Lp ~,\mm
S_{L1} &\equiv& E_0 \slv ~,\mm
S_{L2} &\equiv& E_1 \slv ~.\mm
\vdots &&\qquad\vdots
\ea
\ee
Then, the kinetic part $T$ is expanded by
\be
\ba{ccl}
T &=&\ds \mom^{-1} T^\mi + T^\ze + \mom T^\on + \cdots ~,\mm
T^\mi &\equiv& S_{H-}\otimes S_{L0} ~,\mm
T^\ze &\equiv& S_{H-}\otimes S_{L1} + S_{H0}\otimes S_{L0} ~,\mm
T^\on &\equiv& S_{H-}\otimes S_{L2} + S_{H0}\otimes S_{L1} +
S_{H1}\otimes S_{L0} ~.\mm
\vdots &&\qquad\vdots
\ea\label{eq: T exp}\ee
We expand the quantity $\rho$ defined in \eq{def rho} as
\be
\ba{ccl}
\rho &=&\ds \rho^\ze + \mom \rho^\on + \cdots ~,\mm
\rho_0 &\equiv& \slv \otimes S_{L0} - S_{H-}\otimes \slv~,\mm
       &=& \slv \otimes (\slp - 2m \Lp) + 2m \Lm \otimes \slv ~,\mm
\rho_1 &\equiv& \slv\otimes (S_{L1}-S_{L0})
 - (S_{H0}-S_{H-})\otimes \slv ~,\mm
       &=& 2E_0 \slv \times \slv - \slv\otimes (\slp - 2m\Lp)
 - (\slp + 2m\Lm)\otimes\slv ~,\mm
\vdots &&\qquad\vdots
\ea\label{eq: rho exp}\ee

The BS kernel $K$ is of order one quantity because it does not depend
on the heavy quark mass $M$.
Substituting all the above expansions into the BS equation
\eq{hqbs} or \eq{bra-ket hqbs}, we have equalities order by order in
$m/M$:
\be
\ba{ccl}
\ds\mom^{-1} &:& T^\mi\chi_0 = 0 ~,\mm
\ds\mom^0 &:&\ds T^\mi\chi_1 + T^\ze\chi_0
 = K\chi_0 ~,\mm
\ds\mom^1 &:&\ds T^\mi\chi_2 + T^\ze\chi_1
 + T^\on\chi_0 = K\chi_1~,\mm
\vdots &&\qquad\vdots
\ea\label{eq: hqexp}
\ee
We solve the set of the BS equation \num{hqexp} in the descending way.
The most important parts are of order $O(m/M)^{-1}$ and $O(m/M)^0$.
In the heavy quark limit only these two parts survive and describe the
light degrees of freedom.

\subsection{Heavy Quark Spin-Flavor Symmetry}

Since $S_{H-} \propto \Lm$ and $S_{L0}$ is invertible, the -1st order
equation (\ref{eq: hqexp}) reads
\be
\Lm\chi_0 = 0 ~.\label{eq: pos en cond}
\ee
We call this the positive energy condition.
This condition implies
\be
\chi_0 = \Lp\chi_0 ~. \label{eq: chi 0 cond}
\ee
This means that the leading BS amplitude $\chi_0$ involve no negative
energy state of the heavy quark.
Hence $\chi_0$ is decomposed into two invariant amplitudes $A, B$
appearing in \eq{inv hqBS amp}.

When we project the zeroth order BS equation (\ref{eq: hqexp}) onto
the positive energy states, we have
\[
\Lp T^\ze \chi_0 = \Lp K\chi_0 ~,
\]
where we use $\Lp T^\mi=0$.
Using \eq{chi 0 cond}, the explicit form of $T^\ze$ in \eq{T exp} and
$\Lp\gmu\Lp = \hq^\mu\Lp$, we finally obtain
\be
(\pv - E_0)\chi_0 (\slp-2m\Lp) = K_0 \chi_0 ~,\label{eq: leading hqbs}
\ee
where we define the kernel $K_0$ as
\be
K_0\chi = \int \hq^\mu \chi \gnu D_{\mu\nu} ~. \label{eq: K0}
\ee
This is the leading BS equation for the $B$ meson.

Now, we show that the leading BS equation has so-called the heavy
quark spin-flavor symmetry.
We can easily see that the equation possesses the heavy flavor symmetry
because it includes no heavy quark mass.
The heavy spin symmetry\cite{IWb} is also shown to be exist from a fact.
The multiplication from the left by any $2\times2$ matrix remains
invariant the leading BS equation \num{leading hqbs}.
We consider the case where those matrices have inverses.
The positive energy condition \num{pos en cond} should be kept, and
all matrices which commute with the projection operator $\Lm$ generate
the symmetry of the system.
In other words, the transformations which keep the positive energy
condition forms a group.
This group is found to be the spin rotation symmetry $SU(2)$ of the
heavy quark.\cite{KMY}

Let us show that the pseudoscalar and vector meson form doublet of the
spin symmetry.
As stated above the leading order BS amplitude for the pseudoscalar
meson has two invariant amplitudes:
\be
\chi_{0P} = \Lp\gf(A_P - B_P \slbp) ~.\label{eq: ps decomp}
\ee
While, the BS amplitude for the vector meson has eight invariant
amplitudes:
\be
\chi_V = \Lp\left[ \sle ( A_V - B_V \slbp )
+ \epsilon\!\cdot\! p ( C_V - D_V \slbp ) \right]
+ \Lm\left[ \sle ( A'_V - B'_V \slbp )
+ \epsilon\!\cdot\! p ( C'_V - D'_V \slbp ) \right] ~,
\ee
where $\epsilon_\mu$ is the polarization vector of the vector meson,
which satisfies $\epsilon^2 = -1$ and $\epsilon\!\cdot\! v = 0$.
The BS equation for the vector meson (as well as other type of mesons)
is exactly the same form as that for the pseudoscalar meson in
\eq{bra-ket hqbs}.
Then, the $-1$st order BS equation reads
\be
\Lm \chi_{0V} = 0 ~.
\ee
This implies that $A'_V$, $B'_V$, $C'_V$ and $D'_V$ vanish for the
leading BS amplitude.
The leading BS amplitude $\chi_{0V}$ of the vector mesons also
satisfies the leading BS equation in \eq{leading hqbs}.
This equation for the vector meson can be derived from the
pseudoscalar BS equation by virtue of the heavy quark spin-flavor
symmetry.
Let us explain this.
When we multiply \eq{leading hqbs} by $\sle\gf$ which keeps the
positive energy condition \num{pos en cond} from the left, the
equation itself does not change, but the BS amplitude changes to be
\be
\chi_{0P} = \Lp\gf(A_P-B_P\slbp)\quad
\stackrel{\textstyle\sle\gf}{\longrightarrow} \quad\Lp\sle(A_P-B_P\slbp)
{}~.
\ee
The resulting BS equation is nothing but for the vector meson
and determines the invariant amplitudes of the vector meson as
\be
A_V = A_P,\quad B_V = B_P,\quad C_V = 0,\quad D_V = 0 ~.
\ee
Then, once we obtain the invariant amplitudes $A$, $B$ of, say, the
pseudoscalar meson, we get both the pseudoscalar and vector BS
amplitudes as
\bea
\chi_{0P} &=& \Lp \gf ( A - B\slbp ) ~,\nn
\chi_{0V} &=& \Lp \sle ( A - B\slbp ) ~.
\eea

\subsection{Isgur-Wise Function}

It is enough to consider the matrix element $\bra{B(v')}\overline
b\gmd b\ket{B(v)}$ of the elastic electromagnetic scattering of the
pseudoscalar $B^-$ meson.
The current conservation is expressed by
\be
(v-v')^\mu \bra{B(v')}\overline b\gmd b\ket{B(v)} = 0 ~.\label{eq: cc}
\ee
Then, the matrix element is described in terms of a single elastic
form factor $f(t)$ as
\be
\bra{B(v')}\overline b\gmd b\ket{B(v)} = (v+v')_\mu~f(t)
{}~,\label{eq: el ff}
\ee
where $t=\vpv$.
The elastic form factor $f(t)$ approaches to the Isgur-Wise function
in the heavy quark limit.
When the matrix element is rewritten using the BS amplitude $\chi$ of
the $B$ meson, its general form includes the current inserted
five-point vertex function.
Here we use the improved ladder approximation for the BS amplitude,
while the current conservation \eq{cc} should be incorporated.
To satisfy this requirement we find\cite{KMY}
\be
\bra{B(v')}\overline b\gmd b\ket{B(v)} = N_c\intdpp
\tr\Big[\overline\chi(p';q') \gmd \chi(p;q) S_L(\mmc)\Big]
{}~,\label{eq: mat ele}
\ee
where the tree current $\gmd$ is used.
We impose the condition $\mmc$ which appears in the argument of the
light quark propagator $S_L$.
The momenta flowing in the light quark lines of the BS and conjugate
BS amplitudes should be equal.
We call this the momentum matching condition.
We expand the both sides of \eq{mat ele} in the inverse power of the
heavy quark mass $M$.
The leading term of the form factor $f(t)$ gives the Isgur-Wise
function $\xi(t)$.
Contracting the matrix element \num{mat ele} with $(v+v')^\mu$, we
have
\be
2(t+1)\xi(t) = N_c\intdpp \tr\Big[ \overline\chi_0(p';q')(\slv+\slvp)
\chi_0(p;q) S_{L0}(\mmcl)\Big] ~.\label{eq: iw1}
\ee

It is known that the Isgur-Wise function is absolutely normalized at
the kinematical end point by
\be
\xi(1) = 1 ~.\label{eq: iw norm}
\ee
This condition is satisfied in our formulation.
At the point $t=1$ \eq{iw1} reduces to
\be
2\xi(1) = N_c\intdp \tr\Big[ \overline\chi_0(p;q)\,\slv\,
\chi_0(p;q) S_{L0}\Big] ~.\label{eq: iw2}
\ee
Whereas, the normalization condition of the leading BS amplitude is
given from \eq{norm cond} as
\be
2 = N_c\intdp \tr\Big[ \overline\chi_0(p;q)\,\slv\,
\chi_0(p;q) S_{L0}\Big] ~.\label{eq: iw3}
\ee
The LHS of both equations \num{iw2} and \num{iw3} is exactly the same,
then we have \eq{iw norm}.
We recognize that the value of the Isgur-Wise function at the
kinematical end point reduces exactly to the normalization of the
leading BS amplitude.

In our formulation the Mandelstam normalization condition is rewritten
by
\be
\zeta \bra{B(v)}\overline b\gmd b\ket{B(v)}
- \eta\bra{B(v)}\overline u\gmd u\ket{B(v)}
= 2 v_\mu ~.\label{eq: mand2}
\ee
Taking the derivative with $\zeta$ at the point \num{weights}, we have
\be
\bra{B(v)}\overline b\gmd b\ket{B(v)}
+ \bra{B(v)}\overline u\gmd u\ket{B(v)} = 0 ~.\label{eq: mand3}
\ee
Using both \eqs{mand2} and \num{mand3} we obtain
\be
\bra{B(v)}\overline b\gmd b\ket{B(v)} =
- \bra{B(v)}\overline u\gmd u\ket{B(v)} = 2v_\mu
 ~.\label{eq: current norm}
\ee
The equations \num{current norm} and \num{el ff} states that the
Isgur-Wise function $\xi(t)$ and the elastic form factor $f(t)$ is
absolutely normalized by the non-renormalization theorem of conserved
current.
We have shown that this mechanism is realized in our formulation.
These interplay between the Mandelstam normalization condition and the
current conservation tells us what Feynman diagram we should use to
calculate the current matrix element.

The rest frame $v'{}^\mu = (1,\vec 0)$ is used for the calculation of
\eq{iw1}.
Then, the frame of the initial state is moving frame, and the
four-velocity $v^\mu$ takes the form
\be
v^\mu = (t, - \sqrt{t^2-1}, 0, 0) ~.
\ee
The integration variables $u'$, $x'$, $\ct$ are the components
of the relative momentum $p'$, and are given by
\be
p'{}^\mu = (iu',~ x'\ct,~ x'\sin\theta\sin\varphi,~
x'\sin\theta\cos\varphi) ~,
\ee
where the angle $\varphi$ is trivially integrated to give the overall
factor $2\pi$ because of the rotation symmetry in \eq{iw1}, so we may
put $\sin\varphi=1$ without loss of generality.
Let us move to the $v^\mu$-rest frame, where we have
\bea
v^\mu &=& (1,~ 0,~ 0,~ 0) ~,\nn
v'{}^\mu &=& (t,~ \sqrt{t^2-1},~ 0,~ 0) ~,\nn
p'{}^\mu &=& (iu't+x'\ct\sqrt{t^2-1},~ x't\ct+iu'\sqrt{t^2-1},~
x'\sin\theta,~ 0) ~,\nn
p^\mu &=& p'{}^\mu - m(v'-v)^\mu \nn
&=& (iu't+x'\ct\sqrt{t^2-1}-m(t-1),~ x't\ct+(iu'-m)\sqrt{t^2-1},~
x'\sin\theta,~ 0) ~.\nn
\eea
Thus, we find
\bea
u &\equiv& \pv/i ~~=~  u't
+ i\Big[ m(t-1)-x'\ct\sqrt{t^2-1} \Big] ~,\nn
x &\equiv& \sqrt{\vec p\cdot \vec p} ~=~
 \sqrt{\Big[ x't\ct+(iu'-m)\sqrt{t^2-1}\Big]^2
+ \big(x'\sin\theta \big)^2} ~. \label{eq: ux}
\eea

Let us rewrite \eq{iw1} into a component form.
Substituting \eq{ps decomp} into \eq{iw1} and taking the trace
over the spinor indices, we have
\be
\xi(t) = \frac{N_c}{t+1}\int\frac{du'x'{}^2dx'd\ct}{8\pi^3}~
\left(\ba{c}A(u',x')\\B(u',x')\ea\right)^T
\left(\ba{cc}L_{11} & L_{12} \\ L_{21} & L_{22}\ea\right)
\left(\ba{c}A(u,x)\\B(u,x)\ea\right) ~,
\ee
with the momentum matching condition $p'-mv'=p-mv$.
The $2\times 2$ matrix $L$ is defined by
\bea
L_{ij} &\equiv& \frac{1}{2}\tr\Big[\, \overline\Gamma_i(p';v')
(\slv+\slvp)
\Gamma_j(p;v) S_{L0}(\mmc) \,\Big]~,\nn
L_{11} &=& iu'(t+1) + k_1 ~,\nn
L_{12} &=& -(t+1)x^2 - (t^2-1)iu(iu-m) - k_1^2
+ k_1\Big((t+1)m-(2t+1)iu\Big) ~,\nn
L_{21} &=& -(t+1)x^2 - K_1(iu-2m) ~,\nn
L_{22} &=& (t+1)\Big(iut - (t+1)m\Big)x^2 + k_1^2(iu-2m) \nn
&& \qquad\qquad + k_1\Big[ tx^2 - (t+1)(iu-m)(iu-2m) \Big] ~,
\eea
where $k_1 = p'\cdot(v-v't) = x'\ct\sqrt{t^2-1}$.
At the kinematical end point, the matrix $L$ reduces to the weight
matrix $\rho_0$:
\be
L_{ij}\bigg\vert_{t=1} = 2\rho_{0ij}(u',x') ~,
\ee
and as is stated the calculation of the Isgur-Wise function exactly
gives unity, $\xi(1)=1$, by construction.

We should note that the variables $u$ and $x$ are complex, and we need
the leading BS amplitudes $A(u,x)$ and $B(u,x)$ with the complex
arguments.
At the kinematical end point $t\!=\!1$, the variables $u$ and $x$ are
real and identical to $u'$ and $x'$ respectively.
In order to find the leading BS amplitude $(A(u,x), B(u,x))^T$ with
complex arguments, we make use of the leading BS equation
\num{compo leading hqbs} itself:
\be
\left(\ba{c}A(u,x)\\B(u,x)\ea\right) =
\frac{\rho_0^{-1}(u,x)}{E_0-iu}
 \int\frac{y^2dydv}{8\pi^3} K_0(u,x;v,y)
\left(\ba{c}A(v,y)\\B(v,y)\ea\right) ~,\label{eq: BS itself}
\ee
We substitute values of $u$ and $x$ defined in \eq{ux} into the RHS of
\eq{BS itself}.
We are already know the solution $E_0$ and $(A(v,y), B(v,y))^T$ for
the ground state, then we also substitute this solution into the RHS
of \eq{BS itself}.
After carrying out the momentum integrations, we have the desired
quantity $(A(u,x), B(u,x))^T$.
We should notice that the above prescription to obtain the leading BS
amplitude with complex arguments is not the analytic continuation of
the leading BS amplitude with real arguments.

\subsection{Numerical Calculation}

\subsubsection{Running Coupling}
The running coupling in the one-loop approximation is given by
\be
\alpha(\mu^2) = \frac{g^2(\mu^2)}{4\pi}
= \frac{\alpha_0}{\ln\big(\mu^2/\lqcd^2\big)} ~,
\label{eq: one-loop alpha}
\ee
with
\be
\alpha_0 = \frac{12\pi}{33-2N_f} ~.
\ee
The constant $\alpha_0$ depends on the number $N_f$ of quark flavors.
The dominant part of the support of the leading BS amplitude lies
below the threshold of $c$ quark, so we put $N_f=3$ and then
$\alpha_0=4\pi/9$.

The running coupling \num{one-loop alpha} blows up at the $\lqcd$
scale.
For the numerical calculation we need the running coupling below the
$\lqcd$ scale, and we have to regularize it.
We have no guidance on the form of the running coupling outside the
deep Euclidean region $\mu^2\gg\lqcd^2$.
In what follows we rescale all dimensionful quantities by $\lqcd$.
The most simple prescription will be the Higashijima type:
\be
\alpha(\mu^2) = \frac{\alpha_0}{\ln\big[\max(\mu^2,x_0)\big]} ~,
\label{eq: higa type}
\ee
with a constant $x_0$, but this violates analyticity.
In order to calculate the Isgur-Wise function, the running coupling
should be analytic within the integration region in \eq{BS itself}
needed to evaluate the leading BS amplitude with complex arguments.

For this purpose we adopt the following form:
\be
\alpha(\mu^2) = \frac{\alpha_0}{\ln f(\mu^2)} ~,
\ee
with
\be
f(x) = x + \kappa\ln\left[\, 1
+ \exp\left(\frac{x_0-x}{\kappa}\right)\,\right] ~,
\ee
and use $\alpha(-p^2-k^2)$ for the argument.
The arbitrary constants $\kappa$ and $x_0$ are restricted to
$\kappa>0$ and $x_0>1$.
The function $f(x)$ approaches $x$ for large positive $x$, so that the
asymptotic behavior of the running coupling is correct.
The function $f(x)$ tends to a constant value $x_0$ for large negative
$x$ or $x_0-x\ll\kappa$.

In order to describe the chiral symmetry breaking of the light degrees
of freedom, the value of the running coupling in the infrared region
should be large enough.
We use the three values
\be
x_0 = 1.01\,,\qquad 1.05\,,\qquad 1.10 ~,
\ee
which correspond to $\alpha_{\max} = 140.3$, $28.6$ and $14.6$
respectively.

Next we determine the constant $\kappa$.
The function $f(x)$ has branch point singularities at
\be
x_s \equiv x_0 + i\pi\kappa(2n+1) ~,\qquad n \in {\bf Z} ~,
\ee
with the branch cuts extending along the real axis of $x$ to the left
if we take the principal value of the logarithm.
These singularities give rise to corresponding singularities in
$\alpha(\mu^2)$.
If we take the limit $\kappa\rightarrow0$, then the form of the
running coupling reduces to the Higashijima type \num{higa type}.
But the singularity comes into the region needed to calculate the
Isgur-Wise function.
So, $\kappa$ must be sufficiently large.
{}From the momentum matching condition, we have
\be
-p^2 = -p'{}^2 = 2im(t-1)u' - 2mx'\ct\sqrt{t^2-1} + 2m^2(t-1) ~.
\ee
We can avoid the singularities in the function $f(-p^2-k^2)$ if the
imaginary part of $-p^2-k^2$ is lower than that of $x_s$;
\be
\vert {\rm Im}(-p^2-k^2) \vert = 2m(t-1)u' < \min(x_s) ~,
\label{eq: ineq1}
\ee
in the case when the real part of $-p^2-k^2$ and $x_s$ are equivalent;
\bea
{\rm Re}(x_s) &=& {\rm Re}(-p^2-k^2) \nn
&=& -k^2 + u'{}^2 + \Big(\,x'-m\ct\sqrt{t^2-1}\,\Big)^2
- m^2(t-1)\Big( (t+1)\ct-2 \Big) \nn
&\ge& u'{}^2 - m^2(t-1)^2 ~.\label{eq: ineq2}
\eea
In this case from \eq{ineq2} $u'$ is limited
\be
u' \le u'_{\max} = \sqrt{x_0 + m^2(t-1)^2} ~.
\ee
Thus, from \eq{ineq1} we require
\be
\kappa > \frac{2m(t-1)}{\pi}\sqrt{x_0 + m^2(t-1)^2} ~,
\label{eq: kappa range}
\ee
for the range of the value of $t$ \num{t range}.
This requirement \num{kappa range} avoids all singularities and branch
cuts stem from the running coupling.

\subsubsection{Light Quark Mass $m$ and $\lqcd$}
It turns out that the leading order calculation need no particular
value of the heavy quark mass $M$, but we have to fix the light quark
mass $m$.
There is no unique definition of the constituent quark mass.
In our calculation the value of the light quark mass $m$ should be a
typical value of the mass function $\Sigma(-p^2)$ which is calculated
from the Schwinger-Dyson equation with the improved ladder
approximation in the chiral limit:
\be
\Sigma(x) = \frac{3C_2}{4\pi}\int_0^\infty\!\!ydy
\frac{\alpha(x+y)}{\max(x,y)} \; \frac{\Sigma(y)}{y+\Sigma(y)^2} ~.
\ee
We work with the following two definitions:\cite{GP}
\be
\ba{lllll}
m &=& \Sigma(m^2) &\qquad& \mbox{ type I } ~,\\
m &=& \Sigma(4m^2) &\qquad& \mbox{ type II } ~,
\ea
\ee
where we call them type I mass and type II mass respectively.

In addition, we have to know the value of the unit scale $\lqcd$ for
the evaluation of the dimensionful observables such as $F_B$ and
$E_0$.
We notice that the Isgur-Wise function is dimension-less, and it
requires no dimensionful scale unit.
We calculate the pion decay constant $F_\pi(PS)$ using the obtained
mass function $\Sigma(x)$ and the Pagels-Stokar formula\cite{PS} which
reads
\be
F_\pi(PS)^2 = \frac{N_c}{2\pi^2}\int_0^\infty\!\!xdx ~
\frac{\Sigma(x)\Big( \Sigma(x) - x\Sigma'(x)/2 \Big)}
{\big(x+\Sigma^2(x)\big)^2} ~.
\ee
Imposing the value $F_\pi(PS)=93\sqrt{2}$ MeV allows us to fix
$\lqcd$.

As a consistency check of our choice of the running coupling form,
we evaluate the vacuum expectation value of the quark bilinear
$\vev{\overline\psi\psi}_{\rm 1 GeV}$ using the formula
\be
\vev{\overline\psi\psi}_{\rm 1 GeV} = -
\left(\frac{\alpha(\Lambda^2)}{\alpha({\rm 1 Ge^2})}
\right)^\frac{9C_2}{11N_c-2N_f}
\frac{N_c}{4\pi^2}\int_0^{\Lambda^2}\!\!xdx~
\frac{\Sigma(x)}{x+\Sigma(x)^2} ~,
\ee
with $N_f=3$ and sufficiently large ultraviolet cutoff $\Lambda$.
The calculated values are indeed consistent with the value given by
Gasser and Leutwyler\cite{GL}
\be
\vev{\overline\psi\psi}_{\rm 1 GeV} = - \big( 225 \pm 25 {\rm MeV}
\big)^3 ~.
\ee

\begin{table}[hbtp]
\begin{center}
\begin{tabular}{c|cccc}
$x_0$ & $\lqcd$ & $m_{\rm type I}$& $m_{\rm type II}$
& $\big(-\vev{\overline\psi\psi}_{\rm 1 GeV}\big)^\frac{1}{3}$ \\
\hline
1.01 & 638 & 489 & 288 & 212\\
1.05 & 631 & 482 & 286 & 213 \\
1.10 & 625 & 474 & 285 & 214
\end{tabular}
\caption[]{
Parameter values used in our calculation.
Units are MeV.
}\label{tab: parameters}
\end{center}
\end{table}
We show all the results in Table~\ref{tab: parameters}.

\subsubsection{BS Amplitude}
Next, let us consider how we solve the leading BS equation
\num{leading hqbs}.
The notable property is that \eq{leading hqbs} is an eigenvalue
equation of the leading binding energy $E_0$ and leading BS amplitude
$\chi_0$.
In order to show this property more transparently we define the
`Hamiltonian' $H$ as
\be
H = (\pv) - \rho_0^{-1} K_0 ~.
\ee
Then, using $\chi_0(\slp-2m\Lp) = \rho_0\chi_0$, we have
\be
E_0\,\chi_0 = H\,\chi_0 ~. \label{eq: eigen hqbs}
\ee
For the purpose of the numerical calculation we rewrite the eigenvalue
equation \num{eigen hqbs} into the component form.
Multiplying by the conjugate bispinor base $\overline\Gamma_i^\p$ and
taking the trace we find
\be
E_0 \left(\ba{c}A(u,x)\\B(u,x)\ea\right) =
iu \left(\ba{c}A(u,x)\\B(u,x)\ea\right) \nn
+ \rho_0^{-1}(u,x)\int\frac{y^2dydv}{8\pi^3} K_0(u,x;v,y)
\left(\ba{c}A(v,y)\\B(v,y)\ea\right) ~,\label{eq: compo leading hqbs}
\ee
where $\rho(u,x)$ and $K(u,x;v,y)$ are the $2\times 2$ matrices
defined by
\bea
\rho_{0ij}(u,x) &=& \frac{1}{2}\tr \Big[
\overline\Gamma_i^\p(p;v) \rho_0 \Gamma_j^\p(p;v)\Big] \nn
&=& \left(\ba{cc} iu & -x^2 \\ -x^2 & x^2(iu-2m) \ea\right) ~,
\eea
and
\bea
K_{0ij}(u,x;v,y) &=& \int_{-1}^1\!\!d\ct~ \frac{1}{2}\tr
 \Big[ \overline\Gamma_i^\p(p;v) K_0(p;k) \Gamma_j^\p(k;v)\Big] \nn
&=& C_2g^2\left(\ba{cc}
I_1-(u-v)^2I_2 & i(u-v)(y^2I_2-I_2^{\bf pk}) \\
 -i(u-v)(x^2I_2-I_2^{\bf pk}) & I_1^{\bf pk} - (u-v)^2I_2^{\bf pk}
\ea\right) ~.\nn \label{eq: K0ij}
\eea
We define $\ct$ by the angle between the three-vectors of the
momentum $p^\mu$ and $k^\mu$ as
\be
\ct = \frac{\vec p \cdot \vec k}{|\vec p|\cdot| \vec k|} ~.
\ee
The quantities $I_1$, $I_2$, $I_1^{\bf pk}$ and $I_2^{\bf pk}$ are
defined by
\be\ba{lclcl}
I_1 &=& \ds\int_{-1}^1\!\!d\ct~\frac{1}{-(p-k)^2}
&=& \ds\frac{1}{2xy}\ln
\left( \frac{(x+y)^2+(u-v)^2}{(x-y)^2+(u-v)^2} \right) ~,\mm
I_2 &=& \ds\int_{-1}^1\!\!d\ct~\frac{1}{(p-k)^4}
&=& \ds\frac{2}{[(x+y)^2+(u-v)^2][(x-y)^2+(u-v)^2]} ~,\mm
I_1^{\bf pk} &=& \ds\int_{-1}^1\!\!d\ct~
\frac{\bp\cdot\bkk}{-(p-k)^2}
&=& \ds \frac{x^2+y^2+(u-v)^2}{2} I_1 - 1 ~,\mm
I_2^{\bf pk} &=& \ds\int_{-1}^1\!\!d\ct~
\frac{\bp\cdot\bkk}{(p-k)^4}
&=& \ds \frac{x^2+y^2+(u-v)^2}{2}I_2 - \frac{1}{2}I_1
{}~.\label{eq: I12pk}
\ea
\ee

For further development of the numerical calculation, just for saving
the memory in computers, first we restrict the integration region of
the variable $v$ in \eq{compo leading hqbs} to be positive by
replacing the BS kernel part as
\bea
\lefteqn{
\int\!\!dv K_0(u,x;v,y)\left(\ba{c}A(v,y)\\B(v,y)\ea\right)
\longrightarrow}\nn
&&\mathop{\int}_{v>0}\!\!dv
\left[\,K_0(u,x;v,y)\left(\ba{c}A(v,y)\\B(v,y)\ea\right) +
K_0(u,x;-v,y)\left(\ba{c}A(-v,y)\\B(-v,y)\ea\right) \,\right] ~.
\eea
We divide the invariant amplitude $X(u,x)$ for $X=A$, $B$ into the
even and odd part in $u$ as
\be
X(u,x) \longrightarrow
\left(\ba{c}X_e(u,x)\\X_o(u,x)\ea\right) \equiv
\left(\ba{c}X(u,x)+X(-u,x)\\X(u,x)-X(-u,x)\ea\right) ~.
\ee
When the invariant amplitude $X$ is of the real binding energy, this
replacement corresponds to the division into the real and imaginary
part because of \eq{reality}.
Rewriting the leading BS equation \num{compo leading hqbs} in terms
of the even odd components $X_e$ and $X_o$ for $X= A$, $B$,
we finally obtain
\bea
E_0 \left(\ba{c}A_e(u,x)\\A_o(u,x)\\B_e(u,x)\\B_o(u,x)\ea\right) &=&
\left(\ba{cc|cc}0&-u&&\\u&0&&\\\hline&&0&-u\\&&u&0\ea\right)
\left(\ba{c}A_e(u,x)\\A_o(u,x)\\B_e(u,x)\\B_o(u,x)\ea\right)
+ \rho_0^{-1}(u,x)\mathop{\int}_{v>0}\frac{y^2dydv}{8\pi^3}~ \nn
&\times&
\left[\,{\cal K}_0(u,x;v,y) + {\cal K}_0(u,x;-v,y)\Sigma_3 \,\right]
\left(\ba{c}A_e(v,y)\\A_o(v,y)\\B_e(v,y)\\B_o(v,y)\ea\right)
{}~.\label{eq: leading hqbs eo}
\eea
The weight matrix $\rho_0$ now becomes
\be
\rho_0(u,x)
= \left(\ba{cc|cc}0&-u&-x^2&0\\u&0&0&-x^2\\\hline
-x^2&0&-2mx^2&-ux^2\\0&-x^2&ux^2&-2mx^2\ea\right) ~.
\ee
The matrix $\Sigma_3$ is defined as
\be
\left(\ba{c}A_e(-v,y)\\A_o(-v,y)\\B_e(-v,y)\\B_o(-v,y)\ea\right)
=
\left(\ba{c}A_e(v,y)\\-A_o(v,y)\\B_e(v,y)\\-B_o(v,y)\ea\right)
=
\Sigma_3\left(\ba{c}A_e(v,y)\\A_o(v,y)\\B_e(v,y)\\B_o(v,y)\ea\right)
\longrightarrow
\Sigma_3\equiv
\left(\ba{cc|cc}1&0&&\\0&-1&&\\\hline&&1&0\\&&0&-1\ea\right)
{}~.
\ee
The quantity ${\cal K}_0(u,x;v,y)$ is the BS kernel
$K_{0ij}(u,x;v,y)$ in \eq{K0ij} extended to
the $4\times 4$ matrix given by
\be
{\cal K}_0 =
\left(\ba{cc|cc}
K_{011} & 0 & 0 & -{\rm Im}K_{012} \\
0 & K_{011} & {\rm Im}K_{012} & 0 \\
\hline
0 & -{\rm Im}K_{021} & K_{022} & 0 \\
{\rm Im}K_{021} & 0 & 0 & K_{022}
\ea\right) ~.\label{eq: calK0ij}
\ee

The fundamental variables used to solve the leading BS equation
numerically are $U$, $X$, $V$ and $Y$ defined by
\be
u = \exp U, \quad x = \exp X, \quad v = \exp V, \quad y = \exp Y ~.
\ee
We discretize these variables at $N_{BS}=30$ points evenly spaced in
the intervals
\be
U, V \in [\Lambda_{IRU},\Lambda_{UVU}] = [-10.0,2.5] ~,
\quad X, Y \in [\Lambda_{IRX},\Lambda_{UVX}] = [-4.5,4.0]
 ~.\label{eq: region}
\ee
Then, the integration appears in \eq{leading hqbs eo} becomes the
summation:
\be
\mathop{\int}_{v>0}\!\!y^2dydv \qquad\longrightarrow\qquad
 DVDY\sum_{V,Y} VY^3 ~,
\ee
with
\be
DV = \frac{\Lambda_{UVU}-\Lambda_{IRU}}{N_{BS}-1} ~,\qquad
DY = \frac{\Lambda_{UVX}-\Lambda_{IRX}}{N_{BS}-1} ~.
\ee
The BS kernel ${\cal K}_0(u,x;v,y)$ in \eq{calK0ij} has an logarithmic
integrable singularity at $(u,x) = (v,y)$, so we avoid it by the
four-point average prescription given by
\bea
\lefteqn{
{\cal K}_0(u,x;v,y) \quad\longrightarrow
 \frac{1}{4}\Big[ {\cal K}_0(u,x;v_+,y_+)
+ {\cal K}_0(u,x;v_+,y_-) }\nn
&& + {\cal K}_0(u,x;v_-,y_+) + {\cal K}_0(u,x;v_-,y_-) \Big] ~,
\eea
with
\be
v_\pm = \exp\left[ V \pm \frac{1}{4}DV \right] ~,\qquad
y_\pm = \exp\left[ Y \pm \frac{1}{4}DY \right] ~.
\ee

Now, we are in the position to solve the leading BS equation
\num{leading hqbs eo} as the eigenvalue equation numerically.
We use {\small FORTRAN} subroutine package for the eigenvalue problem.
The number of the point $N_{BS}$ is selected so that the
discretisation dependences of the decay constant $F_B$ and the binding
energy $E_0$ become small well within 1\%.
We chose the momentum region \num{region} so as for the supports of
the integrand of the normalization condition \num{norm cond} and decay
constant \num{FB} to be covered well enough.
These supports extend much farther into the infrared region than in
the case of the pion\cite{ABKMN}.
Consequently our results depend on the infrared behavior of the
running coupling.
We are therefore forced to regard $x_0$ as an input parameter in our
approach.

When we numerically solve the leading BS equation, we throw away all
abnormal solutions which have non-real eigenvalues, i.e. binding
energies.
We find the solution of the ground state; the largest binding energy
and its corresponding eigenvector.
This solution describes the lowest-lying quartet heavy mesons $B$,
$B^*$, $D$ and $D^*$ of the heavy quark spin-flavor symmetry.
The decay constant is easily calculated by \eq{FB} with $C\!=\!0$.
\begin{table}[hbtp]
\begin{center}
\begin{tabular}[h]{cc|ccccc}
type & $x_0$ & m & $F_B\sqrt{M_B}$ &
$E_0^{(0)}$ & $E_0^{(1)}$ & $E_0^{(0)}-E_0^{(1)}$ \\
\hline
 I & 1.01 & 489 & 2551 & 1795 & 1071 & 724 \\
 I & 1.05 & 482 & 3468 &  935 &  498 & 437 \\
 I & 1.10 & 474 & 4093 &  666 &  319 & 347 \\
\hline
II & 1.01 & 288 & 2052 & 1558 &  972 & 586 \\
II & 1.05 & 286 & 2738 &  799 &  447 & 352 \\
II & 1.10 & 285 & 3205 &  566 &  279 & 287
\end{tabular}
\caption[]{
$x_0$ dependence; $N_{\rm BS}=18$.
Units are MeV except for $F_B\sqrt{M_B}$
which is in (MeV)$^{3/2}$.
}\label{tab: E0FB coarser}
\end{center}
\end{table}
The results of the binding energy and the decay constant using a coarser
discretisation $N_{BS}\!=\!18$ is shown in
Table~\ref{tab: E0FB coarser}.
$E_0^\ze$ and $E_0^\on$ are the binding energies of the ground and
first excited states respectively.
Ideally we would use the excitation energies $E_0^\on\!-\!E_0^\ze$ to
fix the parameter $x_0$ or the light quark mass $m$, but at present
there is no experimental data for the masses of the radially excited
pseudoscalar $B(1st)$, $D(1st)$ or vector $B^*(1st)$, $D^*(1st)$
mesons.
\begin{table}[hbtp]
\begin{center}
\begin{tabular}[h]{c|ccccc}
type & m & $F_{B(5279)}$ &
  $F_{D(1869)}$ & $B(\hbox{1st})-B(5279)$ \\
\hline
 I & 482 & 48.0 & 80.6 & 432  \\
II & 286 & 37.8 & 63.4 & 350
\end{tabular}
\caption[]{
Results for the meson decay constants and
mass difference with $N_{\rm BS} = 30$ and $x_0 = 1.05$.
$B(\hbox{1st})$ denotes first excited state of
pseudoscalar $B$ meson. Units are MeV.
}\label{tab: E0FB finer}
\end{center}
\end{table}
The $B$ and $D$ meson decay constants and mass differences using a
finer discretisation $N_{BS}\!=\!30$ are given in
Table~\ref{tab: E0FB finer}.
As can be seen, our result for $F_B$ is smaller than $F_\pi$, and is
much smaller than other values obtained using QCD sum rules, potential
models and lattice simulations (see tables in Ref.\cite{Rosner}).
We note however that one gluon exchange interactions tend to give
small values\cite{Suzuki}, even these are somewhat larger than our
results.

The fact that the heavy meson decay constant is small means that the
BS amplitude at the origin is suppressed, and the probability density
leaks away from the origin so that the size of the heavy meson becomes
large.
We think that this fact is explained by our special prescription.
We use the improved ladder approximation which does not realize the
confining phenomena.
In general the confining phenomena is realized by linear
potential.\cite{KogutSusskind,Wilson}
However a special coupling choice of the improved ladder approximation
enables us to realize the linear potential in a phenomenological
way\cite{Richardson}.
First, we impose two requirements such that asymptotic freedom and
linear quark confinement be realized upon the non-relativistic quark
potential $V(r)$.
An simple interpolating form which invokes these requirements takes
the form
\be
V(r) = -\frac{4}{3} \int\!\!\frac{d^3\bp}{(2\pi)^3}
e^{-i\bp\cdot\mbox{\boldmath $r$}}
\frac{\alpha_s(\bp^2)}{\bp^2} ~,\label{eq: richa pote}
\ee
with
\be
\alpha_s(\bp^2) = \frac{4\pi}{9}\frac{1}{\ln(1+\bp^2/\Lambda^2)}
{}~.\label{eq: richa a}
\ee
Then if we use the form \num{richa a} in the improved ladder
approximation, we will have confining force to tighten the size of the
heavy meson.

Let us here comment on the occurrence of complex binding energies in
our calculation.
The leading BS equation reads
\be
E_0 \rho_0\kket{\chi_0} = {\cal A}\kket{\chi_0} ~,
\ee
where ${\cal A}=\rho_0 H$.
Both quantities $\rho_0$ and ${\cal A}$ are hermitian in the norm
defined by $(\chi',\chi)\equiv\bbra{\chi'}\rho_0\kket{\chi}$.
The special feature of the heavy quark system is that the natural norm
which emerges from the leading BS equation \num{eigen hqbs} itself is
exactly the same as of the Mandelstam formula in the heavy quark limit.
It is not difficult to find the relation
\be
(E_0-E'_0{}^*)\bbra{\chi'}\rho_0\kket{\chi} = 0
{}~,\label{eq: eigen val-vec rel}
\ee
where $\chi'$ and $\chi$ are the corresponding eigenvectors of the
binding energies $E'_0$ and $E_0$ respectively.
Then, the binding energy is real otherwise the corresponding
eigenvector has zero norm $\bbra{\chi}\rho_0\kket{\chi}=0$.
We actually encounter the non-real solutions for the binding energy and
its corresponding null eigenvectors.
These are abnormal solutions.
We think that this is stemmed from the non-positivity of the weight
matrix $\rho_0$.
The sight violation of the hermiticity of ${\cal A}$ by the four-point
splitting prescription will be not important.
If the weight matrix $\rho_0$ would be positive definite matrix, the
norm would be positive definite $\bbra{\chi}\rho_0\kket{\chi}>0$.
In this case we would never encounter the non-real solutions because
of \eq{eigen val-vec rel}.
We note that the abnormal solution has zero eigenvalue for the charge
of the conserved current $\overline b\gmd b$.

\subsubsection{Isgur-Wise Function}

The final form for the evaluation of the Isgur-Wise function needs
five dimensional integrations.
The $\ct$ integration is evaluated using the Gauss-Legendre
integration formula which gives rather precise results than
interpolatory integration formulae.
We save the discretisation point as $N_{GL}=10$.
\begin{figure}[hbtp]
\begin{center}
\ \epsfbox{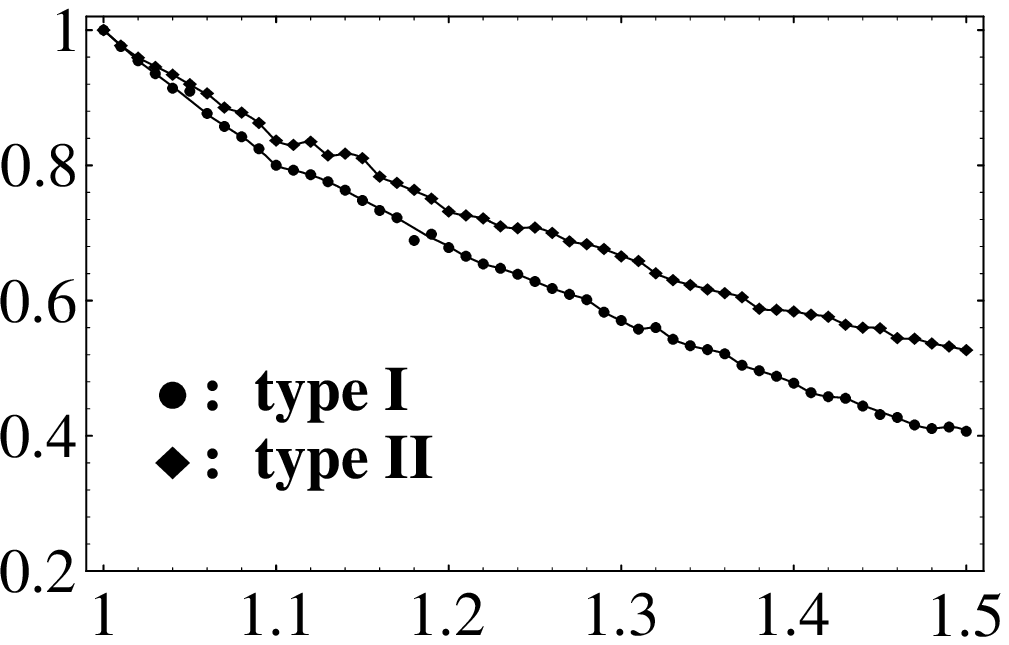}
\vspace{-15pt}
\caption[]{Isgur-Wise functions calculated with $N_{BS}=30$,
$N_{GL}=10$ and $x_0=1.05$ for both type I and type II masses.
}
\label{fig: main result}
\end{center}
\end{figure}
We show our main result in Ref.\cite{KMY} for the two types of light
quark masses in Fig.\ref{fig: main result}.

\begin{figure}[hbtp]
\begin{center}
\ \epsfbox{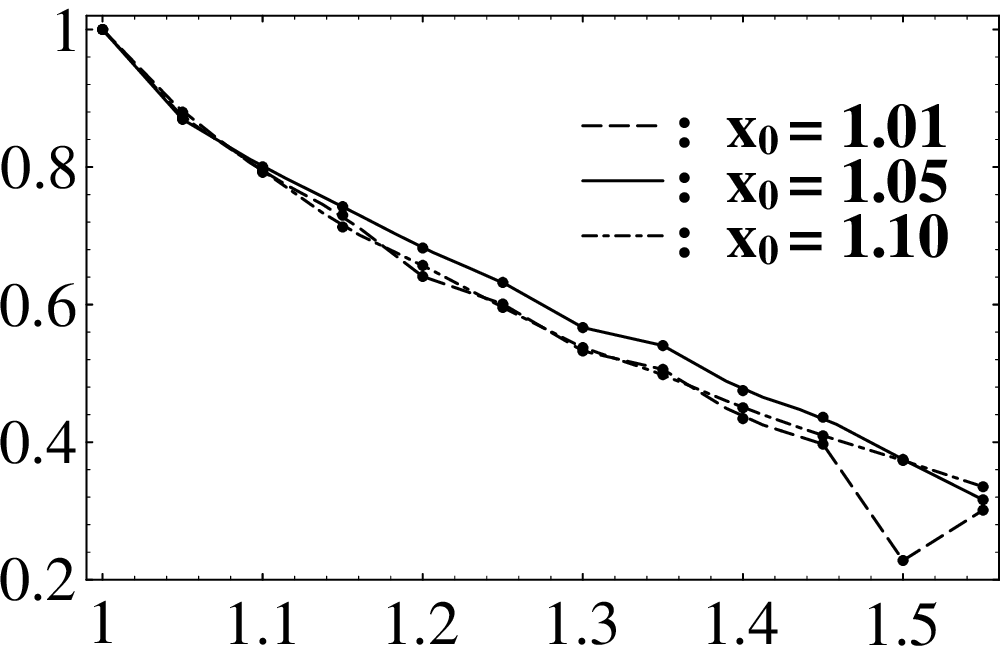}
\vspace{-15pt}
\caption[]{$x_0$ dependence of the Isgur-Wise function with type I
mass, $N_{BS}=18$ and $N_{GL}=10$.
}
\label{fig: x0depI}
\end{center}
%\end{figure}
%
%\begin{figure}[hbtp]
\begin{center}
\ \epsfbox{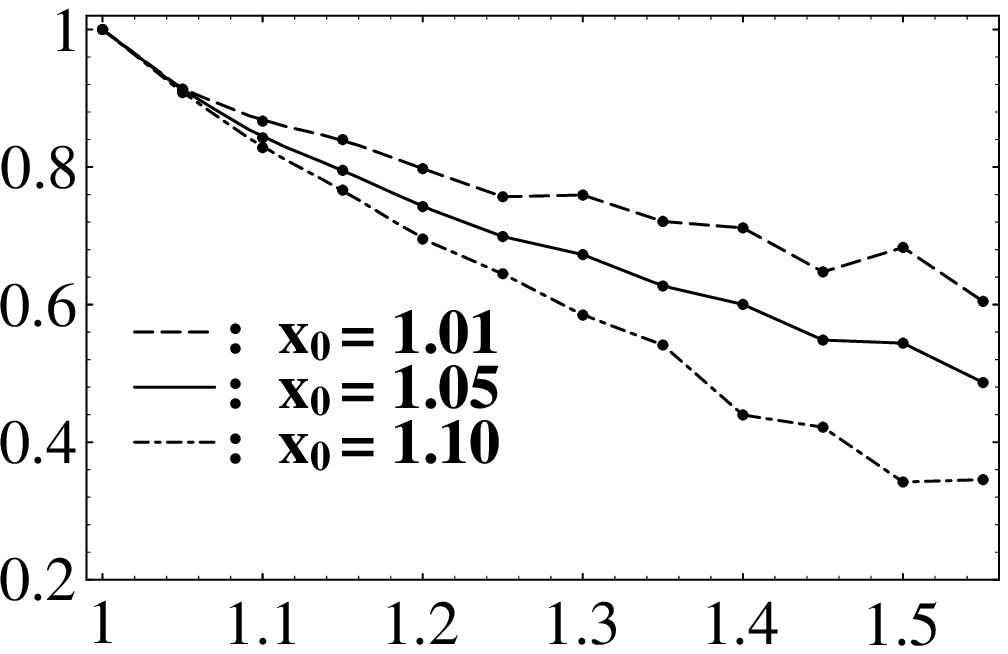}
\vspace{-15pt}
\caption[]{$x_0$ dependence of the Isgur-Wise function with type II
mass, $N_{BS}=18$ and $N_{GL}=10$.
}
\label{fig: x0depII}
\end{center}
\end{figure}
In Figs.~\ref{fig: x0depI} and \ref{fig: x0depII} we show how the
result depends on the parameter $x_0$, namely the low energy behavior
of the running coupling for both the type I and type II masses.
These data are evaluated from the leading BS amplitude with
$N_{BS}=18$ points.
We see that the result is almost independent of $x_0$ for the type I
case, whereas there is some dependence for the type II case.

Some kind of fluctuations on our Isgur-Wise function are observed.
The phenomena is explained in detail in Ref.~\cite{KMY}.
We think that these fluctuations might stem from the violation of the
analyticity of the leading BS amplitude.
The momentum integrations $v$ and $y$ in \eq{BS itself} cross over the
branch cut of the quantity $I_1$ appears in the BS kernel
$K_0(u,x;v,y)$.
The quantity $I_1$ defined in \eq{I12pk} has logarithmic singularity
at the point $(v,y)=(u,x)$ and the branch extends from the point.
We introduce the principal cut for the function $\ln(x)$.

The notable feature of our Isgur-Wise function is that the slope
parameter $\rho^2$ is quite insensitive to the behavior of the running
coupling in the low energy region as is shown in
Figs.~\ref{fig: x0depI} and \ref{fig: x0depII}.
The slope parameter slightly depends only on the value of the light
quark mass.
The values of the slope parameter $\rho^2$ for types I and II are as
large as 2.0 and 1.8 respectively, which are somewhat larger than the
known predictions.
The estimated values of the slope parameter $\rho^2$ in literatures
are given by $0.7\pm0.2$\cite{Neubert94}, $1.0\sim1.1$\cite{JHD}

There is some controversy over the upper bound of the slope parameter.
In Ref.~\cite{BMR} the upper bound is given by $\rho^2 \le 1.42$.
Recently de Rafael and Taron\cite{RT} have argued that the form factor
$F(q^2)$ of the $B$ meson elastic scattering must satisfy the
constraint
\be
F_-(q^2) < F(q^2) < F_+(q^2) ~,
\ee
where $F_\pm(q^2)$ are given functions.
As a result, they give a bound
\be
\left.\frac{dF(q^2)}{dq^2}\right\vert_{q^2=0}
= \frac{1}{2(D-1)}\langle r^2 \rangle
\le \frac{1}{16M_B^2} \left[ D - \frac{5}{2} + \sqrt{\frac{2^5}{\pi C}-1}
\right] ~,
\ee
where $D$ is the dimension of the spacetime and
\be
C = \frac{15\sqrt{\pi}}{2^{D/2+4}N_c
\Gamma\bigg(\ds\frac{D+1}{2}\bigg)\Gamma\bigg(\frac{6-D}{2}\bigg)} ~.
\ee
In Refs.~\cite{GM,FLW} they argue that it is impossible to obtain
any such constrains at all when we carefully reconsider the heavy
quark and light antiquark bound states lying below the heavy meson
pair production threshold.
\begin{figure}[hbtp]
\begin{center}
\ \epsfbox{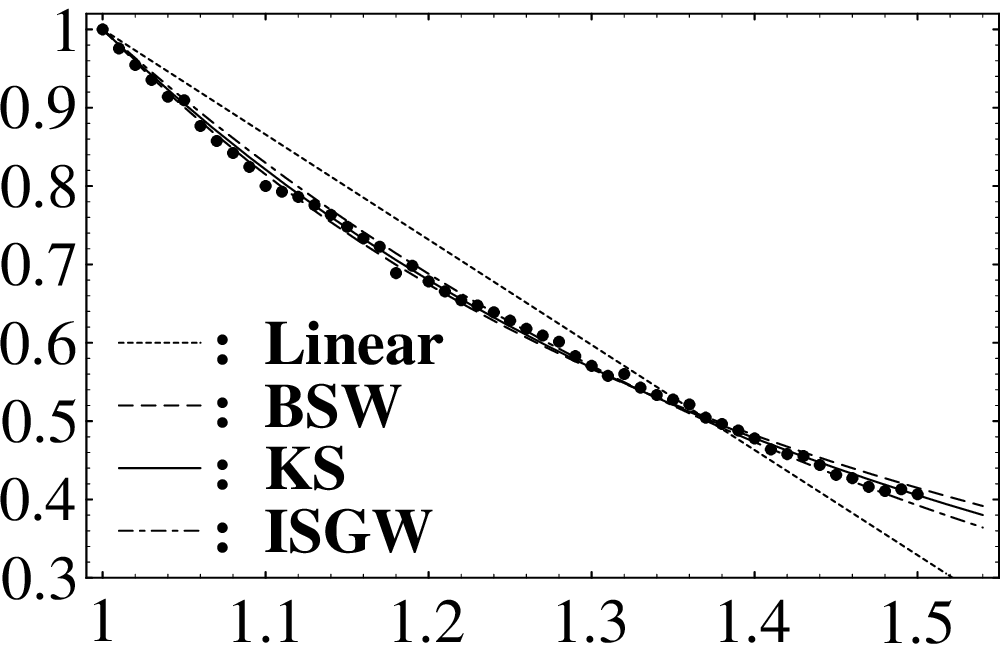}
\vspace{-15pt}
\caption[]{Phenomenological forms of the Isgur-Wise function.
The slope parameters are extracted from our Isgur-Wise function in
Fig.~\ref{fig: main result} with type I mass.
}
\label{fig: modelfitI}
\end{center}
%\end{figure}
%
%\begin{figure}[hbtp]
\begin{center}
\ \epsfbox{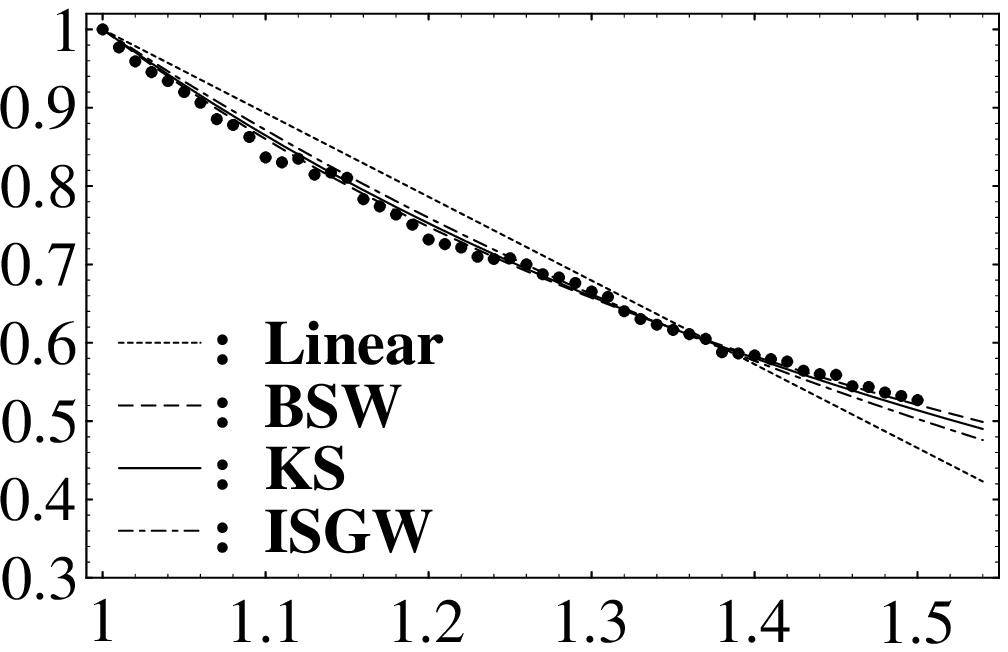}
\vspace{-15pt}
\caption[]{Phenomenological forms of the Isgur-Wise function.
The slope parameters are extracted from our Isgur-Wise function in
Fig.~\ref{fig: main result} with type II mass.
}
\label{fig: modelfitII}
\end{center}
\end{figure}

We use our Isgur-Wise function in Fig.~\ref{fig: main result} with
type I and II masses to estimate the slope parameter $\rho^2$ in four
phenomenological models by the $\chi^2$ fitting method as shown in
Figs.~\ref{fig: modelfitI} and \ref{fig: modelfitII}.
\begin{table}[hbtp]
\begin{center}
\begin{tabular}{|c|cc|}
\hline
\multicolumn{3}{|c|}{ type I } \\
\hline
model & $\rho^2$ & $\chi^2$ \\
\hline
Linear & 1.34 & 0.083 \\
BSW    & 2.14 & 0.0038 \\
ISGW   & 1.87 & 0.0055 \\
KS     & 2.02 & 0.0028 \\
\hline
\end{tabular}
\qquad
\qquad
\begin{tabular}{|c|cc|}
\hline
\multicolumn{3}{|c|}{ type II } \\
\hline
model & $\rho^2$ & $\chi^2$ \\
\hline
Linear & 1.07 & 0.058 \\
BSW    & 1.57 & 0.0033 \\
ISGW   & 1.38 & 0.012 \\
KS     & 1.49 & 0.0055 \\
\hline
\end{tabular}
\caption[]{The values of the slope parameter extracted from our
Isgur-Wise function.
}
\label{tab: slopes}
\end{center}
\end{table}
The values of the slope parameter are given in
Table~\ref{tab: slopes}.
The forms of the Isgur-Wise functions in the four models are given in
\eq{models}.
The values of the slope parameter from our Isgur-Wise function with
type I mass are relatively larger than from that with type II mass.
KS and BSW models give the least $\chi^2$ for the type I and type II
Isgur-Wise functions respectively.
The other models except the linear form give also good fits.

Finally, we fit our Isgur-Wise function of Fig.~\ref{fig: main result}
to the CLEO data by adjusting $\Vcb$ so as to minimize $\chi^2$.
\begin{figure}[hbtp]
\begin{center}
\ \epsfbox{Vcbxi.eps}
\vspace{-15pt}
\caption[]{$\xi(t)\Vcb$ versus $t$.
Our Isgur-Wise functions in Fig.~\ref{fig: main result} are used for
the determination of $\Vcb$.
}
\label{fig: Vcbxi}
\end{center}
\end{figure}
The result is shown in Fig.~\ref{fig: Vcbxi}.
The experimental error bars in Fig.~\ref{fig: Vcbxi} are only
statistical.
The error bar at the kinematical end point is magnified five times,
because of the systematic uncertainty stem from the soft pion
emission.
As is explained our Isgur-Wise function satisfies the condition
$\xi(1)=1$ by construction.
Thus, the intercept at $t=1$ gives $\Vcb$.
We find the absolute values of the Kobayashi-Maskawa flavor mixing
matrix element $\Vcb$ as
\be
\begin{tabular}[h]{c|cc}
\spc type & $\Vcb$ & $\chi^2$ \\
\hline
\spc  I & 0.0454 & 7.47 \\
\spc II & 0.0401 & 0.979 \\
\end{tabular}\label{eq: main result}
\ee
The Isgur-Wise function with type II mass gives a better overall fit
to the CLEO data as is seen from the value of $\chi^2$ in
\eq{main result}.
In other words, we are better to use the type II mass to describe the
experiment.

\section{Next-Leading Corrections of the Heavy-Light Quark Systems}
\reseteqnum

In this section we consider the $1/M_Q$ next leading order corrections
to the various form factors of the exclusive semi-leptonic
$B\rightarrow D^{(*)}$ decays.

\subsection{BS Amplitude}

Let us first expand the normalization condition \num{norm cond} in
terms of $m/M$.
Substituting \eq{rho exp} into \eq{norm cond}, we have
\bea
2 &=& \bbra{\chi_0}\rho_0\kket{\chi_0} \nn
&+& \mom \Bigg\{ \bbra{\chi_0}\rho_1\kket{\chi_0}
 + \bbra{\chi_1}\rho_0\kket{\chi_0}
+ \bbra{\chi_0}\rho_0\kket{\chi_1} \Bigg\} \nn
&+& O\mom^2 ~.
\eea
Thus, the normalization of the BS amplitude is determined order by
order as
\be\ba{ccl}
\ds\mom^0 &:& 2 = \bbra{\chi_0}\rho_0\kket{\chi_0} ~,\mm
\ds\mom^1 &:& 0 = \bbra{\chi_0}\rho_1\kket{\chi_0}
 + 2{\rm Re}\bbra{\chi_0}\rho_0\kket{\chi_1} ~,\mm
\ea\label{eq: norm exp}\ee
where we use $\bbra{\chi_1}\rho_0\kket{\chi_0} =
\bbra{\chi_0}\rho_0\kket{\chi_1}^*$.
The leading order condition is previously used for the leading BS
amplitude.
The next-leading condition fixes the normalization of the next-leading
amplitude $\chi_1$ in terms of the normalized leading BS amplitude.
We notice that the total BS amplitude $\chi=\chi_0+(m/M)\chi_1
+\cdots$
depends on the heavy quark mass although the next-leading BS amplitude
$\chi_1$ does not depend on the heavy quark mass.

Next let us consider the BS equation in the next-leading order.
Projecting the zeroth order BS equation \num{hqexp} onto the
negative energy states, we find
\be
2m \Lm\chi_1(\slp-2m\Lp) = \Lm \slp \chi_0 (\slp-2m\Lp)
- \Lm K\chi_0 ~.\label{eq: nnextl hqbs}
\ee
This equation determines the negative energy part of the
next-leading BS amplitude $\Lm\chi_1$ from the leading BS amplitude
$\chi_0$.

On the other hand, projecting the first order BS equation \num{hqexp}
onto the positive energy states, we obtain
\[
\Lp T^\ze\chi_1 + \Lp T^\on \chi_0
= \Lp K\chi_1 ~.
\]
This gives
\bea
(T_0 - K_0) \Lp\chi_1 &=& -E_0 (\pv-E_0)\chi_0\slv
 + E_1 \chi_0 (\slp-2m\Lp) \nn
&+& \Lp \slp \Lm\chi_1 (\slp-2m\Lp)
+ \Lp K\Lm\chi_1 ~,\label{eq: pnextl hqbs}
\eea
where we use $\chi_1=\Lp\chi_1+\Lm\chi_1$ and put
\be
T_0 \equiv (\pv - E_0)\otimes(\slp-2m\Lp) ~. \label{eq: T0}
\ee
This equation \num{pnextl hqbs} gives the positive energy part of the
next-leading BS amplitude $\Lp\chi_1$ in terms of the leading BS
amplitude $\chi_0$.

First, \eq{pnextl hqbs} determines the next-leading binding energy
$E_1$.
Pre-multiplying \eq{pnextl hqbs} by $\bbra{\chi_0}$, using
$\bbra{\chi_0}(T_0\!-\!K_0)=0$, \eq{norm exp} and the
non-renormalization theorem for the light quark current
\be
\bbra{\chi_0}S_{H0}\otimes\slv\kket{\chi_0} = -1 ~,
\ee
we have
\be
E_1 = - E_0 - \bbra{\chi_0}K\kket{\Lm\chi_1} ~.\label{eq: E1}
\ee
The light quark current $\overline u\gmd u$ is also conserved in our
framework, and the non-renormal\-i\-za\-tion theorem is hold.
This is easily shown along with the same logic in the case of the
heavy quark current $\overline b\gmd b$.

Next let us calculate the amplitude $\Lp\chi_1$ from \eq{pnextl hqbs}.
The operator $(T_0\!-\!K_0)$ has zero eigenvalue and is not
invertible because $(T_0\!-\!K_0)\chi_0=0$ is nothing but
\eq{leading hqbs}.
To avoid the divergence comes from the zero eigenvalue let us divide
$\Lp\chi_1$ by the parallel and perpendicular parts to the $\chi_0$ as
\be
\Lp\chi_1 = c\chi_0 + \chi_{1\perp} ~,
\ee
where
\be
\bbra{\chi_0}\rho_0\kket{\chi_{1\perp}} = 0 ~,\label{eq: perp chi1}
\ee
and  $c$ is the constant determined by the normalization condition of
the BS amplitude.
We note that the leading BS amplitude includes the positive energy
part only; $\chi_0 = \Lp\chi_0$.
{}From \eq{norm exp} we have
\be
c = -\frac{1}{2}\bbra{\chi_0}\rho_1\kket{\chi_0} ~,
\ee
where we use $\bbra{\chi_0}\rho_0\Lm\kket{\chi_1} =
\bbra{\Lm\rho_0\chi_0}1\kket{\chi_1} = 0$.
Let us introduce the parameter $s$ and the self-conjugate operator
$P_s$ defined by
\be
P_s = \rho_0\kket{\chi_0}\bbra{\chi_0}\rho_0 ~.
\ee
This self-conjugate operator has the property
\be
P_s\chi_{1\perp} = 0, \qquad \overline P_s = P_s
{}~.\label{eq: property Ps}
\ee
Then, the LHS of \eq{pnextl hqbs} is rewritten as
\bea
(T_0-K_0)\Lp\chi_1 &=& (T_0-K_0)\chi_{1\perp} \nn
&=& (T_0-K_0-sP_s)\chi_{1\perp} ~.
\eea
We properly chose the parameter $s$ such that the operator
$(T_0-K_0-sP_s)$ becomes invertible and the orthogonality
\eq{perp chi1} or the projection operator $P_s$ satisfies
\eq{property Ps}.
As a result, the next-leading BS equation for $\chi_{1\perp}$
becomes
\bea
(T_0 - K_0-sP_s) \chi_{1\perp} &=& -E_0 (\pv-E_0)\chi_0\slv
 + E_1 \chi_0 (\slp-2m\Lp) \nn
&+& \Lp \slp \Lm\chi_1 (\slp-2m\Lp)
+ \Lp K \Lm\chi_1 ~,\label{eq: pnl hqbs}
\eea

Similarly we can calculate the $n$th-leading BS amplitude and
binding energy after
obtaining the $(n\!-\!1)$th-leading BS amplitude and binding
energy.
This situation reminds us with the bound state perturbation expansion
for the quantum mechanics.

\subsection{Form Factors of the Semi-Leptonic $B$ Decays}

In the heavy quark limit, the semi-leptonic $B\rightarrow D^{(*)}$
decays are expressed in terms of the Isgur-Wise function.
In the $1/M_Q$ next-leading, we have six different kinds of the form
factors.

The semi-leptonic decay $B\rightarrow Dl\overline\nu$ has two form
factors:
\be
\bra{D(q')}\overline c\gmd b\ket{B(q)}
= \xi_+(t)(\hq+\hq')_\mu
+ \xi_-(t)(\hq-\hq')_\mu ~,\label{eq: BgD}
\ee
where $t\equiv\hq'\cdot\hq$.
While the semi-leptonic decay $B\rightarrow D^*l\overline\nu$ has one
plus three form factors:
\be
\bra{D^*(q',\epsilon)}\overline c\gmd b\ket{B(q)}
= i \varepsilon_{\mu\nu\rho\sigma}
\epsilon^\nu\hq'^\rho\hq^\sigma \xi_V(t) ~,\label{eq: BgD*}
\ee
where $\varepsilon^{0123}=-\varepsilon_{0123}=1$ and
\be
\bra{D^*(q',\epsilon)}\overline c\gmd\gf b\ket{B(q)}
= - (t+1)\epsilon_\mu\xi_{A1}(t)
+ \epsilon\!\cdot\!\hq\bigg[
 \hq'_\mu\xi_{A2}(t) + \hq_\mu\xi_{A3}(t) \bigg] ~.\label{eq: Bg5D*}
\ee

The above three matrix elements can be expressed in terms of the BS
amplitudes of the $B$, $D$ and $D^*$ mesons.
Let us use the symbols $B$, $D$ and $D^*$ as the corresponding BS
amplitudes; e.g. the BS amplitude of the $B$ meson is $B(p;q)$.
The matrix elements in \eq{BgD}, \eq{BgD*} and \eq{Bg5D*} become
\be
\bra{D}\overline c\gmd b\ket{B} = N_c \int\trace{
\overline D(p';q')\gmd B(p;q)  S_L(\mmcg)} ~,\label{eq: BgD amp}
\ee
and
\bea
\bra{D^*}\overline c\gmd b\ket{B} &=& N_c \int\trace{
\overline D^*(p';q',\epsilon)\gmd B(p;q)
 S_L(\mmcg) } ~,\label{eq: BgD* amp}\\
\bra{D^*}\overline c\gmd\gf b\ket{B} &=&  N_c \int\trace{
\overline D^*(p';q',\epsilon)\gmd\gf B(p;q)
S_L(\mmcg) } ~.\label{eq: Bg5D* amp}
\eea
In what follows, we use the short hand notation for the integration
\be
\int \equiv \intdpp ~,
\ee
and use the definitions
\bea
\zeta = \frac{M_b}{M_b+m}~,\qquad \eta = \frac{m}{M_b+m} ~,\nn
\zeta' = \frac{M_c}{M_c+m}~,\qquad \eta' = \frac{m}{M_c+m} ~.
\eea
We adopt the tree vertices to the current operators
$\overline c\gmd b$ and $\overline c\gmd\gf b$.
The higher order corrections to the tree vertices will be of order
$O(m/M)^2$.
In this article we study quantities up to $O(m/M)$.

Multiplying $(\hq\pm\hq')^\mu$ by \eq{BgD} and using \eq{BgD amp},
we have
\be
\xi_\pm(t) = \frac{N_c}{2(t\pm 1)}\int
\trace{\overline D(p';q')(\slvp\pm\slv) B(p;q)
 S_L(\mmcg)} ~.\label{eq: xipm ff}
\ee

Let us multiply
$\varepsilon^{\mu\nu\rho\sigma}\epsilon_\nu\hq'_\rho\hq_\sigma$ by
\eq{BgD*} and define $s\equiv\epsilon\!\cdot\!\hq$, at first we have
\[
(t^2-s^2-1)\,\xi_V(t) = \varepsilon^{\mu\nu\rho\sigma}
\bra{D^*}\overline c\gmd b\ket{B}\epsilon_\nu\hq'_\rho\hq_\sigma ~.
\]
We chose the polarization vector $\epsilon_\mu$ so as to satisfy
$s=0$.
Using \eq{BgD* amp} and $i\varepsilon^{\mu\nu\rho\sigma}
\gmd\epsilon_\nu\hq'_\rho\hq_\sigma$ = $\gf[\sle,\slv,\slvp]/6$ where
$[a,b,c] = a[b,c]+b[c,a]+c[a,b]$, we obtain
\be
\xi_V(t) = \frac{N_c}{6(t^2-1)}\int\trace{
\overline D^*(p';q',\epsilon) \:\gf[\sle,\slv,\slvp]\:
B(p;q) S_L(\mmcg) } ~.\label{eq: v ff}
\ee

Next, we also put $s=0$ in order to obtain $\xi_{A1}(t)$.
In this case we see that the matrix element
$\bra{D^*}\overline c\gmd\gf b\ket{B}$ is proportional to the
polarization vector $\epsilon^\mu$ because of \eq{Bg5D*} with
$\epsilon\!\cdot\!\hq=0$.
Multiplying $\epsilon^\mu$ by \eq{BgD* amp} we have
\be
\xi_{A1}(t) = \frac{N_c}{t+1} \int\trace{
\overline D^*(p';q',\epsilon) \sle\gf
B(p;q) S_L(\mmcg)} ~,\label{eq: xiA1 ff}
\ee
with $s=\epsilon\cdot\hq=0$.
Let us consider the general case $s\ne 0$.
Multiplying the three quantities
$(\epsilon^\mu, s(\hq+\hq')^\mu, s(\hq-\hq')^\mu)^T$
by \eq{Bg5D*}, we find
\be
\left(\ba{lll}
(t+1) & 0 & s^2 \\
-s^2(t+1) & s^2(t+1) & s^2(t+1) \\
-s^2(t+1) & s^2(t-1) & -s^2(t-1)
\ea\right)
\left(\ba{c}
\xi_{A1}(t) \\ \xi_{A2}(t) \\ \xi_{A3}(t)
\ea\right) =
\left(\ba{c}
\epsilon^\mu \bra{D^*}\overline c\gmd\gf b\ket{B} \\
s(\hq+\hq')^\mu \bra{D^*}\overline c\gmd\gf b\ket{B} \\
s(\hq-\hq')^\mu \bra{D^*}\overline c\gmd\gf b\ket{B}
\ea\right) ~.\\ \label{eq: xiA123}
\ee
Let us eliminate the polarization vector from \eq{xiA123} by summing
over the polarizations.
Applying the equalities
\be
\sum_\epsilon 1 = 3~,\qquad \sum_\epsilon s^2 = t^2-1 ~,
\ee
to \eq{xiA123}, we finally find
\bea
&&\left(\ba{lll}
3(t+1) & 0 & t^2-1 \\
-(t+1)^2(t-1) & (t+1)^2(t-1) & (t+1)^2(t-1) \\
-(t+1)^2(t-1) & (t+1)(t-1)^2 & -(t+1)(t-1)^2
\ea\right)
\left(\ba{c}
\xi_{A1}(t) \\ \xi_{A2}(t) \\ \xi_{A3}(t)
\ea\right) = ~~~~~~~~~\nn
&&\qquad\qquad\qquad\qquad\qquad\qquad\qquad
\sum_\epsilon\left(\ba{c}
\epsilon^\mu \bra{D^*}\overline c\gmd\gf b\ket{B} \\
s(\hq+\hq')^\mu \bra{D^*}\overline c\gmd\gf b\ket{B} \\
s(\hq-\hq')^\mu \bra{D^*}\overline c\gmd\gf b\ket{B}
\ea\right) ~, \label{eq: xiA123 ff}
\eea
where the RHS is given by \eq{Bg5D* amp}.

Each form factor of $\xi_\pm$, $\xi_V$, $\xi_{A1}$, $\xi_{A2}$,
$\xi_{A3}$, takes the form
\be
f_X\xi_X(t) = \int \trace{ \overline D^{(*)}(p';q') F_X
B(p;q) S_L(\mmc) } ~,
\ee
where $f_X$ is a known function in $t$ and $F_X$ is also a
known function in $N_c$, $v^\mu$, $v'{}^\mu$, $\epsilon^\mu$ and
$\gmu$ for $X=\pm, V, A1, A2, A3$.
We introduce the notation
\be
\Xi(t) = \int \overline D'(p';q') F B(p;q) S_L(\mmc) ~,\label{eq: Xi}
\ee
for $D' = D, D^*$ and $F=F_X$, and we expand this quantity in the
negative power of the heavy quark masses.
We notice that the function $F$ does not depend on the heavy quark
masses.
The BS amplitudes of the $B$ and $D'$ mesons are expanded by
\be
B = \chi_0 + \frac{m}{M_b}\chi_1 ~,\qquad D' = \chi_0' +
\frac{m}{M_c}\chi_1' ~.\label{eq: BD exp}
\ee
The prime denotes the quantity of the $D$ or $D^*$ meson.
So, if we consider the case $B\rightarrow D^*$, the BS amplitude
$\chi'$ is of the vector meson.
The light quark propagator is expanded by
\be
S_L = S_{L0} + \frac{m}{M_b}S_{L0} = S_{L0}' + \frac{m}{M_c}S_{L1}' ~.
\label{eq: SL SL' exp}
\ee
Of course, $S_L'=S_L(p'-\eta'q')$.
Substituting \eqs{BD exp} and \num{SL SL' exp} into \eq{Xi}, we have
\bea
\lefteqn{
\Xi(t) = \int \trace{ \overline\chi_0' F \chi_0 S'_{L0}(\mmc) } }\nn
&& + \frac{m}{M_c}\int \trace{ \overline \chi_1' F \chi_0 S'_{L0}
+ \overline \chi_0' F \chi_0 S'_{L1} }
+ \frac{m}{M_b}\int \trace{ \overline \chi_0' F \chi_1 S'_{L0} }
{}~,\label{eq: Xi exp}
\eea
where we expand $S_L$ by $1/M_c$.
We may expand $S_L$ by $1/M_b$, but there is no essential difference.
The first term of \eq{Xi exp} uses the momentum matching condition
\be
p = p'- \eta' q' = p' - m(v'-v) + mE_0\left( \frac{v'}{M_c} -
\frac{v}{M_b} \right) ~,
\ee
and the other terms are of $O(m/M_Q)$ so that we are enough to use the
momentum matching condition $\mmcl$.
The first term is divided into the leading order quantity which is
calculated under the condition $\mmcl$ and its $1/M_Q$ correction
which is given rise to the difference of the conditions $\mmcl$ and
$\mmc$.
Then, we finally have
\be
\Xi(t) = \Xi^\ze(t) + \Delta\Xi^\ze(t) + \Xi^\on(t) ~,
\ee
with
\bea
\Xi^\ze(t) &=&
 \int\trace{ \overline\chi'_0 F \chi_0 S'_{L0}(\mmcl) }
{}~,\label{eq: Xi0}\\
\Delta\Xi^\ze(t) &=& \int
 \trace{ \overline\chi'_0 F \Delta\chi_0 S'_{L0}(\mmcl) }
\label{eq: DXi0}~,\\
\Xi^\on(t) &=& \frac{m}{M_c}\int\trace{ \overline\chi'_1 F
\chi_0 S'_{L0}(\mmcl) + \overline\chi'_0 F \chi_0
S'_{L1}(\mmcl) } \nn
&& + \frac{m}{M_b}\int\trace{\overline\chi'_0 F \chi_1 S'_{L0}(\mmcl)}
{}~.\label{eq: Xi1}
\eea
The operator $\Delta$ appears in \eq{DXi0} is the differential operator
which causes the $1/M_Q$ order correction
\be
\Delta f(p) = mE_0\left(\frac{v'{}^\mu}{M_c}-\frac{v^\mu}{M_b}\right)
\frac{\partial f(p)}{\partial p^\mu} ~,
\ee
for any function $f(p)$.
Let us write $\chi_0=\Gamma^\p_i\chi_0^i$, i.e. $\chi_0^i=(A_0,B_0)^T$,
then we have
\be
\Delta\chi_0 = \Delta(\Gamma^\p_i\chi_0^i) =
\Delta\Gamma^\p_i\,\chi_0^i + \Gamma^\p_i\,\Delta\chi_0^i ~.
\ee
The quantity $\Delta\chi_0^i$ is evaluated using the leading BS
equation after operating $\Delta$, we find
\be
\Big[ (iu-E_0)\rho_{0ij} - K_{0ij} \Big] \Delta\chi_0^j
= -i\Delta u \rho_{0ij}\chi_0^j -
(iu-E_0)\Delta\rho_{0ij}\chi_0^j + \Delta K_{0ij}\chi_0^j ~.
\ee
This is the inhomogeneous equation.
Solving this equation gives us $\Delta\chi_0^i$.
The $\Delta$ operation on any known function in $u$ and $x$ is easily
calculated, using
\bea
i \Delta u &=& \Delta p\!\cdot\!v =
mE_0\left(\frac{t}{M_c}-\frac{1}{M_b}\right) ~,\nn
2x\Delta x &=& \Delta (-u^2-p^2) = -2u\Delta u
- 2mE_0 \left( \frac{p\cdot v'}{M_c}-\frac{p\cdot v}{M_b} \right) ~.
\eea
So, $\Delta\rho_0(u,x)$ and $\Delta K_0(u,x;v,y)$ are well-defined
quantities.
Let us define
\be
L_{ij} = \trace{ \overline\Gamma'^\p_i(p';v') F \Gamma^\p_j(p;v)
S'_{L0} } ~,
\ee
where the bispinor base $\Gamma^\p_i$ is properly defined for the
pseudoscalar and vector heavy mesons.
Then, we have
\be
\Delta L_{ij} = \trace{ \overline\Gamma'{}^\p_i(p';v') F
\Big(\Delta\Gamma^\p_j(p;v) \Big) S'_{L0} } ~.
\ee
We note that $F$ does not depend on $p$.
Thus after a little algebra, we finally find
\be
\Delta \Xi^\ze(t) = \int \frac{y^2dydv}{8\pi^3} \left[
\chi_0^i(u',x') \Delta L_{ij} \chi_0^j(u,x)
+ \chi_0^i(u',x') L_{ij}\Delta\chi_0^j(v,y) \right]~,
\ee
where we use the fact $\overline\chi_0^i(u',x') = \chi_0^i(u',x')$ if
the binding energy is real.

The detailed expressions for $\Delta\Xi^\ze(t)$ and $\Xi^\on(t)$ are
somewhat complicated, so we omit them.%
\footnote{
The expressions are easily obtained using {\small REDUCE} program.
}

\subsubsection{The Form Factors in the Heavy Quark Limit}

When we take the heavy quark limit $M_Q\rightarrow\infty$ for
$Q=b,c$, the six form factors of the semi-leptonic $B$ decays
are expressed in terms of a single universal function, so called the
Isgur-Wise function $\xi(t)$.
The results are
\bea
&\xi_+(t) = \xi_V(t) = \xi_{A1}(t) = \xi_{A2}(t) = \xi(t)~,& \nn
&\xi_-(t) = \xi_{A3}(t) = 0 ~.&\label{eq: universal}
\eea
Let us prove the above results in our framework.

In the heavy quark limit the BS amplitudes of the $B$, $D$ and $D^*$
mesons are decomposed into two invariant amplitudes $A, B$,
respectively, as
\bea
B_0(p;q) = D_0(p;q) &=& \Lp(A(u,x)+B(u,x)\slbp\;)\,\gf ~,\nn
D_0^*(p;q,\epsilon) &=& \Lp(A(u,x)+B(u,x)\slbp\;)\,\sle ~.
\label{eq: leading BDD*}
\eea
The two invariant amplitudes are common to pseudoscalar
and vector mesons.
This is the consequence of the heavy quark symmetry.
When we use $\Lp'(\slvp-\slv)\Lp$ in \eq{xipm ff}, we immediately find
$\xi_-(t)=0$.
Let us define the form of the Isgur-Wise function as
\be
\xi(t) = \frac{N_c}{t + 1}\int
\trace{\overline D_0(p';q')(\slvp+\slv) B_0(p;q)
 S_{L0}(\mmcl)} ~.\label{eq: isgwis ff}
\ee
Then, $\xi_+(t)$ is identical to the Isgur-Wise function.
In other words, the Isgur-Wise function is defined by $\xi_+(t)$ in
the leading order of $1/M_Q$.

Next we consider the form factors of the decay $B\rightarrow D^*$.
Let us consider the case $s=0$ for evaluating $\xi_V$.
The quantities $\slv$ and $\slvp$ become anti-commute with $\sle$.
{}From \eq{leading BDD*} we have $
\overline D_0^*\gf = - \overline D_0 \sle$.
Eliminating $\sle$, \eq{v ff} is reduced to be
\be
\xi_V(t) = \frac{N_c}{t^2-1}\int\tr\Big[
\overline D_0(p';q',\epsilon) \:[\slv,\slvp]\:
B_0(p;q) S_{L0}(\mmcl)\Big]
{}~.\label{eq: v ff2}
\ee
Using the equalities
\[
\overline D_0[\slv,\slvp]B_0 = (t-1)\overline D_0 B_0
= \frac{t-1}{2}\overline D_0(\slvp+\slv)B_0 ~,
\]
we find $\xi_V(t) = \xi(t)$.

Finally, let us consider the form factors $\xi_X(t)$ ($X=A1,A2,A3$).
We write the RHS of \eq{Bg5D* amp} in the case of the heavy quark
limit as $F_\mu$ for convenience.
{}From \eq{leading BDD*} we see
\[
\overline D_0^*\sle\gf B_0 = \overline D_0 B_0
= \frac{1}{2} \overline D_0 (\hq+\hq') B_0 ~.
\]
Hence from \eq{isgwis ff} we have
\be
2\epsilon^\mu F_\mu=(t+1)\xi(t) ~.\label{eq: F1}
\ee
Similarly, from \eq{leading BDD*} we have
\[
\overline D_0^*\slvp B_0 = \epsilon^\mu \overline D_0 \gmd B_0
\]
Then, from \eq{BgD} with $\xi_+=\xi$ and $\xi_-=0$ we obtain
\be
\hq'{}^\mu F_\mu = \frac{1}{2}\epsilon^\mu(\hq+\hq')_\mu \xi
= \frac{\epsilon\cdot\hq}{2} \xi ~.\label{eq: F2}
\ee
Using \eq{F1} and \eq{F2} we eliminate $\xi$ to make
$(\epsilon\cdot\hq)(\epsilon\cdot F) =(t+1)(\hq'\cdot F)$, and
we have
\beann
0 &=& \{ (\epsilon\cdot\hq)\epsilon_\mu - (t+1)\hq'_\mu \} F^\mu \nn
&=& \{ s\epsilon_\mu - (t+1)\hq'_\mu \}
\{ - \frac{t+1}{2}\epsilon^\mu \xi_{A1}
+ s( \hq'^\mu\xi_{A2} + \hq^\mu\xi_{A3}) \} ~.
\eeann
Thus,
\[
0 = s(t+1)\xi_{A1} - s(t+1)\xi_{A2} + s(s^2-t(t+1))\xi_{A3} ~.
\]
This equality holds for any $s$, then all coefficients of $s^3$ and
$s$ have to vanish:
\[\ba{ccl}
s^3 &:& \xi_{A3} = 0 ~,\\
s   &:& (t+1)(\xi_{A1}-\xi_{A2}) - t(t+1)\xi_{A3} = 0 ~.
\ea\]
Therefore, we finally find
\be
\xi_{A1} = \xi_{A2} ~,\qquad \xi_{A3} = 0 ~.
\ee
It is easy to find $\xi_{A1}=\xi$ from \eq{F1}; i.e. $(t+1)\xi =
2\epsilon\cdot F = (t+1)\xi_{A1}$.

Now, we have shown the results \num{universal}.
As a result, in the heavy quark limit the three matrix elements for
the semi-leptonic decay becomes
\bea
\bra{D(q')}\overline c\gmd b\ket{B(q)} &=& \frac{1}{2}
(\hq+\hq')_\mu \xi(t) ~,\nn
\bra{D^*(q',\epsilon)}\overline c\gmd b\ket{B(q)}
&=& \frac{i}{2} \varepsilon_{\mu\nu\rho\sigma}
\epsilon^\nu\hq'^\rho\hq^\sigma \xi(t) ~,\nn
\bra{D^*(q',\epsilon)}\overline c\gmd\gf b\ket{B(q)}
&=& \frac{1}{2}\left((\epsilon\cdot\hq)\,\hq'_\mu - (t+1)\epsilon_\mu
\right) \xi(t) ~.
\eea

\subsubsection{On the Luke's Theorem}

At the kinematical end point $t=1$, the two form factors $\xi_+(1)$
and $\xi_{A1}(1)$ out of six are protected from the leading order of
the heavy quark symmetry breaking by Luke's
theorem.\cite{Luke,BoydBrahm,FN,CG}
However, we do not seem to hold such theorem in our formalism.
Here we consider the case of $\xi_+(t)$ for simplicity.
The essential point of the theorem is the following:
The form factor $\xi_+(t)$ of the vector current $\overline c \gmd b$
at the kinematical end point takes the form
\be
\xi_+(1) = \xi(1) + \left(\frac{1}{M_c}+\frac{1}{M_b}\right) L_+(1) ~,
\ee
up to $O(1/M_Q)$.
The conservation of the vector current in the limit $M_c=M_b$ implies
the non-renormalization of the form factor $\xi_+(1) = 1$ while the
Isgur-Wise function is normalized to be $\xi(1)=1$, from which it
follows that
\be
L_+(1)=0 ~.
\ee
Namely we obtain, in the general case $M_c\ne M_b$,
\be
\xi_+(1) = 1 + O\left(\frac{1}{M_Q^2}\right) ~.\label{eq: luke}
\ee
It seems to us that this logic does not take account the fact that the
momentum transfer $(q-q')^\mu$ at the kinematical end point becomes of
order $O(M_Q)$ in the case $M_c\ne M_b$.

Let us explain the case in our formulation.
We take the limit $M_b\rightarrow\infty$ just for simplicity and
consider up to order $O(1/M_c)$.
The form factor takes the form
\be
2\xi_+(1) = \bbra{\chi'_0} \slv\otimes S'_{L0} \kket{\chi_0}
+ \frac{m}{M_c} \Big[ \bbra{\chi'_1} \slv\otimes S'_{L0}\kket{\chi_0}
+ \bbra{\chi'_0} \slv\otimes S'_{L1}\kket{\chi_0}
\Big] + O\left(\frac{m}{M_c}\right)^2~,\label{eq: xi+(1)}
\ee
with the momentum matching condition $p^\mu = p'{}^\mu + mE_0
v^\mu/M_c$.
The leading of the first term in \eq{xi+(1)} reduces to the
normalization condition $\bbra{\chi_0}\rho_0\kket{\chi_0}=2$, and the
next-leading of the first term gives
$\bbra{\chi_0}\rho_0\kket{\Delta\chi_0}$ because of
\be
\trace{ \Gamma^\p_i(p';v)\slv \Gamma^\p_j(p;v) S'_{L0} } = 0 ~.
\ee
Thus,
\be
\bbra{\chi'_0} \slv\otimes S'_{L0}\kket{\chi_0} = 2
+ \bbra{\chi_0}\rho_0\kket{\Delta\chi_0} ~.
\ee
The second term of \eq{xi+(1)}, which is already next-leading order,
is evaluated in $p^\mu=p'{}^\mu$.
We put
\be
I = \bbra{\chi_1}\rho_0\kket{\chi_0}
+ \bbra{\chi_0}\slv\otimes S_{L1}\kket{\chi_0} ~,
\ee
which appears in the second term of \eq{xi+(1)}.
The next-leading order part of the normalization condition
\num{norm exp} implies that
\be
\bbra{\chi_1}\rho_0\kket{\chi_0} = - \bbra{\chi_0}\rho_1\kket{\chi_0}
{}~,
\ee
where we use the reality of the matrix element.
Substituting the expression for $\rho_1$ and using the conservation
lows $\bbra{\chi_0} \slv\otimes S_{L0}\kket{\chi_0} = -
\bbra{\chi_0}S_{H0}\otimes \slv \kket{\chi_0} = 2$, we have
\be
I = \bbra{\chi_0} \slv\otimes S_{L1} \kket{\chi_0} ~.
\ee
Using $\Delta S_{L0} = mE_0\slv/M_c$, after a little algebra, we
obtain
\be
2 \xi_+(1) = 2 + \bbra{\chi'_0}\slv\otimes S_{L0}\kket{\chi_0}
- \bbra{\chi'_0}\slv\otimes S'_{L0}\kket{\chi'_0} ~.
\ee
If we ignore the matching condition $p^\mu=p'{}^\mu+mE_0v^\mu/M_c$,
this equation would imply \eq{luke}.

\section{Analyticity of the Isgur-Wise Function}
\reseteqnum

Our prescription \eq{BS itself} for evaluating the BS amplitude with
complex arguments using the BS equation itself does not yield the
analytic BS amplitude, although we use the coupling constant which
does not violate the analyticity of BS amplitude.
The Isgur-Wise function $\xi(t)$ calculated from such BS amplitude is
not analytic.
The origin which violates the analyticity is that the branch cut of
the BS kernel $K_0(u,x;v,y)$ is crossed over by the momentum
integration carried out in \eq{BS itself}.

In this section, we argue that the analyticity of the BS amplitude and
the Isgur-Wise function.
For simplicity, we first study the analyticity of the mass function
$\Sigma(-p^2)$ in the improved ladder approximation.
Next, we generalize the result obtained for the mass function to the
case of the BS amplitude.

The analyticity of the running coupling is very important, but we do
not consider about it here.
It may be plausible to consider as the case of the fixed coupling
constant.
In the following, we concentrate on the structure of the branch cut
stem from the gluon propagator.

\subsection{Analyticity of Mass Function}

The Schwinger-Dyson (SD) equation in the ladder approximation reads
\be
\Sigma(x) = \frac{3C_2}{4\pi}\int_0^\infty ydy
\frac{\alpha}{\max(x,y)} \frac{\Sigma(y)}{y+\Sigma(y)^2} ~.
\label{eq: sdeq}
\ee
Hereafter we replace $3C_2\alpha/(4\pi)$ to $\alpha$.
The function $\max(x,y)$ appears from the gluon propagator after
carrying out the angle integration.
We should notice that the function $\max(x,y)$ is not analytic at all,
but does not spoil the analyticity of the mass function.

The analytic continuation of the SD equation \num{sdeq} should be
performed from outside of the integration.
Then, we obtain the analytic expression
\be
\Sigma(x) = \Bigg[\, \mathop{\int}_{C(0,x)} dy\frac{y}{x}
+ \mathop{\int}_{C(x,\infty)} dy ~\Bigg]
\frac{\alpha\Sigma(y)}{y+\Sigma(y)^2} ~,\label{eq: anal sdeq}
\ee
where $C(a,b)$ represents a contour from $a$ to $b$.

There are two ways to solve \eq{anal sdeq}.
The first one is how we solve by converting \eq{anal sdeq} to the
differential equation:\cite{FK,AB}
\be
\left(\frac{\Sigma'(x)}{(\alpha/x)'}\right)' =
\frac{x\Sigma(x)}{x+\Sigma(x)^2} ~,
\ee
with appropriate boundary conditions.

The second one is to solve the integral equation \num{anal sdeq} as it
is by the iterations.
In order to obtain the mass function along a curve
$C_\Sigma(0,\infty)$, we identify the integration contour in
\eq{anal sdeq} with this curve $C_\Sigma(0,\infty)$ and repeat the
iterations until the form of the mass function converges.
This method will be the most suitable, because it is not always
possible that every type of SD equations is converted to the
differential equation.
Moreover, the second method will be the only one which is applicable
to the BS equation.

It is useful to know the structure of singularities stemmed from the
gluon propagator.
For this purpose we consider the SD equation before carrying out the
angle integration, which is give by
\be
\Sigma(x) = \frac{2}{\pi} \int_0^\infty ydy \int_0^\pi d\theta
\left[ \frac{\sin^2\theta}{x+y-2\sqrt{xy}\ct} \right]
\frac{\alpha\Sigma(y)}{y+\Sigma(y)^2} ~.\label{eq: before sdeq}
\ee
The pole in the integrand of \eq{before sdeq} from the view of $\ct$
yields if
\[
x+y-2\sqrt{xy}\ct = 0
\]
\[\Updownarrow\]
\be
\ct = c(z) \equiv
 \frac{1}{2}\left( \sqrt{z} + \frac{1}{\sqrt{z}} \right) ~,
\qquad z \equiv y/x ~.\label{eq: sing sdeq}
\ee
The singularity in the case of $y=x$, i.e. $\ct=1$, is irrelevant
because of the suppression by the factor $\sin^2\theta$ in
\eq{before sdeq}.
The contour $\ct\in[-1,1]$ of the $\theta$-integration lies in the real
axis.
The $y$-integration begins with at the point $y=0$.
The variable $y$ increases to pass the value $x$, at which $c(y/x)$
becomes the maximum value $1$.
And the variable $y$ goes to infinity.
As far as we solve the SD equation in the real axis $x,y>0$, we have
$\ct \ge 1$, and we never encounter the singularity \num{sing sdeq}.

Let us consider the case when the variables $x$ and $y$ become complex
number, in which the problem is not so simple.
Invoking the Cauchy's theorem, the contours of the $\theta$- and $y$-
integrations can be deformed.
However, the point $c(y/x)$ which moves together with the integration
variable $y$ should not cross the contour of the
$\theta$-integration.
Crossing the contour gives us unexpected contribution from the pole
$\ct=c(y/x)$.

Now, let us consider the singularities on the $y$-plane where the
analyticity of the mass function is investigated.
\begin{figure}[hbtp]
\begin{center}
\ \epsfbox{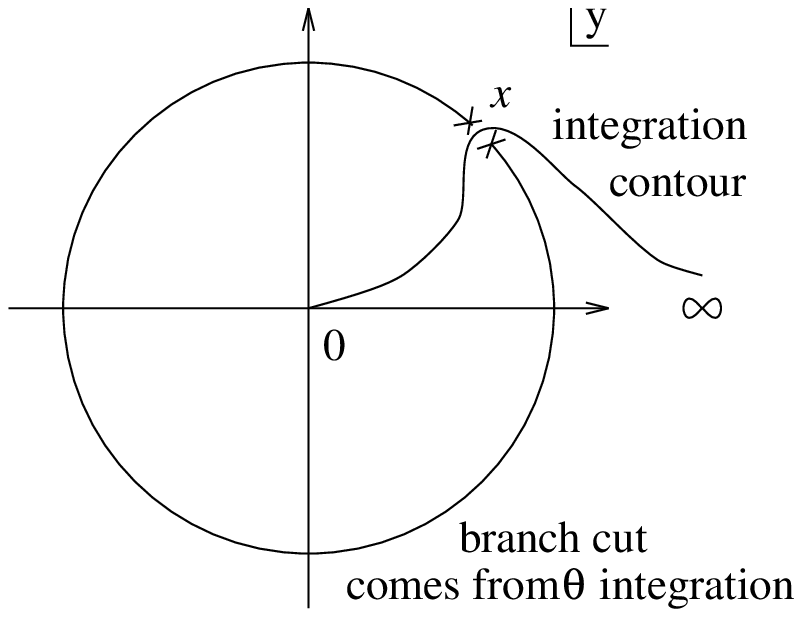}
\vspace{-15pt}
\caption[]{The structure of the branch cut in the SD equation.
}
\label{fig: cut sdeq}
\end{center}
\end{figure}
The set of singularities \sloppy $\{y\in{\bf C} | \ct=c(y/x) \mbox
{ for } \ct  \in C(-1,1)\}$ in $y$-plane yields the branch cut as is
shown in Fig.~\ref{fig: cut sdeq}.
The contour of the $y$-integration is surrounded and pinched by the
branch cut which has the branch point at $x$.
This implies that the contour must involve the point $x$.
Namely, in the case when the variable $x$ becomes complex number, the
contour of the $y$-integration inevitably leaves off the real axis,
and moves to the complex plane.
Therefore, the contour is given by $C(0,x)\cup C(x,\infty)$ as in
Fig.~\ref{fig: cut sdeq} and, as a result, the SD equation takes
the form \num{anal sdeq}.

Finally let us consider the case when the curve on which the mass
functions are requested does not include the origin $x=0$.
For the concreteness, we take the curve as $\{x=(a+iM)^2+b^2~|~0\le a
<\infty \}$ with real constants $b$ and $M$.
The integration region in the SD equation should involve the origin,
so we attach the line $[-M^2+b^2,0]$ to the curve, and use this total
region as the integration region.
Thus, the SD equation becomes
\be
\Sigma(x) = \Bigg[\, \int_0^{-M^2+b^2} dy\frac{y}{x}
+ \int_{-M^2+b^2}^x dy\frac{y}{x}
+ \int_x^\infty dy ~\Bigg]
\frac{\alpha\Sigma(y)}{y+\Sigma(y)^2} ~,\label{eq: complex sd}
\ee
We solve this equation by the iterations, and the desired mass
function is a part of the solution.

\subsection{Analyticity of BS Amplitude}

The BS kernel $K_0(u,x;v,y)$ defined in \eq{K0} consists of four
functions $I_1$, $I_2$, $I_1^{\rm pk}$, $I_2^{\rm pk}$.
The function $I_1$ has logarithmic branch cut.
The other functions $I_2$, $I_1^{\rm pk}$, $I_2^{\rm pk}$ are
expressed in terms of rational functions and $I_1$.
Then, the problem for the analytic continuation is solely given rise
to by the function $I_1$.

Now, let us consider the branch cut structure of $I_1$ which is given
by the $\ct$ integral of the gluon propagator as
\be
I_1 = \int \frac{d\ct}{-(p-k)^2}
= \int_{-1}^1 \frac{d\ct}{(u-v)^2 + x^2+y^2 - 2xy\ct} ~.
\ee
We regard $u$, $v$, $x$ as arbitrary constants for a while.
The singularity in the integrand is given by
\[
(u-v)^2 + x^2+y^2 - 2xy\ct = 0
\]\[\Updownarrow\]
\be
\ct = \frac{z}{2}
+ \frac{1}{2z}\left( 1 + \frac{(u-v)^2}{x^2} \right)
\ge \sqrt{1+\frac{(u-v)^2}{x^2}} ~,
\qquad z \equiv y/x ~.
\ee
In this case the $\ct$ contour lies apart from the singularity,
otherwise $u=v$.
Let us map these singularities to $z$-plane, by which we know the
branch cut in $z$-plane.
\begin{figure}[hbtp]
\begin{center}
\ \epsfbox{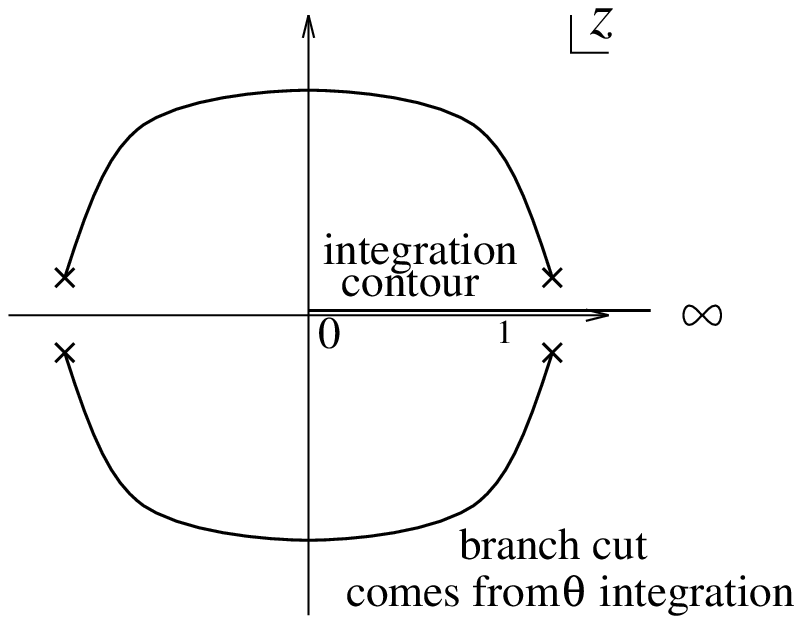}
\vspace{-15pt}
\caption[]{The structure of the branch cut in the BS equation.
}
\label{fig: cut bseq}
\end{center}
\end{figure}
As is shown in Fig.~\ref{fig: cut bseq}, the cut is given by
\be
\left\{ z = \ct \pm i\sqrt{\sin^2\theta + (u-v)^2/x^2 }
{}~\bigg\vert~ 0 \le \theta < \pi \right\} ~,
\ee
for a $\theta$-integration contour.
The $y$-integration region corresponds to the positive part of the
real axis $z\equiv y/x>0$ if $x$ is real.

When $u=v$, the $y$-integration contour is pinched at $z=1$.
This means that the $(v,y)$-integration surface must pass through the
point $(v,y)=(u,x)$.
Other part of the integration surface can be deformed by virtue of the
Cauchy's theorem.
However, we have to obtain all BS amplitudes in the region specified
by $(u,x)$ defined in \eq{ux}.
Thus, we are let to the situation that the BS equation
\num{compo leading hqbs} have to be solved on the complex surface
parameterized by \eq{ux} for the evaluation of the
Isgur-Wise function.

Here we make a comment.
As for the usual BS equation \num{compo leading hqbs}, the integration
region involves the origin $(v,y)=(0,0)$ as a boundary.
However, when we solve the BS equation on the complex plane
$\{(u,x)\}$ parameterized by \eq{ux}, this plane does not involve the
origin in general.
Only in the case $t=1$, the BS equation on the complex plane exactly
reduces to the usual one.
We will be able to overcome this difficulty like the case of the SD
equation on complex plane, but it is rather tedious work.

%  Appendix
\begin{flushleft}
\LARGE Appendix
\end{flushleft}

\appendix

\section{Matrix Representations for the Pion BS Equation}
\label{Matrix Representations for the Pion BS Equation}
\reseteqnum

First we write down the matrix representations for the kinetic part
$T$ and the BS kernel $K$ in the homogeneous BS equation for the pion.
Next we show the relation between the BS amplitude and its truncated
one.

The matrix representations for $T^0$, $T_\mu$ and $K$ are given by the
following:
%  T0
\bea
T^0_{44}(x) &\equiv& \Sp{\overline\gf\: T^0(p) \:\gf} \nn
&=& - (x+\Sigma(x)^2) ~,
\eea
where $\gf$ is regarded as the fourth component of the bispinor base.
The matrix elements of the kinetic parts $T^0$, $T_\mu$ are given by
\bea
T^0_{ij}(x) &\equiv&
 \Sp{\overline\Gamma_i^\mu(p)\:T^0(p)\:\Gamma_{j\mu}(p)} \nn
&=& \Sp{\overline\Gamma_i^\mu(p)\:
(\slp+\Sigma(-p^2))\Gamma_{j\mu}(p)(\slp+\Sigma(-p^2))} \nn
&=& \left(\begin{array}{ccc}
x^2(x+\Sigma(x)^2) & -x(x+\Sigma(x)^2)  &                 0 \\
 -x(x+\Sigma(x)^2) & -2(x-2\Sigma(x)^2) &       6x\Sigma(x) \\
                 0 &        6x\Sigma(x) & 3x(x-\Sigma(x)^2)
\end{array}\right)~.
\eea
%  \lambda
\be
\lambda_i = \Sp{\overline\Gamma_i^\mu\:T_\mu\:\gf} =
- \left(\begin{array}{c}
x(\,2x\Sigma'(x)-\Sigma(x)\,) \\
2(\,2\Sigma(x)-x\Sigma'(x)\,) \\
                       3x
\end{array}\right)~,
\ee
where we define $x=-p^2$ and $y=-k^2$.
The matrix elements of the BS kernel $K$ are found to be
%  K
\bea
K_{44}(x,y) &\equiv&
\frac{1}{8\pi^3}\intdt~\Sp{\overline\gf\:K(p,k)\:\gf} \nn
&=& K_\Sigma(x,y) ~,
\eea
where the SD kernel $K_\Sigma$ is given by \eq{SD kernel}, and
\bea
K_{ij}(x,y) &\equiv&
\frac{1}{8\pi^3}\intdt~
\Sp{\overline\Gamma_i^\mu(p)\:K(p,k)\:\Gamma_{j\mu}(k)} \nn
&=&
\frac{1}{8\pi^3}\intdt~
C_2g^2(p,k)D_{\mu\nu}(p-k)~
\Sp{\overline\Gamma_i^\lambda(p)\:\gmu\Gamma_{j\lambda}(k)\gnu} \nn
&=&
\left(\begin{array}{ccc}
K_{11}(x,y) & K_{12}(x,y)  &     0 \\
K_{21}(x,y) & K_{22}(x,y)  &     0 \\
    0  &       0 & K_{33}(x,y)
\end{array}\right)~,
\eea
where
\be
D_{\mu\nu}(l) = \frac{1}{-l^2}\left( g_{\mu\nu} - \eta(-l^2)
\frac{l_\mu l_\nu}{l^2} \right) ~.
\ee
The functional form of $K_{ij}$ depends on the choice of the running
coupling as well as the gauge function $\eta(-l^2)$.

In the type (I) or (II) choice (see \eq{coupling choice})
of the coupling and in the Landau gauge $\eta(-l^2) \equiv 1$ the
angle integration is carried out analytically,
and we find (see also the appendix in Ref.\cite{ABKMN})
\bea
K_{11}(x,y) &=& F(x,y) \left[\; -3I_1^2(x,y)
+ 2\Big(x^2y^2I_2^1(x,y)-I_2^3(x,y)\Big) \;\right] ~,\nn
K_{12}(x,y) &=& F(x,y) \left[\; 3xI_1(x,y)
- 2\Big(x^2y^2I_2(x,y)-I_2^2(x,y)\Big) \;\right] ~,\nn
K_{21}(x,y) &=& K_{12}(y,x) ~,\nn
K_{22}(x,y) &=& -6F(x,y) I_1(x,y) ~,\nn
K_{33}(x,y) &=& 2F(x,y) \left[\, 3\Big(I_1^1(x,y)+I_2^2(x,y)\Big)
- \Big(x^2y^2I_2(x,y)-I_2^2(x,y)\Big) \,\right] ~,\nn
\eea
where $F(x,y)=C_2g^2(x,y)/(16\pi^2)$.
We introduce the quantities $I_n^m(x,y)$ made of the $\ct$
integration of the gluon propagator.
The first several members are given by
\be
\ba{lllll}
I_0(x,y)      &\equiv& \ds\frac{2}{\pi}\intdt
&=& \ds 1 ~,\\
I_1(x,y)      &\equiv& \ds\frac{2}{\pi}\intdt~\frac{1}{(p-k)_{_E}^2}
&=& \ds\frac{1}{\max(x,y)} ~,\\
I_2(x,y)      &\equiv& \ds\frac{2}{\pi}\intdt~\frac{1}{(p-k)_{_E}^4}
 &=& \ds  \frac{1}{|x-y|\max(x,y)} ~,\\
I_1^1(x,y) &\equiv&
 \ds\frac{2}{\pi}\intdt~\frac{\pdk}{(p-k)_{_E}^2}
 &=& \ds \frac{x+y}{2}I_1(x,y) - \frac{1}{2} ~,\\
I_2^1(x,y) &\equiv&
 \ds\frac{2}{\pi}\intdt~\frac{\pdk}{(p-k)_{_E}^4}
 &=& \ds \frac{x+y}{2}I_2(x,y) - \frac{1}{2}I_1(x,y) ~,
\ea
\ee
where the subscript $E$ denotes the inner product after the Wick
rotation, e.g., $\pdk = p_{_E}\cdot k_{_E} = -(p\cdot k)$.
The higher members are evaluated using the recursion relation
\bea
I_n^m(x,y) &=& \frac{2}{\pi}\intdt~\frac{\pdk^m}{(p-k)_{_E}^{2n}} \nn
&=& \frac{2}{\pi}\intdt~\frac{\pdk^{m-1}}{(p-k)_{_E}^{2n}} ~
\frac{x+y-(p-k)_{_E}^2}{2} \nn
&=& \frac{x+y}{2} I_n^{m-1}(x,y)
- \frac{1}{2}I_{n-1}^{m-1}(x,y) ~,
\eea
for $m,n \ge 1$.

In the consistent gauge (see eq.(\ref{eq: eta diff})) the angle
integration is carried out numerically, and we have
\bea
K_{11}(x,y) &=& \frac{2}{\pi} \intdt~F(z)
   \left[\; -\frac{2+\eta(z)}{z}\pdk^2
+ \frac{2\eta(z)}{z^2} \pdk\pck^2 \;\right] ~,\nn
K_{12}(x,y) &=& \frac{2}{\pi} \intdt~F(z)
   \left[\; \frac{2+\eta(z)}{z}x
- \frac{2\eta(z)}{z^2}\pck^2 \;\right] ~,\nn
K_{21}(x,y) &=& K_{12}(y,x) ~,\nn
K_{22}(x,y) &=& \frac{2}{\pi} \intdt ~F(z)
   \left[\frac{2(\eta(z)-4)}{z} \;\right] ~,\nn
K_{33}(x,y) &=& \frac{2}{\pi} \intdt~2F(z)\eta(z)
   \left[\; \frac{3\pdk}{z} + \frac{3\pdk^2-\pck^2}{z^2} \;\right]~,
\eea
where $z = x+y-2\sqrt{xy}\cos\theta$, $F(z) = C_2g^2(z)/(16\pi^2)$,
$\pdk = \sqrt{xy}\cos\theta$ and
$\pck = \sqrt{x^2y^2-\pdk^2} = \sqrt{xy}\sin\theta$.
It is better to use the Gauss-Legendre integration formula for the
angle integration.

Next we show the relation between the BS amplitude and its truncated
one.
Let us expand the definition \num{inv BS amp} of the (truncated) BS
amplitude in powers of $q_\mu$
\[
\wh\chi^0 + q^\mu \wh\chi_\mu = (T^0 + q^\mu T_\mu)
 (\chi^0 + q^\mu\chi_\mu) + O(p\cdot q)^2 ~,
\]
we obtain
\be
\begin{array}{rl}
O(1) : & \wh\chi^0 = T^0\:\chi^0  ~,\nn
O(q_\mu) : & \wh\chi_\mu =
 T^0\chi_\mu + T_\mu \chi^0 ~.
\end{array} \label{eq: chih compo}
\ee

{}From the above equalities (\ref{eq: chih compo}) we know the
relations between the BS amplitude and its truncated one.
The $O(1)$ equality in eq.(\ref{eq: chih compo}) gives
\be
-\wh S(x) = (x+\Sigma(x)^2)S(x) \qquad \longleftrightarrow \qquad
S(x) = - \frac{\wh S(x)}{x+\Sigma(x)^2} ~.
\ee
It is convenient to introduce the ``metric''
\be
g_{ij}(x) \equiv \Sp{\overline\Gamma_i^\mu \Gamma_{j\mu}} =
\left(\begin{array}{ccc}
-x^2 &  x & 0 \\
   x & -4 & 0 \\
   0 &  0 & 3x
\end{array}\right) ~.
\ee
The $O(q_\mu)$ equality in eq.(\ref{eq: chih compo}) gives
\be
g_{ij}(x)\,\wh\chi^j(x) =
 T^0_{ij}(x)\,\chi^j(x) + \lambda_i(x)S(x) ~.
\ee

%%%%%%%%%%%%%%%%%%%%%%%%%%%%%%%%%%%%%%%%%%%%%%%%%%%%%%%%%%%%%%%%%%%%%%
\newpage

%  Acknowledgements

\begin{center}
\large\bf Acknowledgements
\end{center}
I am grateful to T. Kugo for introducing an interesting field to me.
I would like to thank M. Harada, T. Kugo and M.G. Mitchard for I made
a fruitful works with them.
I would like to thank H. Hata for encouragements.
I also thank all of my colleages in the Department of
Physics in Kyoto University.

% References
%
\newcommand{\PR}[1]{{\it Phys.~Rev.}~{\bf #1}}
\newcommand{\PRL}[1]{{\it Phys.~Rev.~Lett.}~{\bf #1}}
\newcommand{\PRep}[1]{{\it Phys.~Rep.}~{\bf #1}}
\newcommand{\PL}[1]{{\it Phys.~Lett.}~{\bf #1}}
\newcommand{\MPL}[1]{{\it Mod.~Phys.~Lett.}~{\bf #1}}
\newcommand{\NP}[1]{{\it Nucl.~Phys.}~{\bf #1}}
\newcommand{\SJNP}[1]{{\it Sov.~J.~Nucl.~Phys.}~{\bf #1}}
\newcommand{\AP}[1]{{\it Ann.~Phys.}~{\bf #1}}
\newcommand{\PTP}[1]{{\it Prog.~Theor.~Phys.}~{\bf #1}}
\newcommand{\NC}[1]{{\it Nuovo~Cim.}~{\bf #1}}
\newcommand{\ZP}[1]{{\it Z.~Phys.}~{\bf #1}}
\newcommand{\ibid}[1]{{\it ibid.}~{\bf #1}}

\end{document}